\documentclass[aps,prd,twocolumn,floatfix,superscriptaddress]{revtex4-2}
\usepackage{tabularx}
\usepackage{hyperref}
\hypersetup{hypertex=true,
	colorlinks=true,
	linkcolor=blue,
	anchorcolor=blue,
	citecolor=blue}
\usepackage{graphicx}
\usepackage{epstopdf}
\usepackage{amsmath}
\usepackage{amssymb}
\usepackage{comment}
\usepackage{xcolor}
\usepackage{xspace}
\usepackage{multirow}
\usepackage{natbib}
\usepackage{subfigure}	% 子图功能包
\usepackage[english]{babel}
\usepackage{url}
\usepackage{makecell}
\usepackage{ulem} %删除线宏包
\usepackage{gensymb}
\usepackage{multirow}
\usepackage{booktabs}

\begin{document}
	\title{Simultaneous search for multi-target Galactic binary gravitational waves}
	\author{Pin Gao}
	\affiliation{Institute for Frontiers in Astronomy and Astrophysics, Beijing Normal University, Beijing 102206, China}
	\affiliation{School of physics and Astronomy, Beijing Normal University, Beijing 100875, China}
	\affiliation{International Centre for Theoretical Physics Asia-Pacific,
	University of Chinese Academy of Sciences, 100190 Beijing, China}
	\author{Xi-Long Fan}\email{xilong.fan@whu.edu.cn}
	\affiliation{School of Physics and Technology, Wuhan University, Wuhan 430072, China}
	\author{Zhou-Jian Cao}
	\affiliation{Institute for Frontiers in Astronomy and Astrophysics, Beijing Normal University, Beijing 102206, China}
	\affiliation{School of physics and Astronomy, Beijing Normal University, Beijing 100875, China}
	\affiliation{School of Fundamental Physics and Mathematical Sciences, Hangzhou Institute for Advanced Study, UCAS, Hangzhou 310024, China}

	%%%%%%%%%%%%%%%%%%%%%%%%%%%%%%%%%%%%%%%%%%%%%%%%%%%%%%%%%
	\begin{abstract}
	The search for Galactic binary gravitational waves is a critical challenge for future space-based gravitational wave detectors, such as LISA. We propose an innovative approach to simultaneously explore gravitational waves originating from Galactic binaries by developing a new Local Maxima Particle Swarm Optimization (LMPSO) algorithm. 
	This new approach effectively addresses the inaccuracies often associated with signal subtraction contamination, a challenge for traditional iterative subtraction methods, particularly when dealing with low signal-to-noise ratio (SNR) signals (e.g., SNR $<$ 15). We also account for the effects of overlapping signals and degeneracy noise. To demonstrate the effectiveness of our approach, we use residuals from the LISA mock data challenge (LDC1-4), where 10,982 injected sources with SNR $\ge$ 15 have been removed. For the remaining sources with SNR $<$ 15, our method successfully identifies 6,508 signals, yielding a false alarm rate of $\text{FAS}_{0.8} = 36.8\%$. By focusing on a subset of sources—specifically, those with $f > 3$ mHz and those with $f \le 3$ mHz but SNR $\ge 13$—we identify 3,406 signals, with a reduced false alarm rate of $\text{FAS}_{0.8} = 22.5\%$. We further demonstrate that, within the same detection SNR range, our method achieves a comparable or lower $\text{FAS}$ than other existing methods.
	\end{abstract}
	\maketitle

	%%%%%%%%%%%%%%%%%%%%%%%%%%%%%%%%%%%%%%%%%%%%%%%%%%%%%%%%%%
	\section{Introduction}
	In the evolving landscape of ground-based gravitational wave (GW) detectors, recent observations \cite{Smith_2009, Accadia_2011, PhysRevLett.123.231108} underscore advancements in sensitivity and increased detection distances, resulting in a surge of GW events \cite{PhysRevX.11.021053, article1}. However, the inherent frequency limitations (10 Hz to 1000 Hz) of these detectors introduce a bias toward specific sources, such as stellar-mass binary black hole mergers, binary neutron star mergers, and neutron star-black hole mergers \cite{abbott_observation_2016, PhysRevLett.119.161101, Abbott_2021}. Despite their significance, these detections merely scratch the surface of the comprehensive GW spectrum.
	The quest for detecting GWs at alternative frequencies has prompted the exploration of kilohertz GWs \cite{PhysRevD.98.044044, PhysRevD.99.102004}, lunar GW detection \cite{Lunar}, space-based GW detection \cite{article, unknown}, and the utilization of pulsar timing arrays \cite{Agazie1_2023, Afzal2_2023, refId0}. Pulsar timing arrays, in particular, have yielded compelling evidence of potential signals in the extremely low-frequency range \cite{Xu_2023, Reardon_2023, Agazie_2023}.
	Beyond pulsar timing arrays, several space-based GW detection initiatives, including LISA \cite{article, unknown, Gong_2011}, Taiji \cite{hu_taiji_2017, taiji, the_taiji_scientific_collaboration_chinas_2021, taiji1}, and Tianqin \cite{luo_tianqin_2016, tianqin, mei_tianqin_2020, luo_first_2020}, among others, are in preparation, expanding the horizons of gravitational wave research.
	
	The Taiji program \cite{hu_taiji_2017, taiji, the_taiji_scientific_collaboration_chinas_2021, taiji1} and the LISA program \cite{article, unknown, Gong_2011} are positioned to establish space-based GW detectors strategically, with one located 20 degrees ahead and the other 20 degrees behind Earth's orbit. Configured with arm lengths of approximately 3 million kilometers and 2.5 million kilometers, respectively, each detector consists of three independent satellites interconnected by laser links. Operating on the same fundamental principles as ground-based GW detectors, these instruments will measure minute spatial displacements between pairs of satellites for GW detection.
	By evading ground-based seismic noise and possessing substantial arm lengths, space-based GW detectors exhibit sensitivities several orders of magnitude lower than their ground-based counterparts, covering a frequency range from about $10^{-4}$ Hz to $10^{-1}$ Hz. Within this frequency band, an array of GW sources is anticipated, including supermassive black hole mergers, Galactic binaries, and extreme mass ratio inspirals, among others \cite{robson_construction_2019}. Galactic binaries, especially double white dwarf (DWD) binaries, are expected to be abundant. According to current Galactic models, tens of millions to billions of Galactic binaries could emit nearly monochromatic GW signals within the sensitivity frequencies of LISA and Taiji \cite{edlund_simulation_2005}. The prevalence of these signals poses a significant challenge in signal resolution due to potential overlap. Additionally, as Galactic binary gravitational wave signals persist stably throughout the detection period, they have the potential to impact the analysis of other types of signals.
	
	In recent decades, many data processing methods have emerged for analyzing GW signals from Galactic binaries \cite{mohanty_tomographic_2006, crowder_extracting_2007, Prix_2007, crowder_solution_2007, littenberg_detection_2011, blaut_mock_2010, littenberg_global_2020, zhang_resolving_2021, PhysRevD.106.062003, Lu_2023, PhysRevD.106.102004, gao_fast_2023, lackeos_lisa_2023, PhysRevD.106.062003, 2023PhRvD.108j3018S}. 
	Among these algorithms, LDC1-4 data (Radler) were utilized for testing in three different algorithms. The first algorithm employed RJMCMC's global fit approach \cite{littenberg_global_2020, lackeos_lisa_2023}, while the second utilized MCMC and GPU-accelerated techniques \cite{PhysRevD.106.062003, 2023PhRvD.108j3018S}, and the third implemented a PSO-based iterative-subtraction algorithm combined with a cross-validation approach \cite{zhang_resolving_2021, gao_fast_2023}. When applied to the LDC1-4 data, the first algorithm identified 826, 1819, 4356, and 7255 sources over periods of 1.5, 3, 6, and 12 months, respectively \cite{lackeos_lisa_2023}. The second algorithm detected 3407, 6251, and 10363 sources over periods of 6, 12, and 24 months, respectively \cite{2023PhRvD.108j3018S}. The third algorithm identified 10341 sources over 24 months \cite{zhang_resolving_2021}. Additionally, the performance of the third algorithm was also evaluated on 3 and 6 months data, revealing the detection of 3010 and 5418 sources (this evaluation was conducted without cross-validation, which could reduce the false sources but lead to the loss of total numbers) \cite{gao_fast_2023}. Despite the efforts of these algorithms to target sources with SNR $>7$ or SNR $>10$, there are still many sources that remain unsearched, given that there are approximately 27,000 injection sources with SNR $>7$ and 18,000 with SNR $>10$ for two-year detection. This scenario may arise due to several factors. Firstly, the presence of high SNR sources may overshadow the weaker low SNR signals during multi-source fitting, resulting in their neglect. Secondly, as the number of searched signals increases, the potential for inaccurate subtraction contamination from the iterative-subtraction algorithm in the data residual grows, impacting the detection of low SNR signals. Thirdly, the more serious overlap of low SNR signals poses a more significant challenge compared to high SNR signals, particularly in the absence of sufficient priors regarding the number of sources and the approximate parameter ranges. To alleviate these challenges, especially the second, we propose a non iterative-subtraction strategy and evaluate its performance in single-source fitting scenarios. We believe that our method holds advantages in low SNR and low-frequency regions characterized by higher source densities compared to high SNR and high-frequency regions.
	
	Our objective is to accurately identify signals with SNR $<$ 15 (with respect to instrument noise) during the two-year detection period. While the iterative-subtraction strategy proves effective in scenarios where signals have SNR $\ge$ 15 \cite{zhang_resolving_2021, gao_fast_2023}, benefiting from relatively sparse signal density and high SNR, its efficacy gradually diminishes as the SNR decreases.
	We propose a novel search strategy capable of simultaneously detecting multiple Galactic binaries, and implement it using LDC1-4 residual data. Before initiating the search, assuming a method exists to eliminate signals with SNR $\ge$ 15, we begin by subtracting all injection sources with SNR $\ge$ 15 from the LDC1-4 data. Subsequently, we proceed with our method within the residual data. Our algorithm is designed to identify as many local maxima of the $\mathcal{F}$-statistic surpassing a specified threshold as possible in parameter space. Following this, we analyze the results and extract signal parameters from these local maxima.
	
	The structure of this article is organized as follows:
	In Sec.~\ref{sec:Data_description}, we provide an introduction to the gravitational waveform of DWD binaries and the response function of the LISA detector, both based on the LDC1-4 data.
	Sec.~\ref{sec:F_distribution} explores the distribution of the $\mathcal{F}$-statistic when single or multiple signals coexist in LISA data, which leads to the generation of degeneracy noise, and discusses the presence of overlapping signals within the LDC1-4 injection sources.
	Sec.~\ref{sec:Search_method} outlines the LMPSO-CV method utilized for searching the local maxima of the $\mathcal{F}$-statistic, incorporating a specially developed Local Maxima Particle Swarm Optimization (LMPSO) algorithm and a technique for reducing parameter space to prevent redundant searches for the same position in parameter space—referred to as ``create voids" (CV).
	In Sec.~\ref{sec:Comprehensive_analysis}, we elaborate on how to analyze these local maxima with the ``find-real-$\mathcal{F}$-statistic-analysis", focusing on removing a significant amount of degeneracy noise and mitigating instrument noise.
	Sec.~\ref{sec:Search_results} presents the results in terms of the correlation coefficient and SNR, along with the parameter errors and computational time.
	Conclusions are drawn in Sec.~\ref{sec:Conclusion}.
	In Appendix \ref{sec:case_study}, we present the results of the case searches.

	\section{Data description}
	\label{sec:Data_description}
	We utilize the LISA mock data challenge codenamed Radler (LDC1-4) \cite{noauthor_ldc-manual-sangriapdf_nodate}. This dataset encompasses around 30 million binary systems, accompanied by simulated LISA instrument noise, and employs a signal sampling period of 15 seconds. The complete dataset spans a total duration of two years, equivalent to 62,914,560 seconds.
	
	\subsection{Galactic binaries distribution}
	Galactic binary gravitational waves in the source frame are defined as:
	\begin{equation}
		\begin{aligned}
			h_{+}(t)&=\mathcal{A}\left(1+\cos ^{2} \iota\right) \cos \Phi(t), \\ 
			h_{\times}(t)&=-2 \mathcal{A} \cos \iota \sin \Phi(t), \\
			\Phi(t)&=\phi_{0}+2 \pi f t+\pi \dot{f} t^{2},
		\end{aligned}
	\end{equation}
	where $\mathcal{A}$ represents the GW amplitude, $\iota$ represents the Galactic binary orbit to the line of sight from the Solar System barycentric (SSB) origin, $\phi_{0}$ represents the initial phase. $f$ is the GW frequency, and $\dot{f}$ is the first order derivative of $f$. Incorporating the polarization, denoted as $\psi$, from $h_{+}$ and $h_{\times}$, along with two positional parameters—Ecliptic Latitude $\beta$ and Ecliptic Longitude $\lambda$—constitutes a total of eight parameters that characterize a Galactic binary system.
	
	There are 29857650 Galactic binaries in LDC1-4 data, but only 27680 with SNR $>$ 7 in two-year detection. Among these, there are 10982 binaries with SNR $>$ 15, which are sources that most of them can already solved good by current search methods. For the remaining 16698 sources, detection is more difficult because they are mainly concentrated in low ($f < 1\times10^{-3}$ Hz) and medium ($1\times10^{-3}$ Hz $< f <$ $4\times10^{-3}$ Hz) frequencies with a high density and are more susceptible to instrument noise. 
	
	\begin{figure}
		\centering
		\includegraphics[width=1\linewidth]{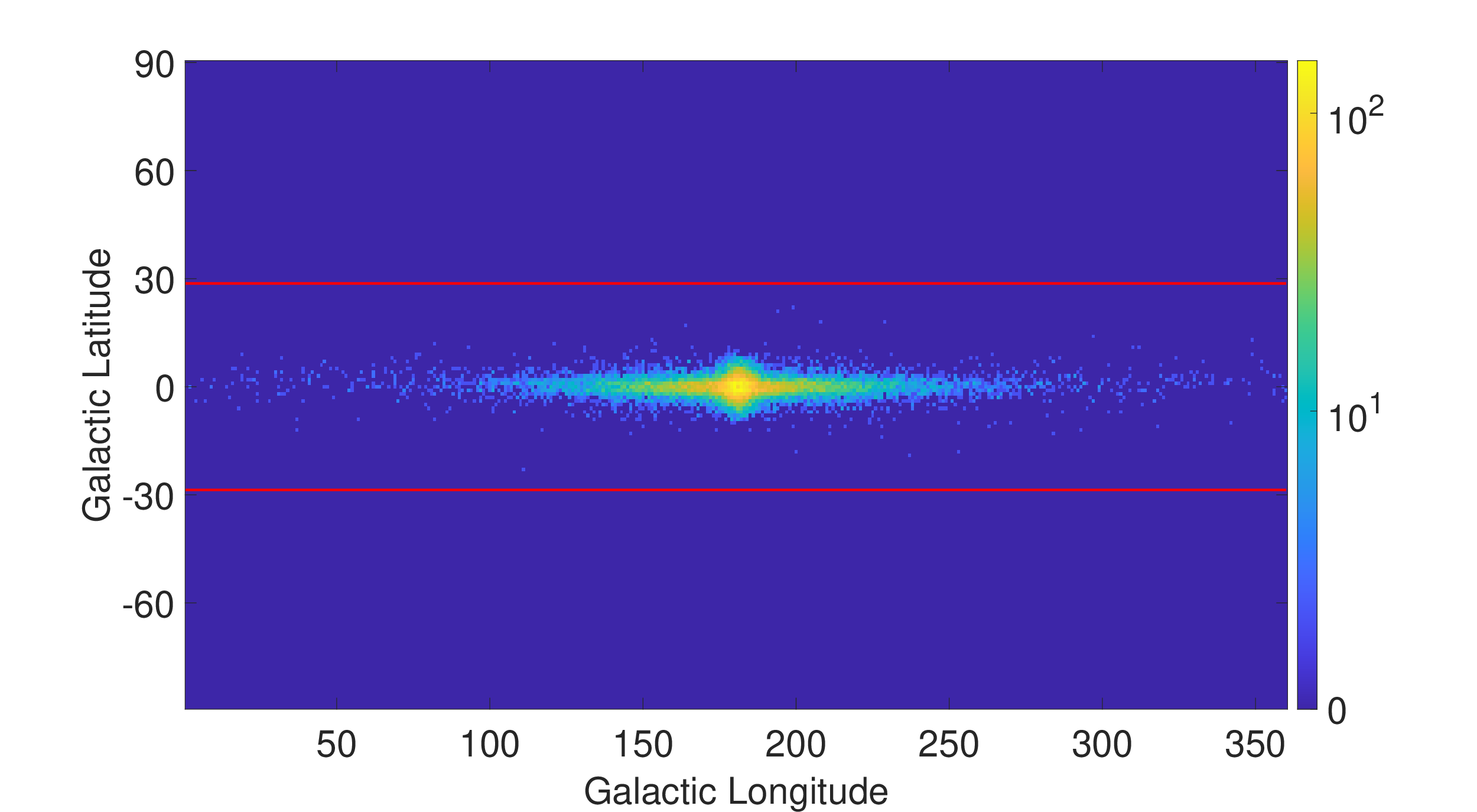}
		\caption{Galactic binaries distribution in the galactic coordinate system of LDC1-4 (SNR $>$ 7). The color bar represents the number of binaries within one square degree on the celestial sphere. The two red lines mark the area where the Galactic Latitude is between -0.5 rad and 0.5 rad ($-28.65 \degree \sim 28.65 \degree$), and there are 27,436 binaries in this area out of 27,680 binaries in total. This prior information will be used in the data analysis (See Sec.~\ref{sec:Restrict_the_Galactic_latitude} for details).}
		\label{fig:galacticdistribution}
	\end{figure}
	\begin{figure}
		\centering
		\includegraphics[width=1\linewidth]{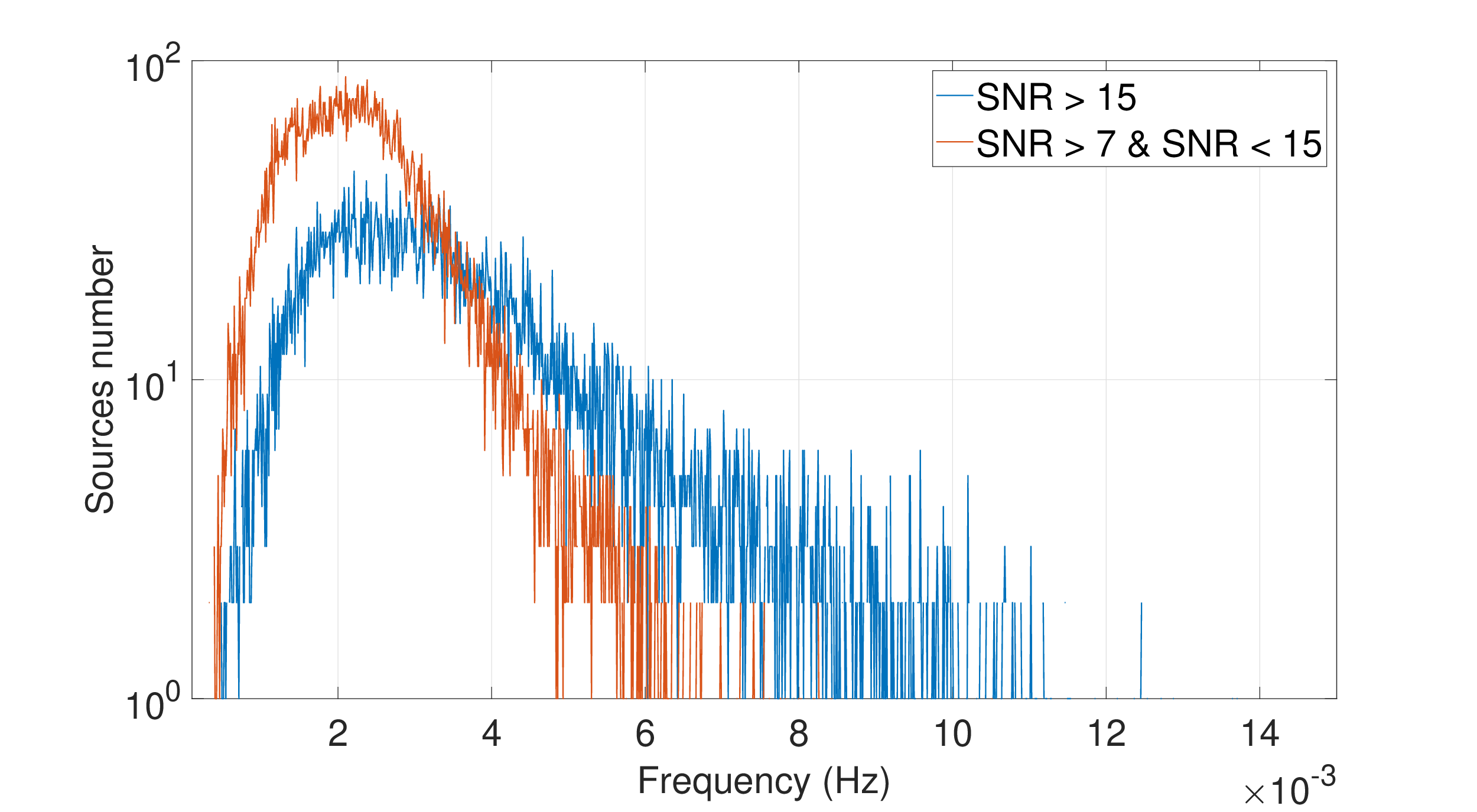}
		\caption{Distribution of Galactic binaries in different frequency bins of the LDC1-4 data. Each bin has a width of 0.01 mHz. We categorize the signals into two groups based on SNR: those with SNR $>$ 15 and those with 7 $<$ SNR $<$ 15. In the figure, we observe that binaries are more concentrated at low frequencies ($f < 1 \times 10^{-3}$ Hz) and medium frequencies ($1 \times 10^{-3}$ Hz $<$ $f$ $<$ $4 \times 10^{-3}$ Hz) when considering signals with 7 $<$ SNR $<$ 15.}
		\label{fig:frequencydistribution}
	\end{figure}
	
	Fig.~\ref{fig:galacticdistribution} illustrates the distribution of binaries in the galactic coordinate system, with the majority congregating on the Galactic Disk. Less than one percent of binaries exhibit a Galactic Latitude outside the range of -0.5 rad to 0.5 rad. In Fig.~\ref{fig:frequencydistribution}, we depict the frequency distribution of binaries. Most binaries with SNR $>$ 15 are readily resolvable using current search methods. In contrast, binaries with 7 $<$ SNR $<$ 15 exhibit higher densities at low frequencies ($f < 1 \times 10^{-3}$ Hz) and medium frequencies ($1 \times 10^{-3}$ Hz $<$ $f$ $<$ $4 \times 10^{-3}$ Hz), posing challenges for effective discrimination. Given our primary focus on signals with low SNR (SNR $<$ 15), we initiate the process by subtracting binary waveforms with SNR $>$ 15 from the raw data.
	
	\subsection{Space-based GW detector response}
	Due to the significant influence of laser phase noise on GW signals, space-based GW detectors employ a technique known as time-delay interferometry (TDI). This method involves the combination of data from different arms to alleviate the impact of laser phase noise. In the LDC1-4 data, the first-generation TDI Michelson combinations, denoted as $X$, $Y$, $Z$, are utilized. This first-generation TDI assumes constant distances between spacecraft, treating them as a rigid body \cite{tinto_time-delay_2002, tinto_time-delay_2014, armstrong_timedelay_1999, tinto_time_2004}. We recombine the three channels $X$, $Y$, $Z$ into $A$, $E$, $T$ channels to ensure that instrument noises are uncorrelated:
	\begin{equation}
		\begin{gathered}
			A=\frac{Z-X}{\sqrt{2}}, \\
			E=\frac{X-2 Y+Z}{\sqrt{6}}, \\
			T=\frac{X+Y+Z}{\sqrt{3}}.
		\end{gathered}
	\end{equation}
	
	The development of a comprehensive response for a space-based gravitational wave detector is a complex task, requiring consideration of intricate details related to the detectors' orbits. In this regard, we rely on the formulation presented in Ref.~\cite{blaut_mock_2010}, a model that has demonstrated high effectiveness in previous studies \cite{blaut_mock_2010, zhang_resolving_2021, gao_fast_2023}.
	
	While the gravitational wave (GW) emitted by a Galactic binary inherently exhibits a monochromatic waveform, the Doppler effect introduces frequency broadening due to the orbital motion of the detector. The Earth, moving at a velocity of approximately 30 km/s around the SSB, shares a similar speed with LISA. As expressed by the Doppler effect formula, $f_1=f\times\frac{c\pm v}{c}$ (where c is the speed of light), the GW frequency undergoes a shift of approximately $0.01\%$ in one frequency direction (this is a maximum, and the specific shift depends on the location of the GW source).

	\section{Parameter degeneracy}
	\label{sec:F_distribution}
	\subsection{$\mathcal{F}$-statistic}
	The Galactic binary GW of the first-generation TDI Michelson observable $X$ is given by a linear combination of the four time-dependent functions $h^{(k)}(t)$.
	\begin{equation}
		h(t)=2 \omega L \sin (\omega L) \sum_{k=1}^4 a^{(k)} h^{(k)}(t),
	\end{equation}
	where $\omega$ is the angular GW frequency and $L$ is the arm length of LISA. $a^{(k)}=\left(a^{(1)}, a^{(2)}, a^{(3)}, a^{(4)}\right)$ is the reparametrization of four extrinsic parameters $\mathcal{A}$, $\phi_{0}$, $\psi$ and $\iota$. $h^{(k)}(t)$ is a matrix of template waveforms and it depends on the four remaining intrinsic parameters: $f$, $\dot{f}$, $\beta$ and $\lambda$.
	
	The $\mathcal{F}$-statistic is proposed to 
	search for monochromatic GWs \cite{jaranowski_data_1998}. 
	Supposing there is a Galactic binary GW in the data, the detector's data in $X$ can be expressed as
	\begin{equation}
		s(t) = n(t) + h(t),
	\end{equation}
	where $n(t)$ is the instrument noise and $h(t)$ is LISA response GW. The log likelihood function has the form
	\begin{equation}
		\ln \Lambda=(s \mid h)-\frac{1}{2}(h \mid h),
	\end{equation}
	where the scalar product $(~\mid~)$ is defined by
	\begin{equation}
		(x \mid y):=4 \operatorname{Re} \int_0^{\infty} \frac{\tilde{x}(f) \tilde{y}^*(f)}{S_h(f)} df.
	\end{equation}
	The symbol $\tilde{}$ represents the Fourier transform, ${}^*$ denotes complex conjugation, and $S_h$ represents the one-sided spectral density of the detector's noise.
	
	It maximizes the log-likelihood function $\ln \Lambda$ concerning parameters $a^{(k)}$, by solving
	\begin{equation}
		\frac{\partial \ln \Lambda}{\partial a^{(k)}}=0.
	\end{equation}
	Then it substitutes the maximum likelihood estimators $\hat{a}^{(k)}$ in  
	$\ln \Lambda$ yielding the reduced log-likelihood function denoted by $\mathcal{F}$-statistic.  
	One finds the $\mathcal{F}$-statistic maxima in the intrinsic parameter space
	and then obtains four extrinsic parameters through analytic expression (see 
	\cite{blaut_mock_2010} for more details). The expectation of the $\mathcal{F}$-statistic is 
	\begin{equation}
		E[2\mathcal{F}]=4+\rho^2,
	\end{equation}
	where $\rho$ is SNR, and 
	\begin{equation}
		\label{eq:SNR}
		\rho=\sqrt{T\sum_{I}(\bar{s}^{I}(\theta)*\bar{s}^{I}(\theta))/S_n^I(f)}.
	\end{equation}
	Here, $\theta$ denotes the parameter set of the source, and $I \in \{A, E, T\}$. The waveform $\bar{s}^I(\theta)$ represents the gravitational wave signal corresponding to $\theta$. Its exact expression can be found in Refs.~\cite{blaut_mock_2010, zhang_resolving_2021}. The sampling interval is set to $T = 15~\mathrm{s}$.  
	Since each Galactic binary gravitational wave signal occupies only a narrow frequency band, the associated noise power spectral density $S_n(f)$ is approximated as constant within that band.  
	In this work, we adopt the instrument noise PSD (the ``Instrument noise" curve in Fig.~\ref{fig:residual}) for SNR calculation and refer to the resulting value as the optimal SNR. Unless otherwise specified, all ``SNR" values reported in this paper refer to the optimal SNR.  
	The only exception is in Sec.~\ref{sec:Compared}, where, for comparison with Refs.~\cite{2023PhRvD.108j3018S, lackeos_lisa_2023}, we convert the optimal SNR into a detection SNR using a noise PSD that includes both instrument noise and the foreground noise.
	
	\subsection{The frequency-position parameter degeneracy}
	\begin{figure}
		\centering
		\includegraphics[width=1\linewidth]{"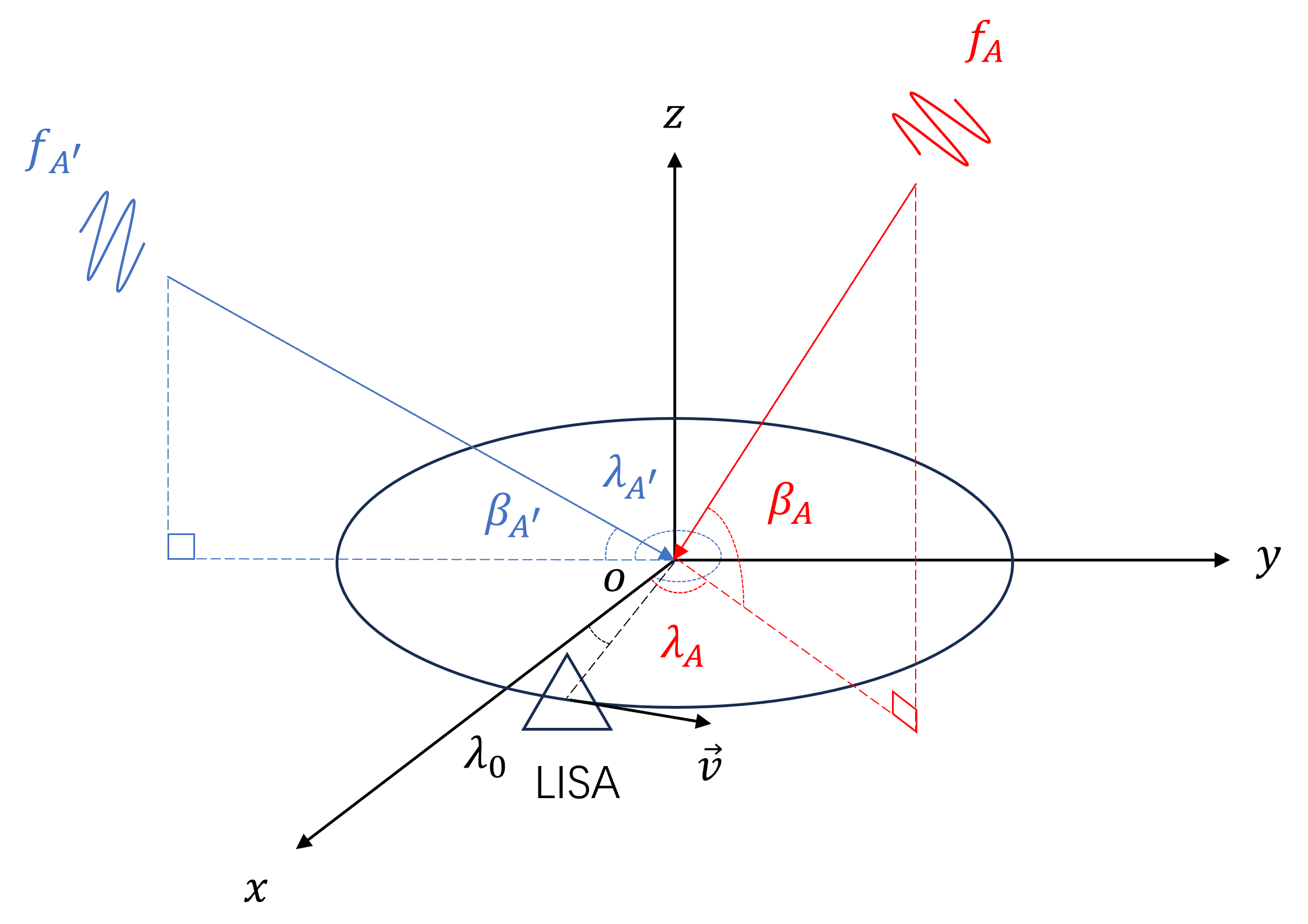"}
		\caption{The $x-y$ coordinate plane represents the ecliptic plane, with the $x$-axis directed toward the vernal equinox. LISA orbits the SSB within the ecliptic plane at a speed denoted by $\vec{\nu}$, and the origin $o$ corresponds to the SSB's position. Within the LISA data, there exists a GW signal originating from binary $A$ with parameters $f_{A}$, $\beta_A$ and $\lambda_A$. Employing a match filtering method, for detection involves assuming that the match waveform corresponds to binary $A^{\prime}$ ($f_{A^{\prime}},~\beta_{A^{\prime}},~\lambda_{A^{\prime}}$).}
		\label{fig:diagram1}
	\end{figure}
	Due to the inherent nature of the $\mathcal{F}$-statistic as a matching filtering method, employing a LISA response GW waveform to correlate with the signal in LISA data yields a $\mathcal{F}$-statistic value that exceeds instrument noise levels if the two waveforms share the same Doppler modulating frequency at the same time. This results in the frequency-position parameter degeneracy.
	In Fig.~\ref{fig:diagram1}, the $x-y$ coordinate plane denotes the ecliptic plane, with the $x$-axis directed toward the vernal equinox. LISA orbits the SSB with a velocity vector $\vec{\nu}$ within the ecliptic plane (this is a simplified model and does not include cartwheel motions of LISA), and the SSB is positioned at the origin ($o$). Assuming binary $A$ is the signal in the LISA data, $f_{A}$ represents binary $A$'s frequency relative to the SSB. Due to the orbital motion of LISA, the frequency $f_{A}$, modulated by the Doppler effect $f_{A_{\text{do}}}$ in the LISA data, can be expressed as
	\begin{equation}
		f_{A_{\text{do}}} = f_{A} \frac{c+\nu_{A}}{c},
	\end{equation}
	where 
	\begin{equation}
		\nu_A = |\vec{\nu}| \sin{(\lambda_A-\lambda_0)} \cos{\beta_A}
	\end{equation}
	represents the velocity of LISA relative to binary $A$, while $\lambda_0$ denotes LISA's Ecliptic Longitude at the given time. Additionally, $\lambda_A$ represents the Ecliptic Longitude of binary $A$, and $\beta_A$ represents its Ecliptic Latitude.
	
	Assuming we use binary $A^{\prime}$ as the match waveform, due to the Doppler effect of LISA orbit motion, $f_{A^{\prime}_{\text{do}}}$ is its LISA response frequency
	\begin{equation}
		f_{A^{\prime}_{\text{do}}} = f_{A^{\prime}} \frac{c+\nu_{A^{\prime}}}{c},
	\end{equation}
	where 
	\begin{equation}
		\nu_{A^{\prime}} = |\vec{\nu}| \sin{(\lambda_{A^{\prime}}-\lambda_0)} \cos{\beta_{A^{\prime}}}.
	\end{equation}
	$\lambda_A^{\prime}$ and $\beta_A^{\prime}$ are the Ecliptic Longitude and Ecliptic Latitude of binary $A^{\prime}$, respectively.
	Making these two modulated frequencies equal ($f_{A_{\text{do}}} = f_{A^{\prime}_{\text{do}}}$), we can get the equation
	\begin{equation}
		\begin{gathered}
			f_{A}[c + |\vec{\nu}| \sin{(\lambda_A-\lambda_0)} \cos{\beta_A}] \\= f_{A^{\prime}}[c + |\vec{\nu}| \sin{(\lambda_{A^{\prime}}-\lambda_0)} \cos{\beta_{A^{\prime}}}]. \\
		\end{gathered}
	\end{equation}
	Let $\theta = \lambda_A-\lambda_0$. Therefore, $\lambda_{A^{\prime}}-\lambda_0 = \theta + (\lambda_{A^{\prime}} - \lambda_A)$. The above equation can be
	\begin{equation}
		\begin{gathered}
			\label{con:3}
			f_{A^{\prime}} = f_{A}\frac{c + |\vec{\nu}| \sin{\theta} \cos{\beta_A}}{c + |\vec{\nu}| \sin{(\theta + \lambda_{A^{\prime}}-\lambda_A)} \cos{\beta_{A^{\prime}}}}.
		\end{gathered}
	\end{equation}
	This equation pertains to the presence of a GW signal from binary $A$ in LISA data. When employing the $\mathcal{F}$-statistic to detect the signal, it describes the distribution of the $\mathcal{F}$-statistic in the parameter space.
	
	Considering the aforementioned equation, if the LISA data contains a signal with the frequency $f_A$ relative to the SSB, the maximum frequency $f_{A^{\prime}_{\text{max}}}$ in the parameter degeneracy is
	\begin{equation}
		\label{con:1}
		f_{A^{\prime}_{\text{max}}} = f_{A}\frac{c + |\vec{\nu}|}{c - |\vec{\nu}|} \approx 1.0002 \times f_{A}
	\end{equation}
	when $\theta=\frac{\pi}{2}$, $\lambda_{A^{\prime}}-\lambda_A=\pi$, and $\beta_A=\beta_{A^{\prime}}=0$.
	Similarly, the minimum frequency $f_{A^{\prime}_{\text{min}}}$ in the parameter degeneracy is
	\begin{equation}
		\label{con:2}
		f_{A^{\prime}_{\text{min}}} = f_{A}\frac{c - |\vec{\nu}|}{c + |\vec{\nu}|} \approx 0.9998 \times f_{A}
	\end{equation}
	when $\theta=-\frac{\pi}{2}$, $\lambda_{A^{\prime}}-\lambda_A=\pi$, and $\beta_A=\beta_{A^{\prime}}=0$.
	It determines the frequency range $[f_{A^{\prime}_{\text{min}}},~f_{A^{\prime}_{\text{max}}}]$ of the parameter degeneracy generated by the GW signal which has frequency $f_A$ relative to the SSB.
	Eq.~\ref{con:1} and \ref{con:2} show the frequency range of the parameter degeneracy generated by the $f_A$ signal is proportional to $f_A$ itself. This results in higher $f_A$ exhibiting a more extensive distribution of the $\mathcal{F}$-statistic, while lower $f_A$ presents a narrower frequency distribution for the $\mathcal{F}$-statistic. 
	
\begin{figure*}
		\centering  %图片全局居中
		\subfigbottomskip=2pt %两行子图之间的行间距
		\subfigcapskip=0pt %设置子图与子标题之间的距离
		\subfigure[\label{fig:degeneracy_noise1}]{
			\includegraphics[width=0.48\linewidth]{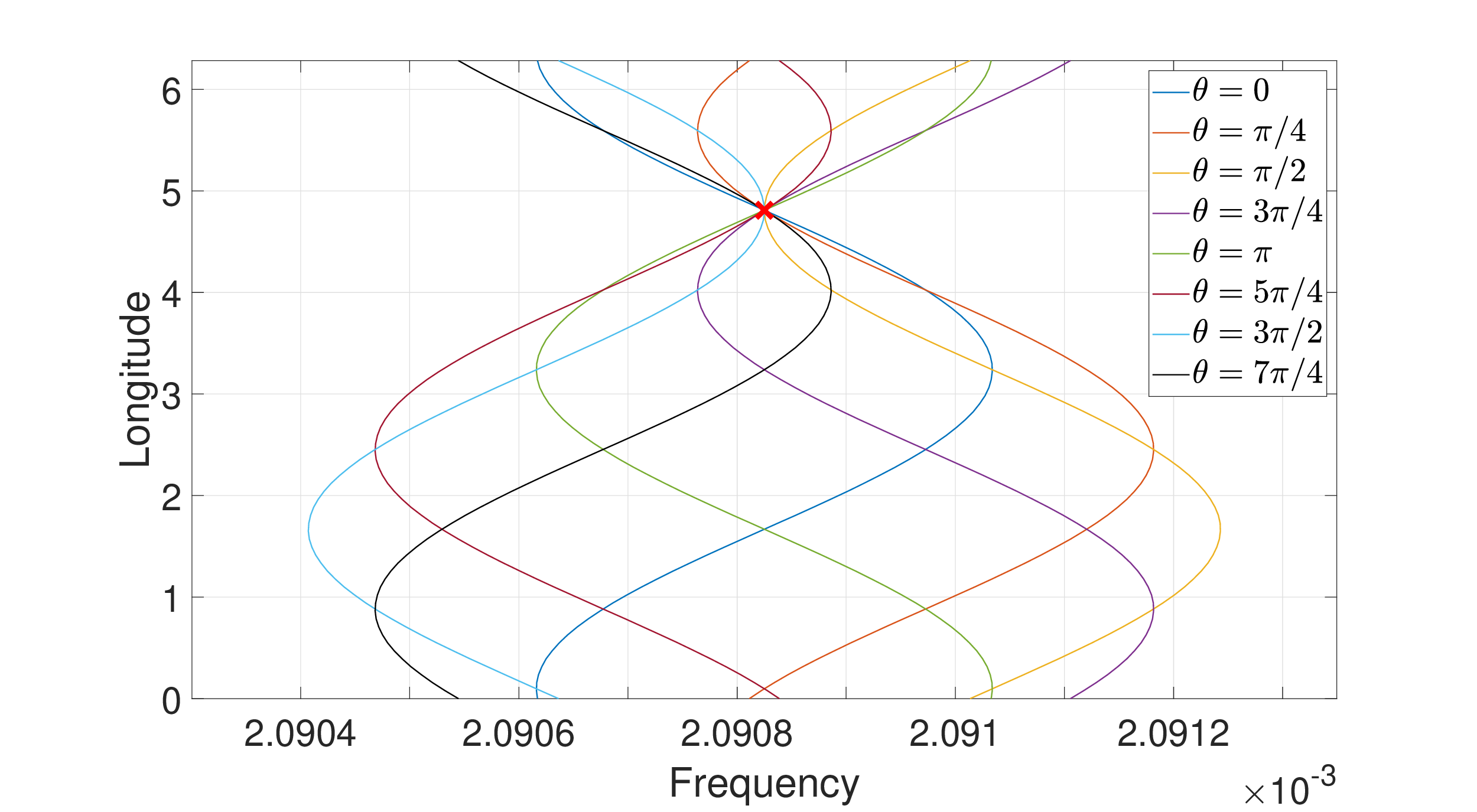}}
		\subfigure[\label{fig:degeneracy_noise2}]{
			\includegraphics[width=0.48\linewidth]{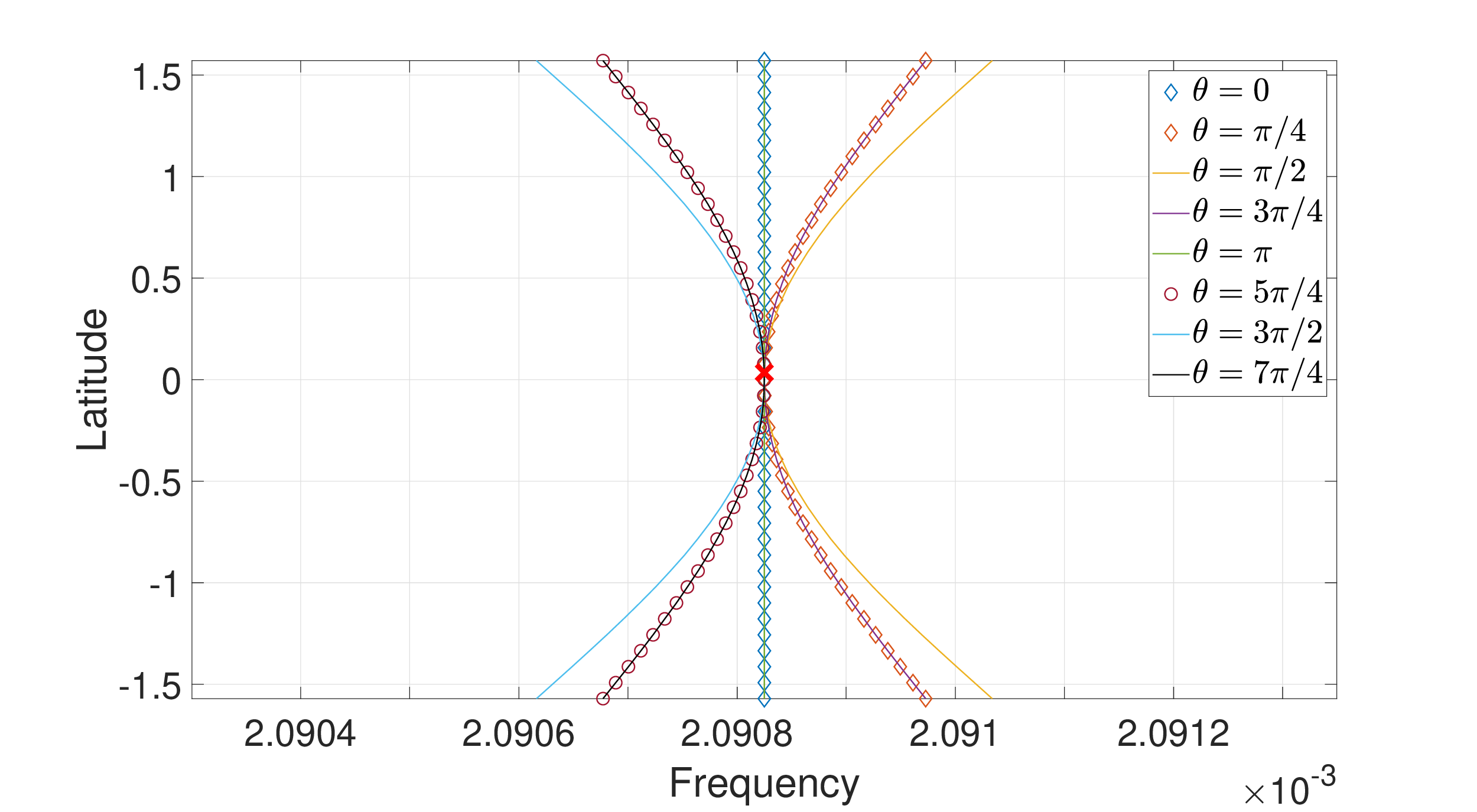}}\\
		\caption{\label{fig:degeneracy_noise}
			The parameter degeneracy from binary $A$ is derived according to Eq.~\ref{con:3}. The parameters of binary $A$—$f_{A},~\beta_A,~\lambda_A$—are known, and the position of binary $A$ is denoted by red cross marks. In Fig.~\ref{fig:degeneracy_noise1}, $\beta_A^{\prime}=\beta_A$, and in Fig.~\ref{fig:degeneracy_noise2}, $\lambda_A^{\prime}=\lambda_A$. The horizontal axis represents the frequency ($f_{A^{\prime}}$) of the matched waveform, binary $A^{\prime}$, while the vertical axis represents its Ecliptic Longitude or Ecliptic Latitude ($\lambda_A^{\prime}$ or $\beta_A^{\prime}$). $\theta$ denotes the angle between LISA and binary $A$, with different $\theta$ values representing various positions of LISA relative to binary $A$.}
	\end{figure*}

	We selected a set of parameters from the LDC1-4 injection sources catalog and assigned them to binary $A$, with the specific parameter values detailed in Tab.~\ref{table:2} (Appendix \ref{sec:case_study}). The parameter degeneracy originating from binary $A$ is plotted in Fig.~\ref{fig:degeneracy_noise}. In Fig.~\ref{fig:degeneracy_noise1}, $\beta_{A^{\prime}}=\beta_A$, and in Fig.~\ref{fig:degeneracy_noise2}, $\lambda_{A^{\prime}}=\lambda_A$. We utilized a spatial resolution of $\pi/40$ radians in the figures. $\theta$ represents the angle between LISA and binary $A$, providing a representation of LISA at various positions in its orbit.
	The parameter degeneracy for binary $A$ spans a frequency range approximately from $0.9998 \times f_{A}$ to $1.0002 \times f_{A}$, corresponding to [$2.090406\times10^{-3}$ Hz, $2.091243\times10^{-3}$ Hz].
	
	\subsection{The degeneracy noise of signal}
	Resulting from the above calculation, the $\mathcal{F}$-statistic for an individual source does not exhibit an isolated peak in parameter space \cite{cornish_lisa_2003, cornish_lisa_2003_1}. A sequence of similar elevated $\mathcal{F}$-statistic value emerges in the parameter space, challenging the maximum likelihood-based statistic searching algorithm, these high values of the $\mathcal{F}$-statistic, apart from the real binary, are termed ``degeneracy noise".
	
	\subsubsection{Individual signal}
\begin{figure*}
		\centering
		\includegraphics[width=1\linewidth]{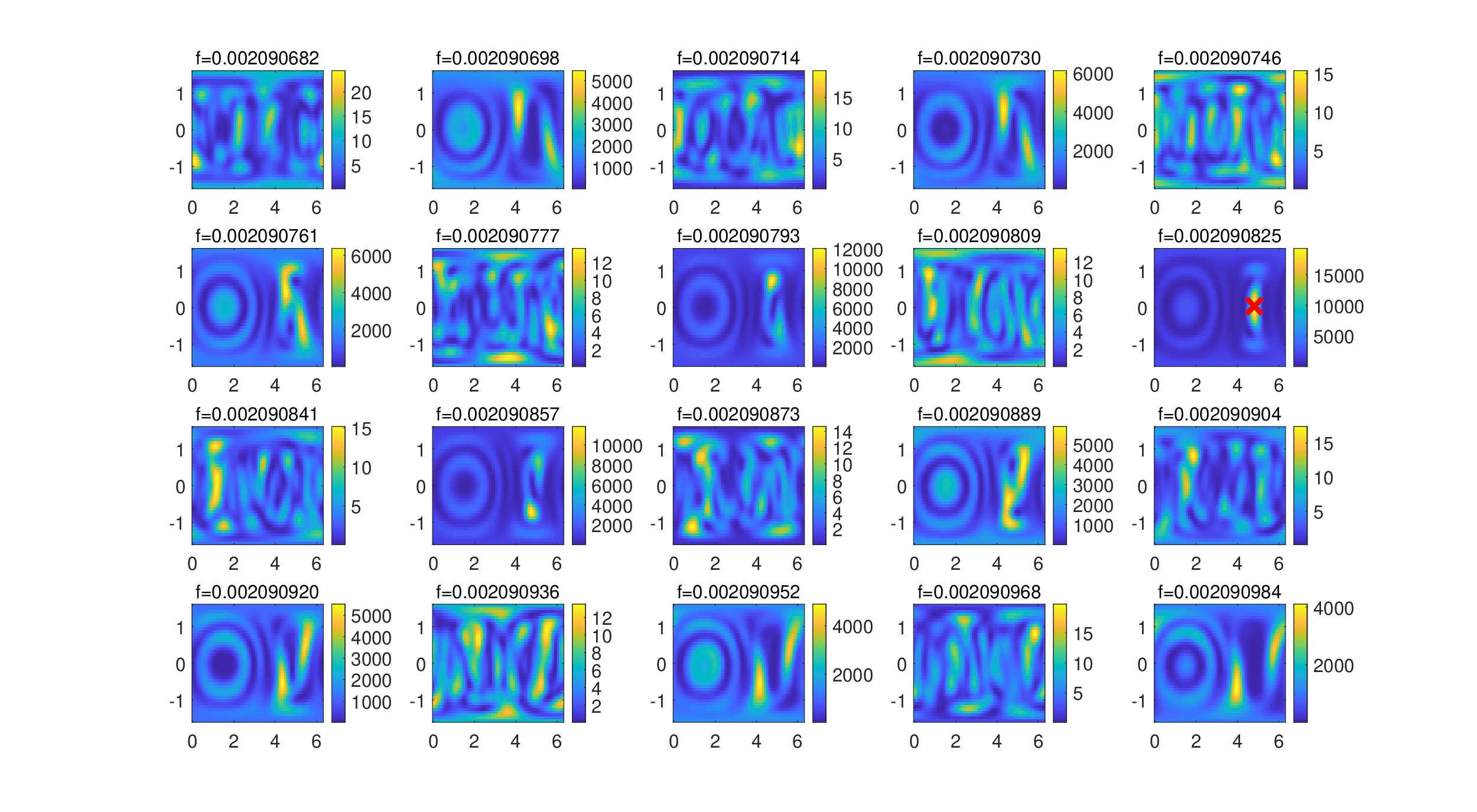}
		\caption{Slices of the $\mathcal{F}$-statistic in three-dimensional parameter space for binary $A$. The parameter position of binary $A$ is marked in the subfigure (2, 5) with a red cross. The frequency intervals between two adjacent subfigures correspond to the minimum frequency resolution $\mathrm{d}f=1.59\times10^{-8}$ Hz. In each subfigure, the horizontal coordinate represents Ecliptic Longitude $\lambda$, and the vertical coordinate is Ecliptic Latitude $\beta$. The figure does not encompass the entire frequency range of the degeneracy noise distribution.}
		\label{fig:fstatisticdistribution}
	\end{figure*}

	We generate a singular LISA response GW signal with the parameters of binary $A$ outlined in Tab.~\ref{table:2}. Subsequently, we introduce instrument noise to acquire mock data and compute the parameter degeneracy for data instances with diverse parameter positions. The intentional selection of binary $A$, distinguished by a high SNR, aims to minimize the influence of instrument noise.
	In Fig.~\ref{fig:fstatisticdistribution}, the distribution of degeneracy noise of binary $A$ across three parameter space ($f$, $\beta$, and $\lambda$) is illustrated. Each subfigure has a spatial resolution of $\pi/40$ radians. The colorbar indicates the $\mathcal{F}$-statistic value calculated at the specific parameter position. The red cross in the (2, 5) subfigure marks the real parameter position of binary $A$. The horizontal coordinate represents Ecliptic Longitude $\lambda$, while the vertical coordinate signifies Ecliptic Latitude $\beta$. The frequency interval between two adjacent subfigures corresponds to the minimum frequency resolution $\mathrm{d}f =1.59\times10^{-8}$ Hz for the two-year detection.
	From the figure, it is evident that, in addition to the ``primary peak" of the $\mathcal{F}$-statistic at the position of binary $A$, numerous ``secondary peaks" with elevated $\mathcal{F}$-statistic values emerge. These secondary peaks represent the degeneracy noise mentioned earlier.
	
\begin{figure*}
		\centering  %图片全局居中
		\subfigbottomskip=2pt %两行子图之间的行间距
		\subfigcapskip=0pt %设置子图与子标题之间的距离
		\subfigure[\label{fig:distribution4}]{
			\includegraphics[width=0.48\linewidth]{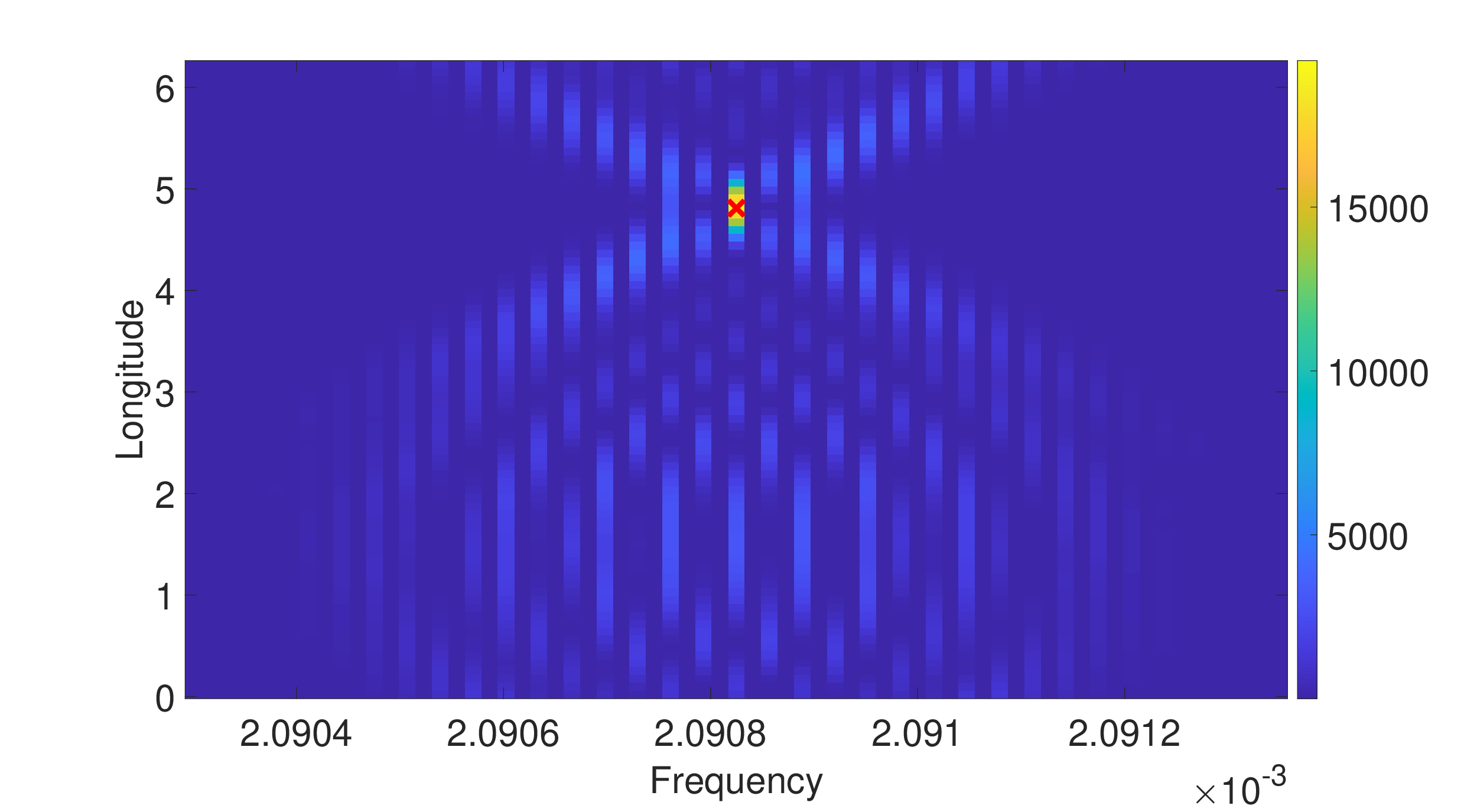}}
		\subfigure[\label{fig:distribution5}]{
			\includegraphics[width=0.48\linewidth]{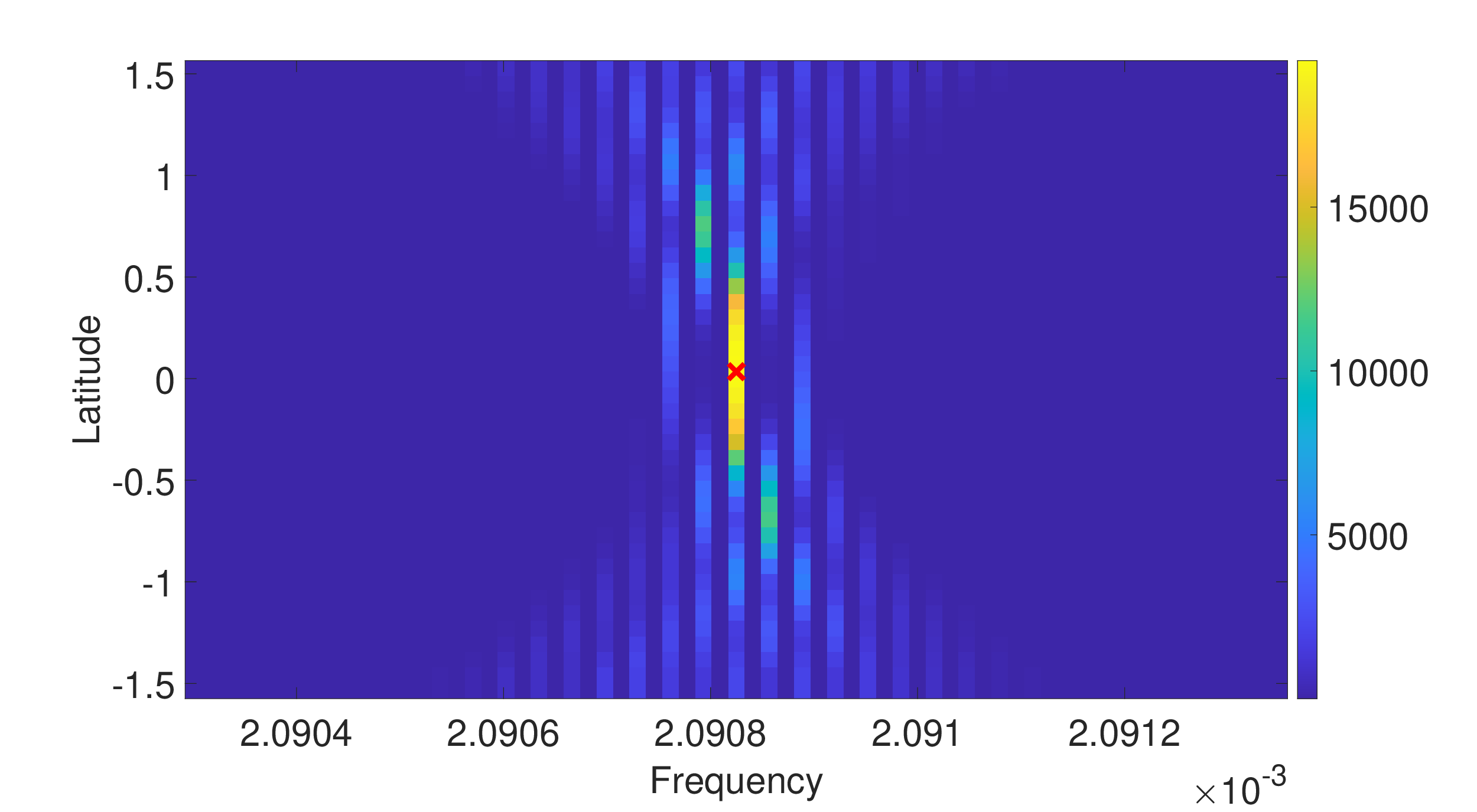}}\\
		\caption{\label{fig:distribution41}The red cross in the figures represent the position of binary $A$, while the colorbar indicate the $\mathcal{F}$-statistic value. The vertical coordinate corresponds to Ecliptic Longitude $\lambda$ or Ecliptic Latitude $\beta$, respectively. The parameter degeneracy for binary $A$ spans a frequency range approximately from $0.9998 \times f_{A}$ to $1.0002 \times f_{A}$, corresponding to [$2.090406\times10^{-3}$ Hz,~$2.091243\times10^{-3}$ Hz].}
	\end{figure*}

	Fig.~\ref{fig:distribution4} and Fig.~\ref{fig:distribution5} illustrate the distribution of the $\mathcal{F}$-statistic in two-dimensional parameter space. The frequency resolution is $\mathrm{d}f$, and the Ecliptic Latitude $\beta$ or the Ecliptic Longitude $\lambda$ resolution is $\pi/40$ radians. Parameters not displayed in the figures remain constant and correspond to binary $A$.
	The parameter degeneracy in both figures (Fig.~\ref{fig:distribution4} and Fig.~\ref{fig:distribution5}) demonstrates excellent consistency with the theoretical predictions illustrated in Fig.~\ref{fig:degeneracy_noise1} and Fig.~\ref{fig:degeneracy_noise2}.
	
	\subsubsection{Two overlapping signals}
	Signals that overlap in parameter space introduce complexity to the distribution of the $\mathcal{F}$-statistic. The proximity of two signals in parameter space can lead to interference, manifesting as a high $\mathcal{F}$-statistic peak at an incorrect parameter position, diminishing the value at the actual binary's parameter position. In scenarios with substantial differences in SNR between the two binaries, the impact of the low SNR signal on the high SNR signal might not be conspicuous, and the low SNR signal could be discerned by subtracting the high SNR signal. However, when overlapping signals exhibit similar SNR, their interference may yield a more pronounced $\mathcal{F}$-statistic peak at an incorrect parameter position. This complicates the iterative-subtraction strategy, as subtracting an incorrect waveform becomes a possibility. In our method, as long as the $\mathcal{F}$-statistic peak at the real binary's parameter position does not disappear or deviate significantly, the removal of degeneracy noise, detailed in Sec.~\ref{sec:Comprehensive_analysis}, can effectively eliminate the erroneous $\mathcal{F}$-statistic peak.
	
\begin{figure*}
		\centering  %图片全局居中
		\subfigbottomskip=2pt %两行子图之间的行间距
		\subfigcapskip=0pt %设置子图与子标题之间的距离
		\subfigure[\label{fig:F_statistic_distribution6}]{
			\includegraphics[width=0.48\linewidth]{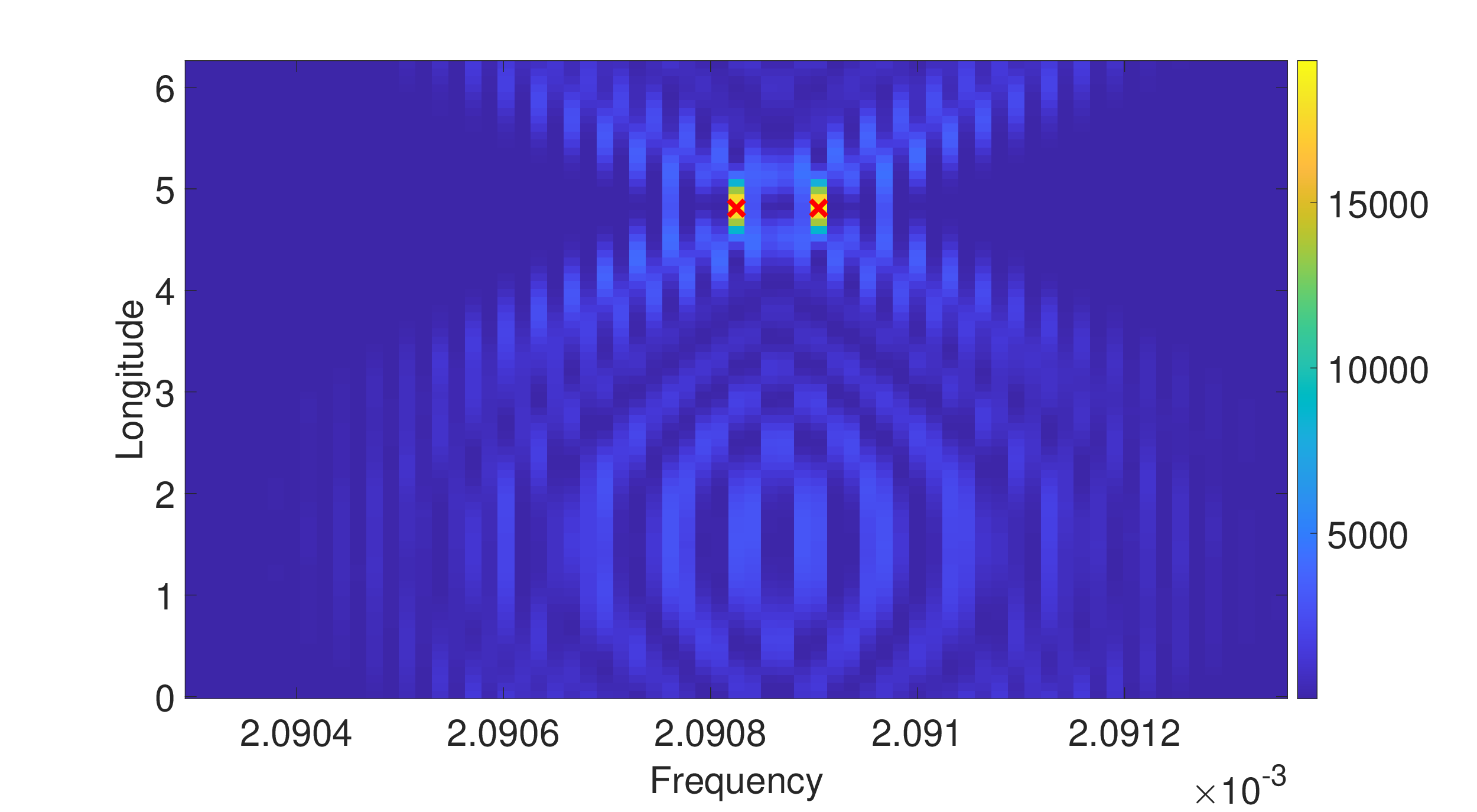}}
		\subfigure[\label{fig:F_statistic_distribution7}]{
			\includegraphics[width=0.48\linewidth]{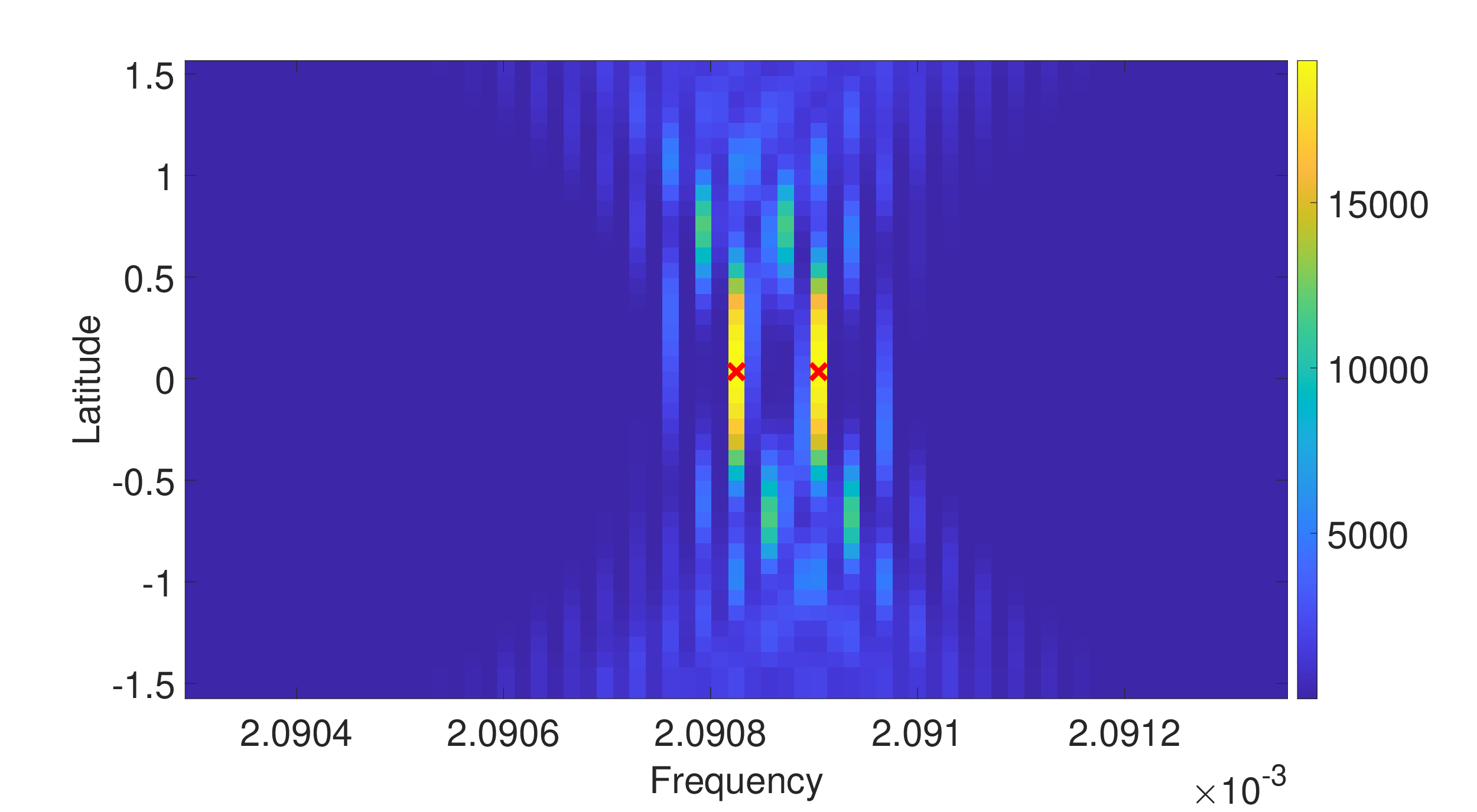}}\\
		\subfigure[\label{fig:F_statistic_distribution8}]{
			\includegraphics[width=0.48\linewidth]{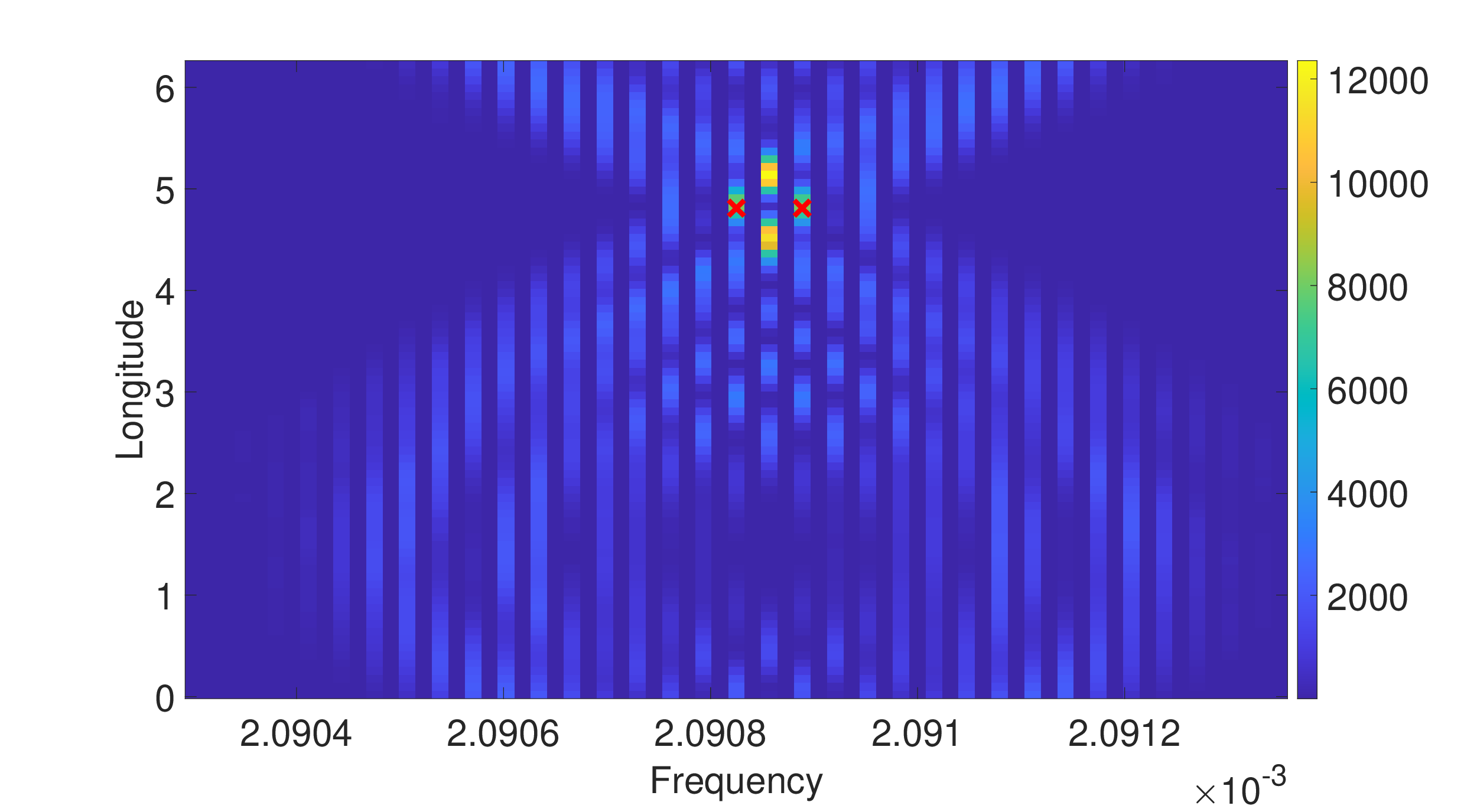}}
		\subfigure[\label{fig:F_statistic_distribution9}]{
			\includegraphics[width=0.48\linewidth]{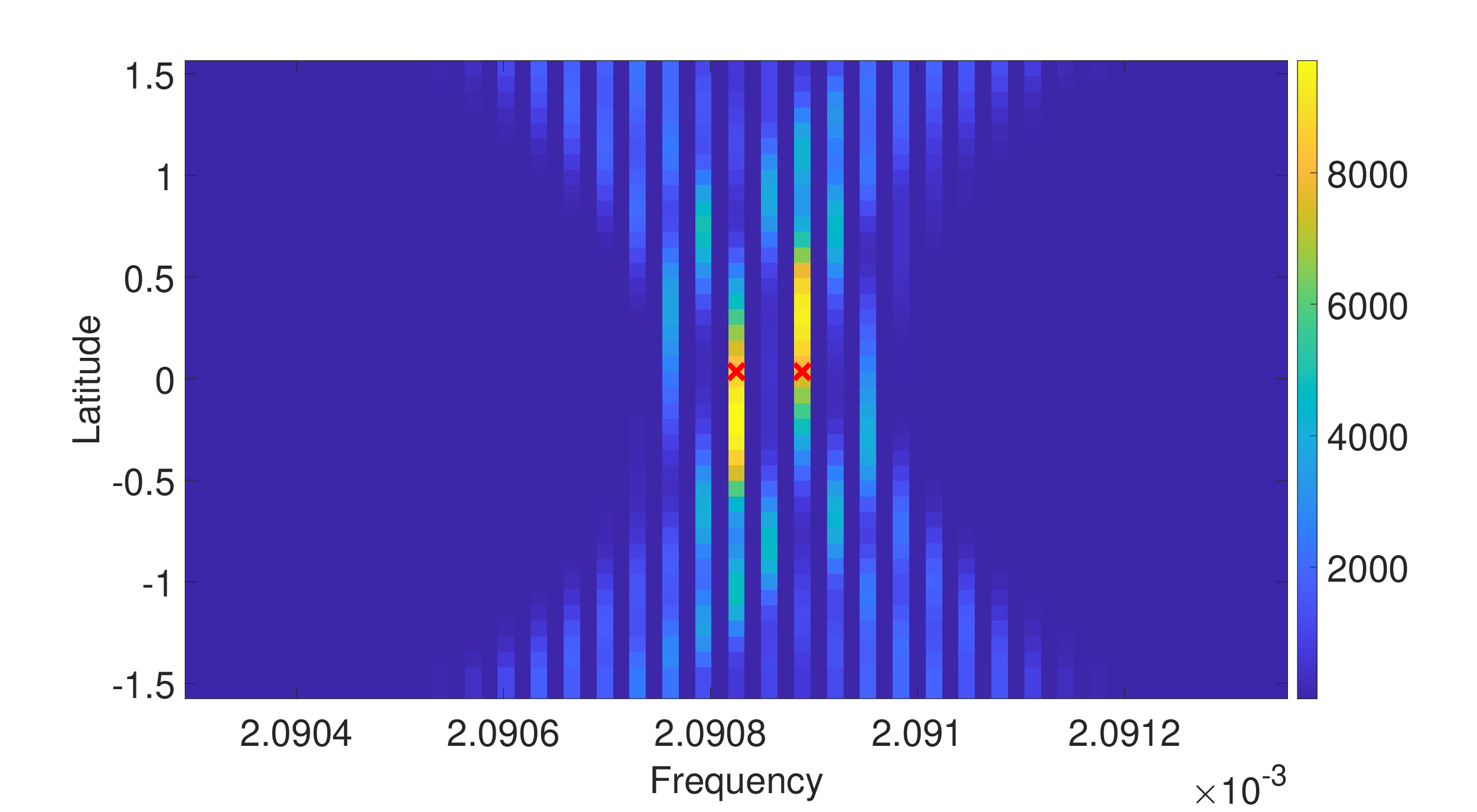}}
		\caption{\label{fig:F_statistic_distribution}The parameter degeneracy when two signals overlap. The frequency interval is $5 \times \mathrm{d}f$ for Fig.~\ref{fig:F_statistic_distribution6} and Fig.~\ref{fig:F_statistic_distribution7}, and $4 \times \mathrm{d}f$ for Fig.~\ref{fig:F_statistic_distribution8} and Fig.~\ref{fig:F_statistic_distribution9}. The red crosses marked the real positions of the two binaries.}
	\end{figure*}

	Building upon binary $A$, we increment its frequency by integer multiples of the minimum frequency resolution $\mathrm{d}f$ to create binary $B$ ($f_B=f_A+ \Delta f=f_A+ N  \times\mathrm{d}f,~N\in Z$). Fig.~\ref{fig:F_statistic_distribution} illustrates the distribution of the $\mathcal{F}$-statistic when two signals coexist. The frequency of binary $B$ in Fig.~\ref{fig:F_statistic_distribution6} and Fig.~\ref{fig:F_statistic_distribution7} is increased $5 \times \mathrm{d}f$ relative to binary $A$, with red crosses denoting the accurate positions of both binaries—binary $A$ on the left and binary $B$ on the right. When the frequency difference $\Delta f$ between two signals is $\ge 5\times \mathrm{d}f$, the ``primary peaks" of the $\mathcal{F}$-statistic for these two signals do not significantly influence each other, each behaving akin to a single source.
	However, for the binary $B$ in Fig.~\ref{fig:F_statistic_distribution8} and Fig.~\ref{fig:F_statistic_distribution9}, where the frequency of binary $B$ is incremented by $4 \times \mathrm{d}f$ relative to binary $A$, the overlap of signals causes a noticeable deviation in their $\mathcal{F}$-statistic behavior compared to the single signal scenario. In Fig.~\ref{fig:F_statistic_distribution8}, two maximal $\mathcal{F}$-statistic peaks emerge outside the parameter positions of binaries $A$ and $B$, substantially reducing the $\mathcal{F}$-statistic values at the true binary positions. Similarly, in Fig.~\ref{fig:F_statistic_distribution9}, the maximal $\mathcal{F}$-statistic peaks deviate from the accurate positions of the binaries. Consequently, in the iterative-subtraction method, incorrect waveforms of signals $A$ and $B$ would be subtracted, and the maximal $\mathcal{F}$-statistic peak located at a wrong parameter position for overlapping signals is referred to as ``the overlapping degeneracy noise".
	
\begin{figure}
		\centering
		\includegraphics[width=1\linewidth]{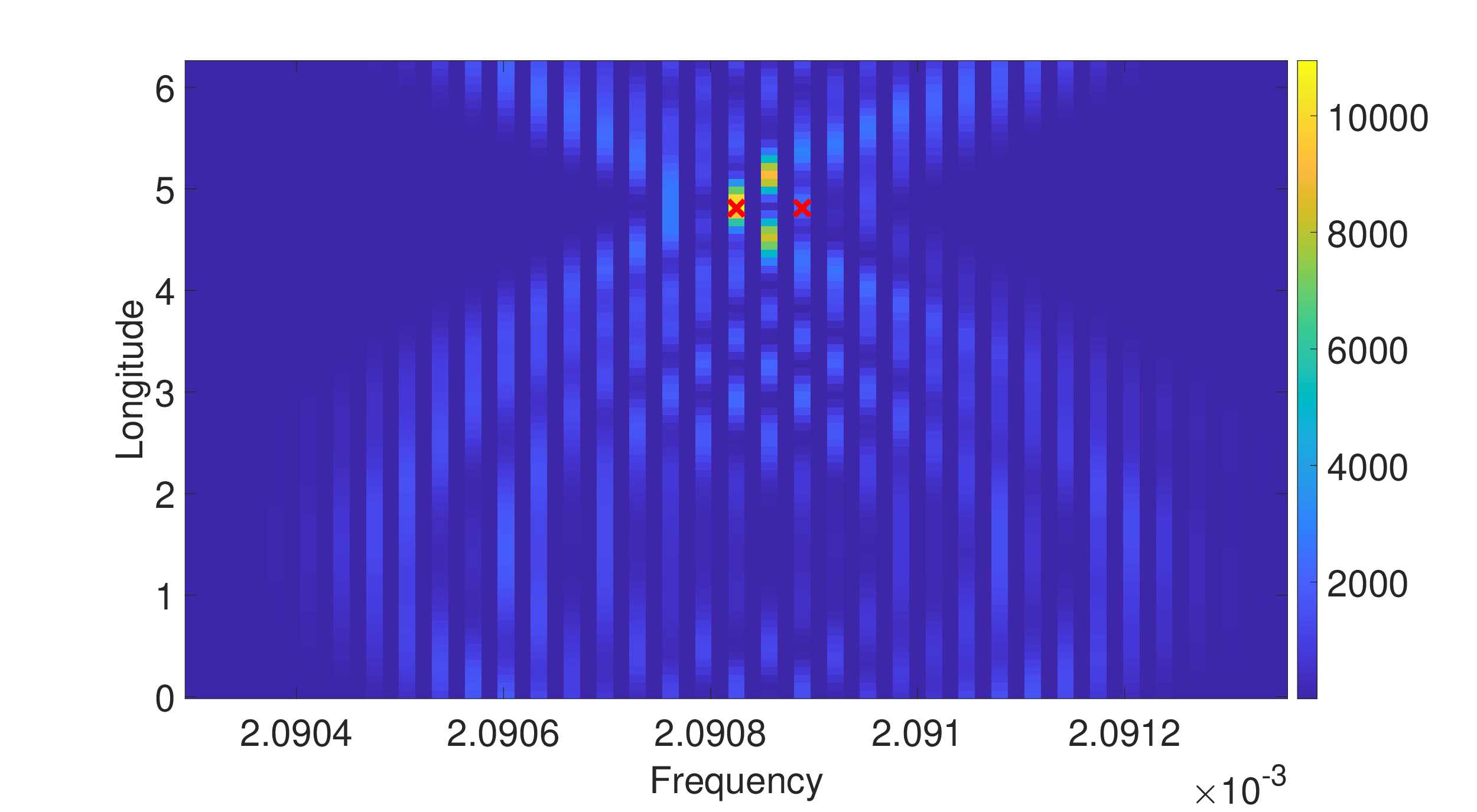}
		\caption{Binary $A$ remains unchanged, while the amplitude $\mathcal{A}$ of binary $B$ is reduced to 0.7 times its initial value, leading to a proportional decrease in SNR, and the $\mathcal{F}$-statistic value is approximately halved.}
		\label{fig:fstatisticdistribution15}
	\end{figure}

	When the two signals possess different SNR, the challenge posed by overlapping signals is mitigated. In Fig.~\ref{fig:fstatisticdistribution15}, we adjust the amplitude $\mathcal{A}$ of binary $B$ to $70\%$ of the value used in Fig.~\ref{fig:F_statistic_distribution8}, resulting in a proportionate decrease in SNR, and the $\mathcal{F}$-statistic value is roughly halved. In this scenario, even though the overlapping signals continue to influence their $\mathcal{F}$-statistic at the two binary positions, the $\mathcal{F}$-statistic value of the two maximal degeneracy noise does not surpass that of binary $A$. Consequently, binary $A$ maintains the maximal $\mathcal{F}$-statistic value in the parameter space.
	This situation proves advantageous for both our method and the traditional iterative-subtraction method.
	
	\subsubsection{Three overlapping signals}
\begin{figure*}
		\centering  %图片全局居中
		\subfigbottomskip=2pt %两行子图之间的行间距
		\subfigcapskip=0pt %设置子图与子标题之间的距离
		\subfigure[\label{fig:F_statistic_distribution11}]{
			\includegraphics[width=0.48\linewidth]{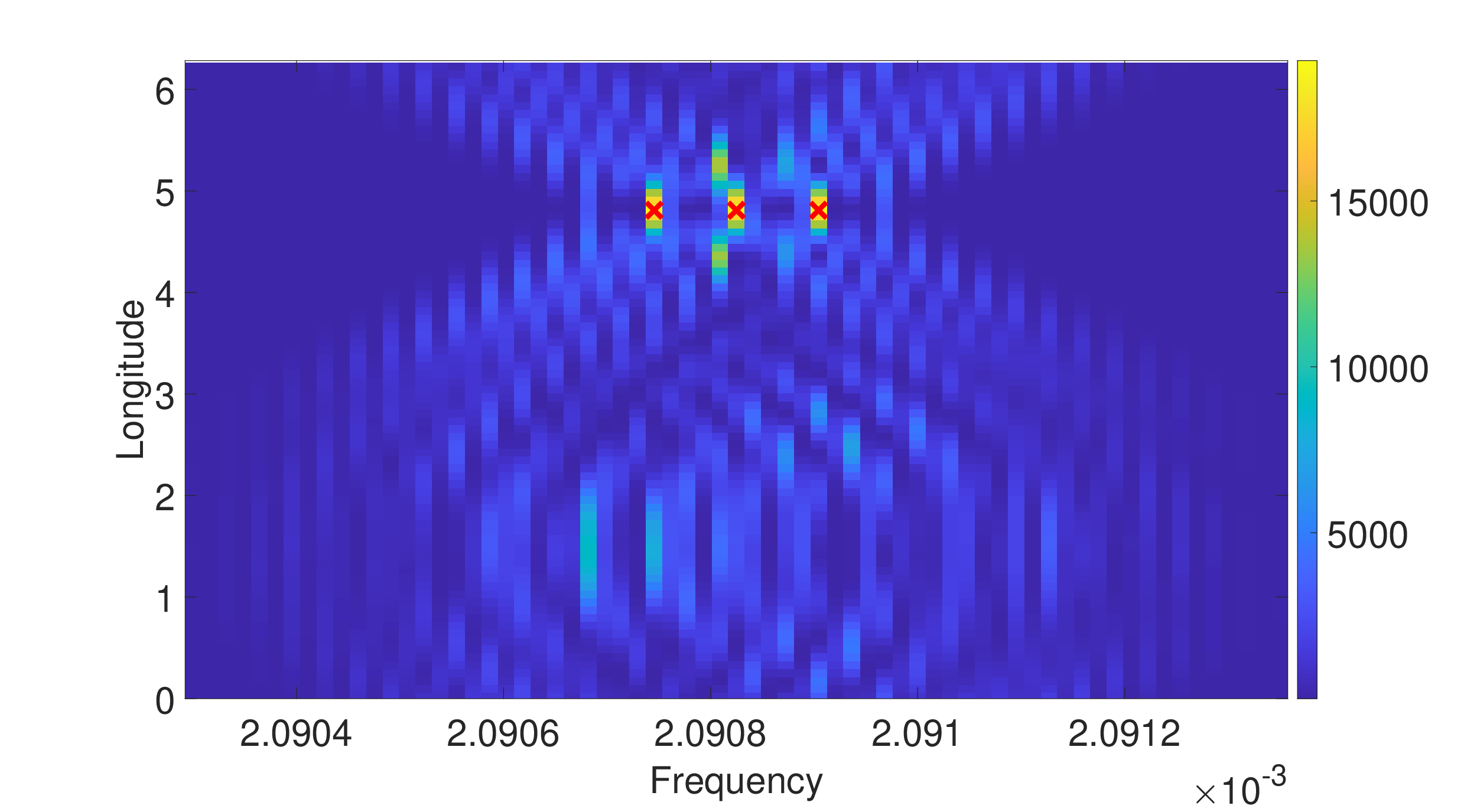}}
		\subfigure[\label{fig:F_statistic_distribution12}]{
			\includegraphics[width=0.48\linewidth]{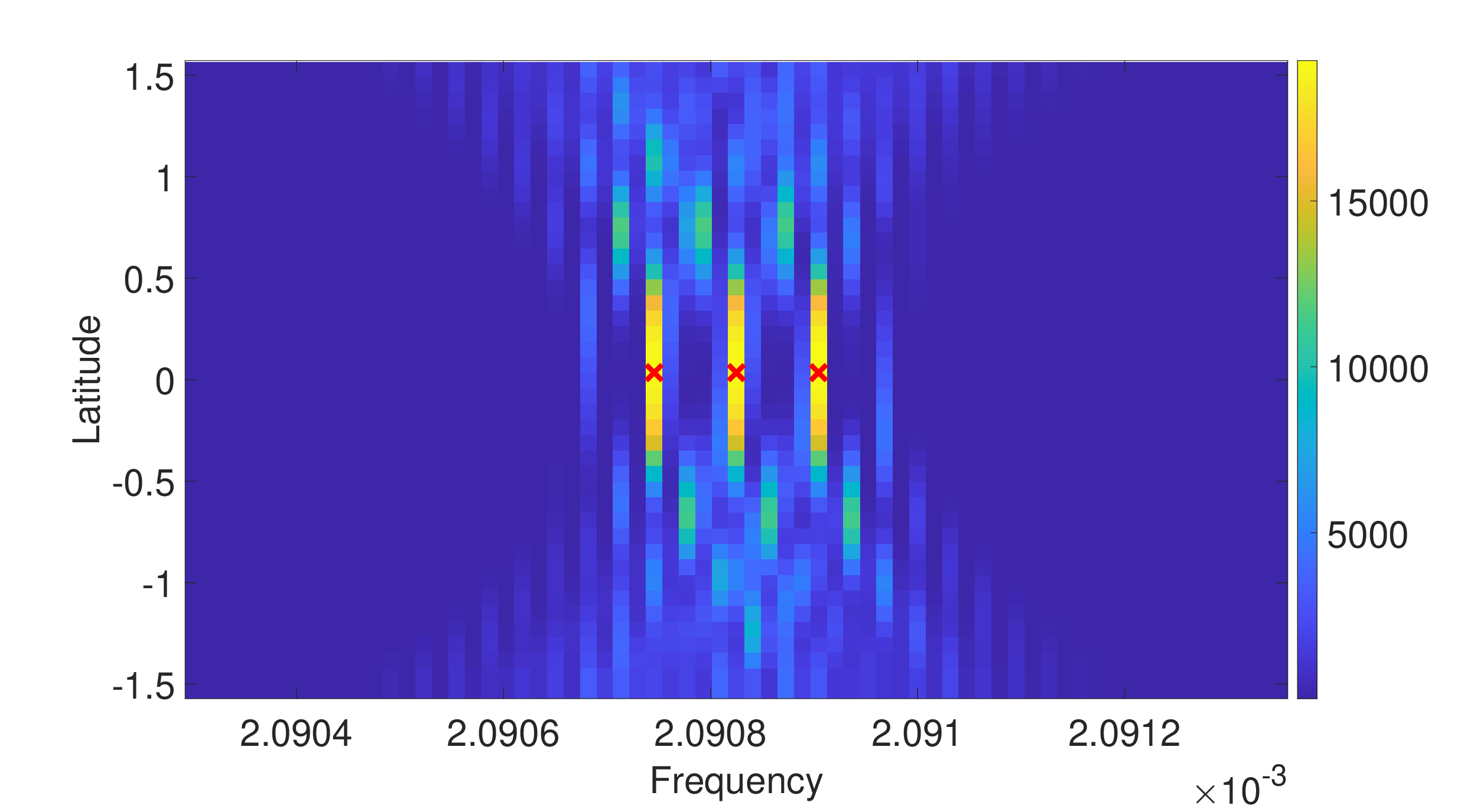}}\\
		\subfigure[\label{fig:F_statistic_distribution13}]{
			\includegraphics[width=0.48\linewidth]{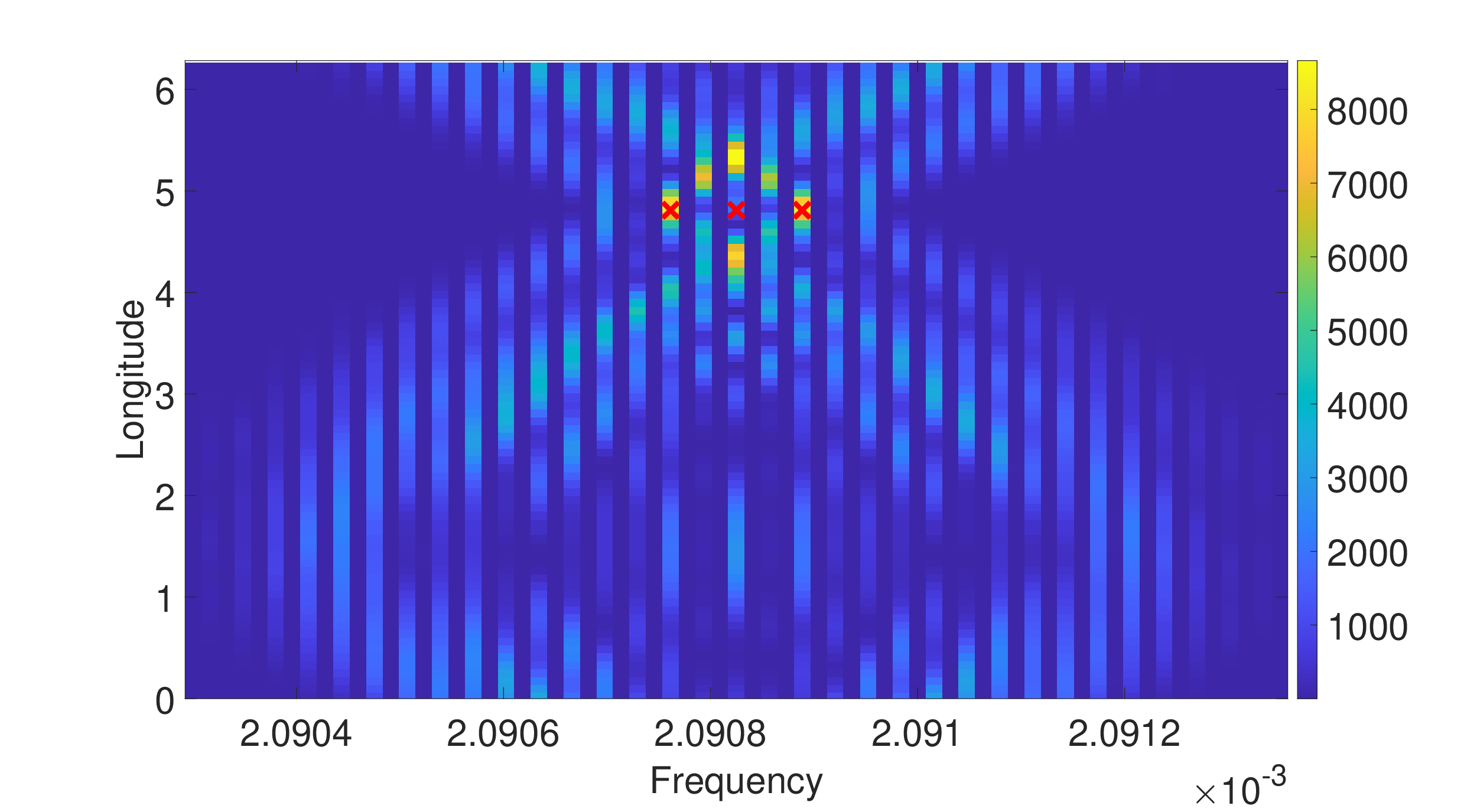}}
		\subfigure[\label{fig:F_statistic_distribution14}]{
			\includegraphics[width=0.48\linewidth]{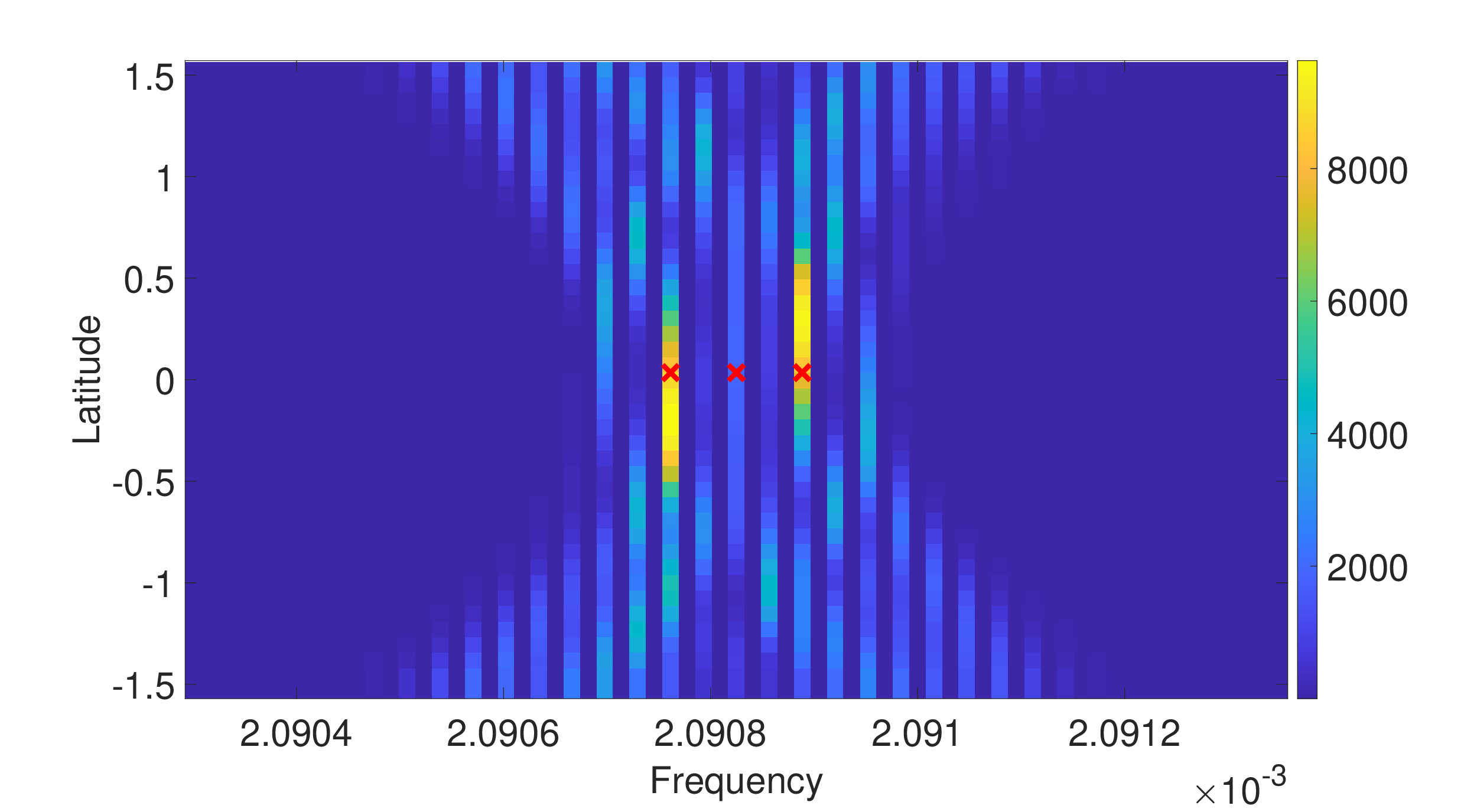}}
		\caption{\label{fig:F_statistic_distribution1}The parameter degeneracy when three signals overlap. The frequency intervals are $5 \times \mathrm{d}f$ for Fig.~\ref{fig:F_statistic_distribution11} and Fig.~\ref{fig:F_statistic_distribution12}, and $4 \times \mathrm{d}f$ for Fig.~\ref{fig:F_statistic_distribution13} and Fig.~\ref{fig:F_statistic_distribution14}. The red crosses marked the positions of the three binaries.}
	\end{figure*}

	It is conceivable for three or more signals with comparable SNR to overlap, especially in scenarios with low SNR and an increased number density of signals. Based on binary $A$, we generate two new binaries—one with a lower frequency (binary $C$) and one with a higher frequency (binary $B$). In Fig.~\ref{fig:F_statistic_distribution11} and Fig.~\ref{fig:F_statistic_distribution12}, the frequency difference of the three signals is $5 \times \mathrm{d}f$, and the interference between these signals is negligible. We anticipate that separating these three signals during the search should be relatively straightforward.
	However, for the case where the frequency difference of these three signals is $4 \times \mathrm{d}f$ (as shown in Fig.~\ref{fig:F_statistic_distribution13} and Fig.~\ref{fig:F_statistic_distribution14}), signal interference extinguishes the $\mathcal{F}$-statistic value of binary $A$ and reduces the values of $B$ and $C$. The overlapping signals also generate two erroneously prominent $\mathcal{F}$-statistic peaks above and below the true position of binary $A$ (in Fig.~\ref{fig:F_statistic_distribution13}). This complicates the search for these three signals, and it is only possible to rediscover binary $A$ after successfully disentangling binaries $B$ and $C$ from the data. How to search out these three overlapping signals is beyond the scope of this article.

	\subsection{The overlapping signals from LDC1-4 injection sources}
	\label{sec:overlap_signal}
	We identified injection sources with SNR $\ge7$ in LDC1-4 data and evaluated the presence of overlapping sources among them for two-year detection. These injection sources were categorized based on SNR thresholds: SNR $\ge15$ or SNR ranging from $7$ to $15$. In Tab.~\ref{table:3}, ``Sole" indicates that no other source occurs within its own $5\times \mathrm{d}f$ range, thus not considered an overlapping source. ``Two" indicates a source overlapping within the $5\times \mathrm{d}f$ range, as illustrated in Fig.~\ref{fig:F_statistic_distribution8} and Fig.~\ref{fig:F_statistic_distribution9}. ``Three" denotes three continuous overlapping sources, where the frequency interval between two sources is less than $5\times \mathrm{d}f$, and the interval between one of the sources and the third source is also less than $5\times \mathrm{d}f$, as depicted in Fig.~\ref{fig:F_statistic_distribution13} and Fig.~\ref{fig:F_statistic_distribution14}. ``Four" indicates four continuous overlapping sources in a sequence, and so on. Among the injection sources with SNR $\ge15$, we found a maximum of five continuous overlapping sources, with non-overlapping sources comprising 73.4\%. For injection sources with SNR between 7 and 15, up to eleven continuous overlapping sources were observed, with non-overlapping sources accounting for only 47.5\%. It is evident that as SNR decreases, the likelihood of overlapping sources increases, impacting the signal processing. In fact, we expect as the detection time increases, more and more signals can be resolved, which means that there will be more and more overlap problems.
	\begin{table*}
		\renewcommand{\arraystretch}{1.5}
		\begin{center}  
			\begin{tabular}{|p{3cm}<{\centering}|p{1cm}<{\centering}|p{1cm}<{\centering}|p{1cm}<{\centering}|p{1cm}<{\centering}|p{1cm}<{\centering}|p{1cm}<{\centering}|p{1cm}<{\centering}|p{1cm}<{\centering}|p{1cm}<{\centering}|p{1cm}<{\centering}|p{1cm}<{\centering}|p{1.5cm}<{\centering}|}
				\hline
				Injection sources' SNR & Sole & Two & Three & Four & Five & Six & Seven & Eight & Nine & Ten & Eleven & Total number \\
				\hline
				$\ge15$ & 8057 & 2182 & 597 & 136 & 10 & 0 & 0 & 0 & 0 & 0 & 0 & 10982 \\
				\hline
				$7\sim15$ & 7932 & 4478 & 2139 & 1096 & 500 & 306 & 98 & 72 & 36 & 30 & 11 & 16698 \\
				\hline
			\end{tabular}
			\caption{The number of overlapping signals from LDC1-4 injection sources is categorized as follows: ``Sole" denotes no overlapping sources. ``Two" indicates that the frequency interval between two sources is less than $5\times \mathrm{d}f$, resulting in their overlap. ``Three" implies that not only the frequency interval between two sources is less than $5\times \mathrm{d}f$, but also the frequency interval between one of the sources and the third source is less than $5\times \mathrm{d}f$, and so forth.}
			\label{table:3} 
		\end{center}   
	\end{table*}

	\section{Search method}
	\label{sec:Search_method}
	\subsection{Data preprocessing}
\begin{figure}
		\centering
		\includegraphics[width=1\linewidth]{"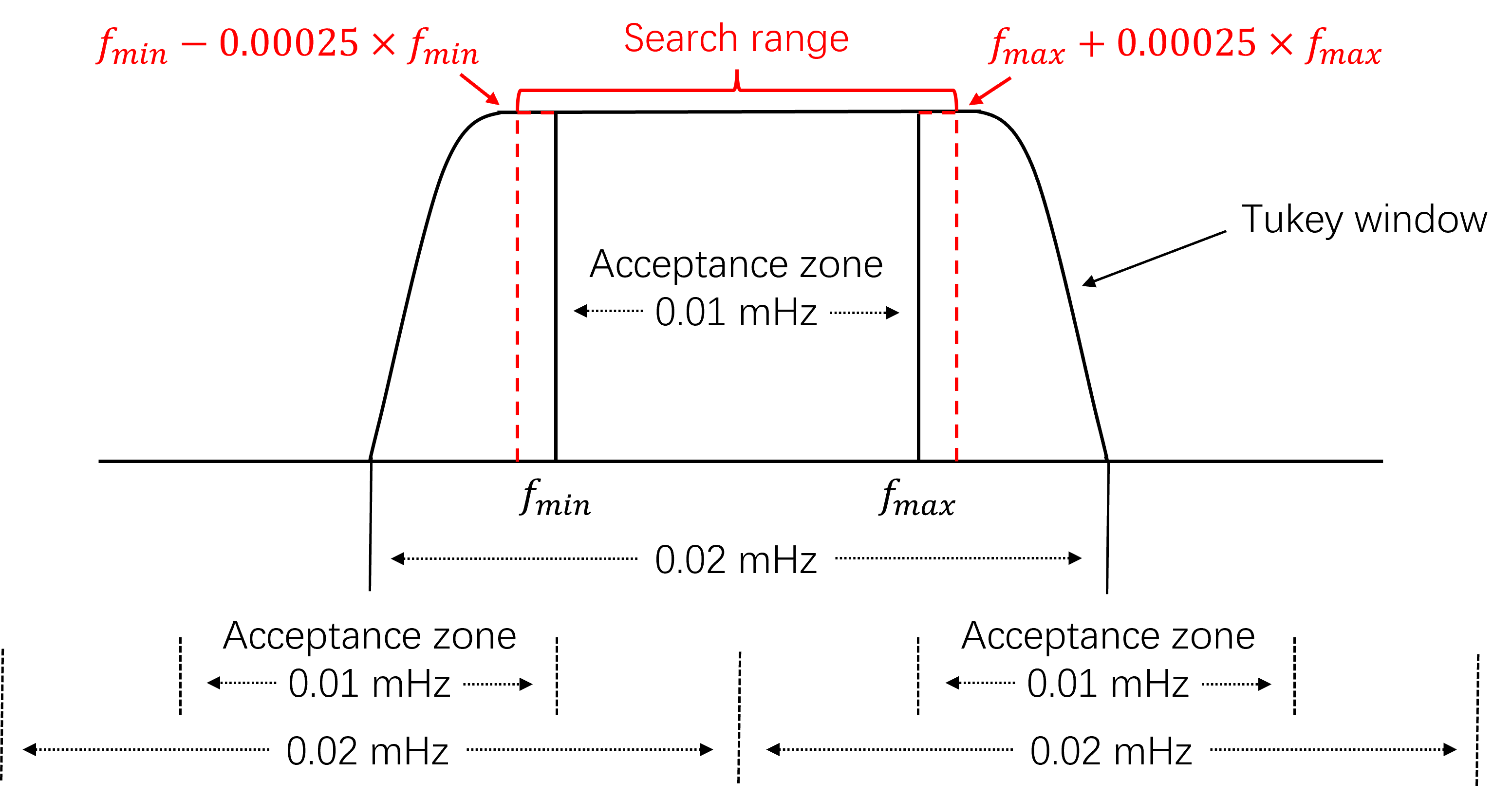"}
		\caption{The schematic diagram illustrates the division of data into frequency bins, following a method similar to Ref.~\cite{zhang_resolving_2021}. Instead of searching the entire range of each bin, we focus on regions where signals may introduce degeneracy noise within the acceptance zone.}
		\label{fig:frequencyrange}
	\end{figure}

	Before initiating the search, it is essential to preprocess the LDC1-4 data. Assuming conventional methods have detected signals with SNR $>$ 15, our initial step involves subtracting all injected Galactic binary signals with SNR $>$ 15 from the raw data, allowing us to focus on the search for lower SNR signals. Then the data undergoes band-pass filtering and is divided into 1491 bins for parallel computation, covering a frequency range from 0.09 mHz to 15 mHz. The band-pass filter method aligns with the one illustrated in Figure 3 of Ref.~\cite{zhang_resolving_2021}, and other techniques, such as undersampling to reduce data length, follow the approach outlined in the same reference.
	As our method avoids subtracting any signal during the search, concerns about inaccurate subtraction contamination are minimized. Our primary concern shifts to addressing the potential presence of degeneracy noise at the edges of each bin. In preparation for subsequent degeneracy noise removal steps, we slightly expand the frequency search range on both sides of the acceptance zone. This extension encompasses 0.025\% times outward from the boundary of the acceptance zone, ensuring coverage of all signals outside the acceptance zone whose degeneracy noise may affect the acceptance zone. This expansion is visually represented in Fig.~\ref{fig:frequencyrange}.
	
	The search ranges for $\beta$ and $\lambda$ cover the whole sky: $\beta\in[-\frac{\pi}{2},~\frac{\pi}{2}]$ and $\lambda\in[0,~2\pi]$. The settings for $\dot{f}$ is $[-10^{-16},~10^{-15}]$ when $f\le4$ mHz and $[-10^{-14},~10^{-13}]$ when $f>4$ mHz.
	
	\subsection{Local maxima PSO}
	The Particle Swarm Optimization (PSO) algorithm, renowned for its efficacy in seeking global maxima of continuous nonlinear functions \cite{488968, gad_particle_2022}, has proven instrumental in the quest for Galactic binaries \cite{zhang_resolving_2021}. Previous methodologies focused on identifying global maxima of the $\mathcal{F}$-statistic within parameter space, employing multiple searches to ensure the robustness of global maxima detection. This process involved iteratively subtracting the detected signal and initiating subsequent search rounds, known as the iterative-subtraction strategy.
	In our LMPSO-CV approach, we streamline the process by concentrating solely on searching for local maxima, enabling us to achieve convergence with fewer steps. To facilitate this, we have tailored the PSO algorithm to swiftly converge on local maxima, optimizing the efficiency of the search process.
	
	We employ a total of 40 particles in our algorithm. The process begins by randomly dispersing these particles throughout the parameter space. Each particle calculates the $\mathcal{F}$-statistic value at its respective position. Based on the calculation results, the positions of the particles undergo iteration. Each particle ``remembers" the best value and the position it explored, denoted as the personal best position ($P$). From the $P$ of all particles, the global best position ($G$) is determined to represent the globally best position found by the swarm. To reach the optimal solution, each particle moves toward both its $P$ and the $G$ within the swarm. The position of each particle is updated in each iteration through the following equation:
	\begin{equation}
		\begin{aligned}
			\mathbf{v}_i^{t+1}&=\omega \mathbf{v}_i^t+c_1 \mathbf{r}_1\cdot\left(P_i^t-\mathbf{x}_i^t\right)+c_2 \mathbf{r}_2\cdot\left(G^t-\mathbf{x}_i^t\right),\\ 
			\mathbf{x}_i^{t+1}&=\mathbf{x}_i^t+\mathbf{v}_i^{t+1}.
		\end{aligned}
	\end{equation}
	In the given equation, $t$ denotes the number of iterations. $\mathbf{v}$ represents the velocity vector. $\omega$ serves as the inertia weight crucial for balancing local exploitation and global exploration. $\mathbf{r}_1$ and $\mathbf{r}_2$ denote random vectors uniformly distributed within the range $[0, 1]^D$ ($D$ is the dimensionality of the searched parameter space). $\mathbf{x}_i$ indicates the position of the $i$-th particle. $c_1$ and $c_2$, referred to as ``acceleration coefficients", are positive constants.
	In particular, when $c_1=c_2=2$ and $\omega=\text{max}(\omega_{1}-\frac{\omega_{2}}{\omega_{3}} t,~\omega_{4})$ (where $\omega_{1}=0.9,~\omega_{2}=0.4,~\omega_{3}=1999,~\omega_{4}=0.2$), this parameter configuration strikes a harmonious balance between $P$ and $G$, rendering the PSO algorithm highly effective in the pursuit of the global maxima.
	
	In the employed PSO algorithm, we initially configure the parameters as described above. Upon reaching an optimal $\mathcal{F}$-statistic surpassing a predefined threshold ($\mathcal{F}$-statistic $>$ 53, corresponding to SNR $>$ 7), we dynamically adjust the parameter configuration by setting $c_1 = \omega = 0$ in the iterative equation. So the iterative equation becomes: 
	\begin{equation}
		\begin{aligned}
			\mathbf{v}_i^{t+1}&=c_2 \mathbf{r}_2\cdot\left(G^t-\mathbf{x}_i^t\right),\\ 
			\mathbf{x}_i^{t+1}&=\mathbf{x}_i^t+\mathbf{v}_i^{t+1}.
		\end{aligned}
	\end{equation}
	This adjustment accelerates the convergence of the PSO algorithm towards a local maxima. To ascertain convergence, we implement an automatic convergence criterion: if the optimal $\mathcal{F}$-statistic value remains constant for 15 consecutive iterations, convergence is determined, and facilitating the transition to search the next local maxima.
	
	The flow chart of LMPSO in Fig.~\ref{fig:flowchart4} illustrates the workflow of the LMPSO algorithm. Following the completion of each local maxima search, we create a void in the parameter space corresponding to the position of the $G$. In the subsequent search, particles are initially assessed to determine if they fall within the voids. If a particle lands in the voids, its value is assigned negative infinity, indicating that the particle is invalid. For the algorithm, this signifies a position that requires no further exploration. During the next iteration, the particle will navigate to other positions in the parameter space. As the maximum $\mathcal{F}$-statistic value surpasses a predefined threshold ($\mathcal{F}$-statistic $>$ 53), the search transitions from global to local maxima. Ultimately, the convergence criterion adjudicates whether the search is concluded.
	
	\subsection{Create voids in the parameter space}
\begin{figure*}
		\centering  %图片全局居中
		\subfigbottomskip=2pt %两行子图之间的行间距
		\subfigcapskip=0pt %设置子图与子标题之间的距离
		\subfigure[\label{fig:Diameter1}]{
			\includegraphics[width=0.48\linewidth]{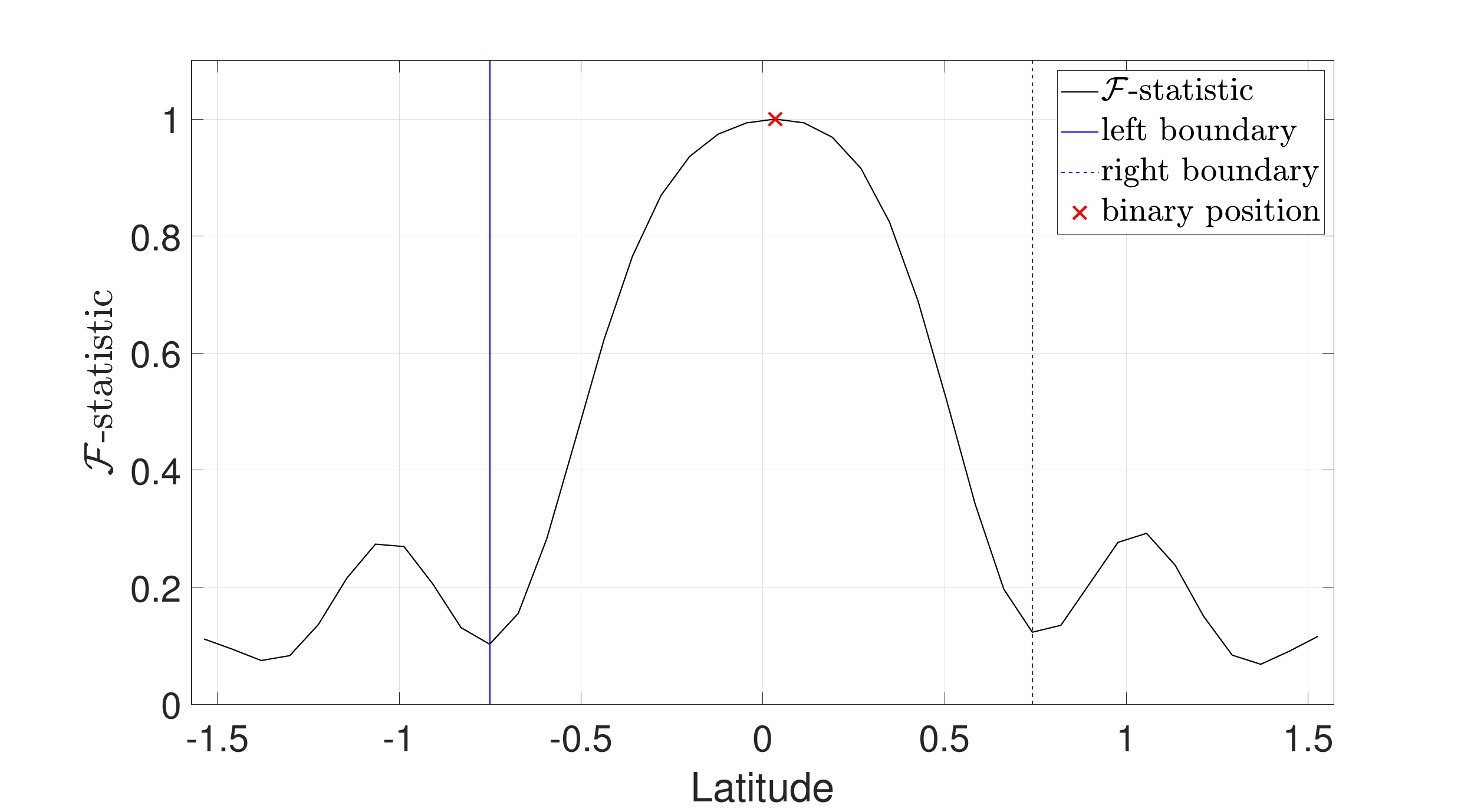}}
		\subfigure[\label{fig:Diameter2}]{
			\includegraphics[width=0.48\linewidth]{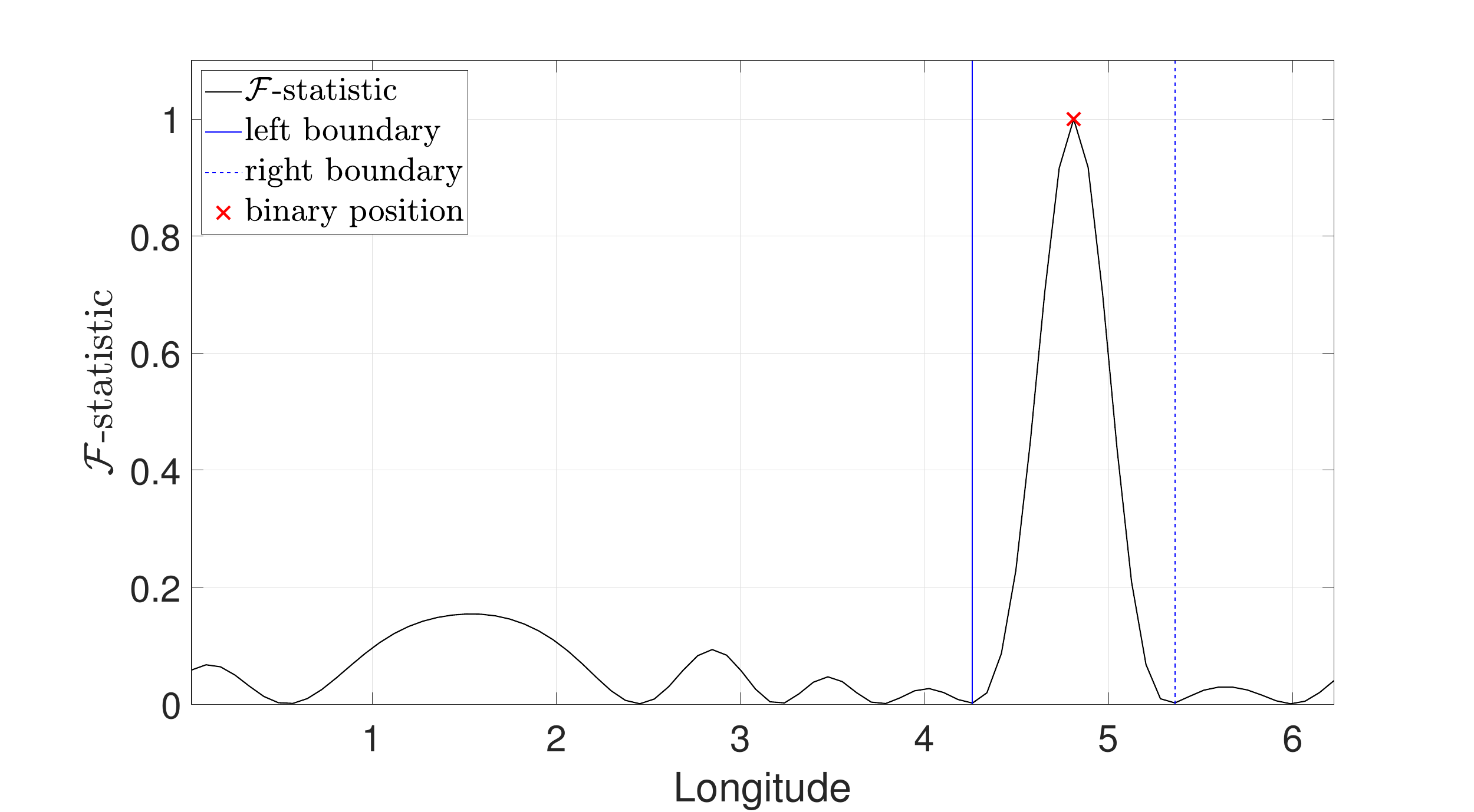}}\\
		\caption{\label{fig:Diameter}We conduct an analysis of the normalized parameter degeneracy to determine the void radii within the Ecliptic Latitude $\beta$ and Ecliptic Longitude  $\lambda$. Throughout this computation, all other parameters are kept constant while one parameter is systematically varied at a time. Specifically, in Fig.~\ref{fig:Diameter1}, we investigate the effects of changing $\beta$, and in Fig.~\ref{fig:Diameter2}, we explore the impact of varying $\lambda$. The spatial resolution employed is $\pi/40$ radians. The determination of the ellipsoidal radius associated with each parameter involves selecting half the distance between the two local minima, identified by the left boundary and right boundary, which are closest to the sought local maxima. The position of the sought local maxima is denoted by the red cross, symbolizing the authentic binary position in our analysis.}
	\end{figure*}

	After identifying a local maxima using LMPSO algorithm, we create a void in the parameter space to prevent redundant searches. The void is modeled as a spheroid described by the ellipsoid equation ($\frac{x^2}{a^2}+\frac{y^2}{b^2}+\frac{z^2}{c^2}=1$), where $a$, $b$, and $c$ represent the radii of the spheroid in $f$, $\beta$, and $\lambda$, respectively. We set $a=df$ to be equal to the minimum frequency resolution. The values for $b$ and $c$ depend on the parameter degeneracy calculated in Fig.~\ref{fig:Diameter}. In this figure, while keeping the other parameters constant, we vary one parameter at a time to observe the trend of the $\mathcal{F}$-statistic in $\beta$ or $\lambda$. Specifically, Fig.~\ref{fig:Diameter1} explores the variation with respect to $\beta$, and Fig.~\ref{fig:Diameter2} with respect to $\lambda$. The resolution of $\beta$ and $\lambda$ is $\pi/40$ radian. We determine the ellipsoidal radius by selecting half the distance between the two local minima closest to the searched local maxima. The positions of these two local minima are denoted by the left boundary and right boundary lines in Fig.~\ref{fig:Diameter1} and Fig.~\ref{fig:Diameter2}, with the red cross marking the real binary position.
	
\begin{figure*}
		\centering  %图片全局居中
		\subfigbottomskip=2pt %两行子图之间的行间距
		\subfigcapskip=0pt %设置子图与子标题之间的距离
		\subfigure[\label{fig:holes1}]{
			\includegraphics[width=0.48\linewidth]{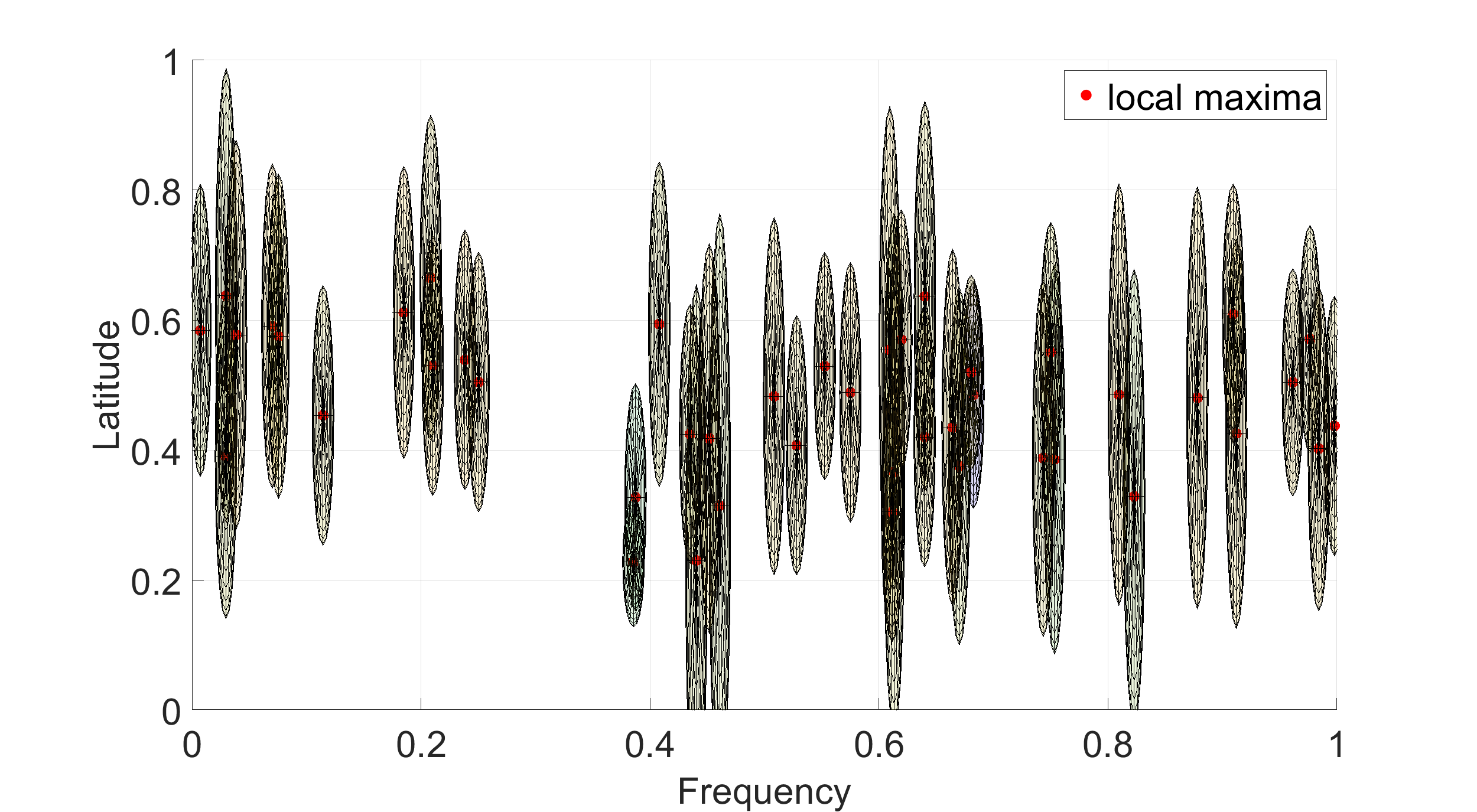}}
		\subfigure[\label{fig:holes2}]{
			\includegraphics[width=0.48\linewidth]{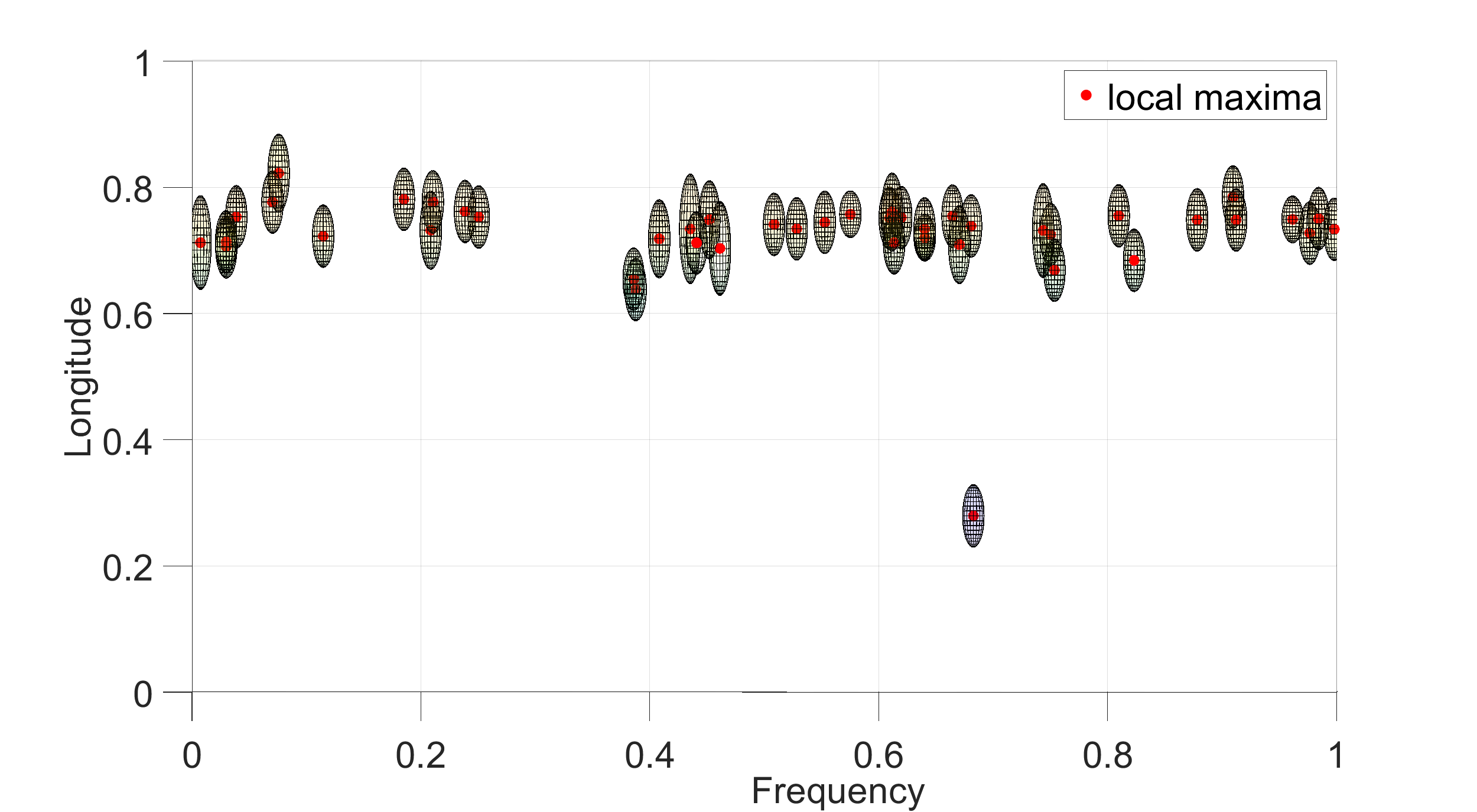}}\\
		\caption{\label{fig:holes}The example for the voids in normalized parameter space. Fig.~\ref{fig:holes1} and Fig.~\ref{fig:holes2} are the two side views. The red points represent the local maxima inside each void. To make the example more obvious, we take the radius on frequency $6\times \mathrm{d}f$ in the figures rather than $1\times \mathrm{d}f$ in real search.}
	\end{figure*}

	In the subsequent search, we must assess whether the particles in the LMPSO fall within the voids. This is determined using the formula
	\begin{equation}
		D=\frac{(x_1-x)^2}{a^2}+\frac{(y_1-y)^2}{b^2}+\frac{(z_1-z)^2}{c^2}
	\end{equation}
	where $x_1$, $y_1$, $z_1$ represent the three parameter positions of the particle, and $x$, $y$, $z$ represent the searched local maxima position at the center of the void. When $D > 1$, it indicates that the particle is outside the void, and we proceed to calculate the $\mathcal{F}$-statistic value of the particle. Conversely, when $D~\leq~1$, it indicates that the particle is inside the void, and we assign it an infinity value, rendering it an invalid particle. It will be repositioned elsewhere in the subsequent iteration.
	Fig.~\ref{fig:holes} illustrates the voids in normalized parameter space. The vertical coordinates in Fig.~\ref{fig:holes1} represent $\beta$, while Fig.~\ref{fig:holes2} represents $\lambda$. Notably, the radius of the void at latitude is larger than that at longitude, suggesting that the $\mathcal{F}$-statistic is more sensitive to longitude. Although the radius in frequency is set at $\mathrm{d}f$, for clarity in the figure, we use $6 \times \mathrm{d}f$ as an example.
	
	Ultimately, we have generated numerous ``voids" in the parameter space, each indicating the presence of a local maxima. The count of local maxima far surpasses the number of binaries. In Sec.~\ref{sec:Comprehensive_analysis}, we will conduct an analysis of these voids, which we call find-real-$\mathcal{F}$-statistic-analysis, to identify real binaries and eliminate noise.    
	
	\subsection{Termination rule of search}
	The criterion for concluding the search can greatly impact the number of voids. We conducted a maximum of 10,000 searches per bin. Alternatively, if the $\mathcal{F}$-statistic values remain below the threshold ($\mathcal{F}$-statistic = 53) for 30 consecutive searches (equivalent to 30 consecutive local maxima), the search is terminated. Ultimately, none of the bins met the first criterion. The frequency range with the highest number of searched local maxima is from 2.305 mHz to 2.315 mHz, resulting in 3206 voids.

	\section{Find-real-$\mathcal{F}$-statistic-analysis}
	\label{sec:Comprehensive_analysis}
	\begin{figure*}
		\centering
		\caption{The flow chart of our method.}
		\includegraphics[width=1\linewidth]{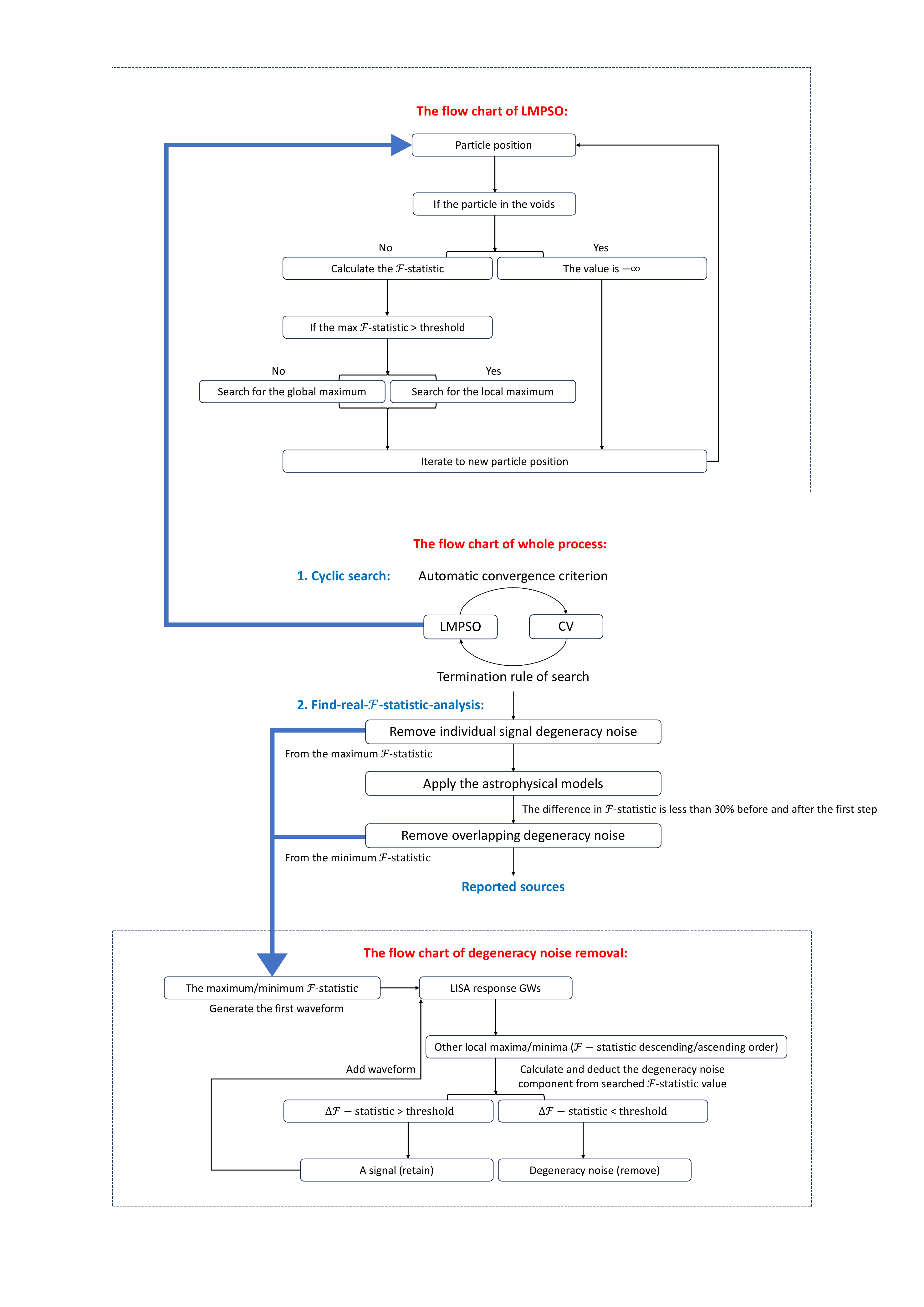}
		\label{fig:flowchart4}
	\end{figure*}
	
	Given the existence of degeneracy noise, the count of local maxima far exceeds the number of real binaries, a situation exacerbated by the influence of instrument noise. Without effective noise removal, the search outcomes become severely contaminated.
	In this section, our primary goal is the removal of both degeneracy noise and instrument noise, striving to achieve the most accurate representation of the search results. We term this comprehensive analysis ``find-real-$\mathcal{F}$-statistic-analysis".
	
	\subsection{Remove individual signal degeneracy noise}
	\label{sec:Comprehensive_analysis1}
	The degeneracy noise produced by an individual signal is the most numerous and easiest to remove, characterized by relatively weaker $\mathcal{F}$-statistic values compared to the binaries from which they originate, so we commence by systematically eliminating individual signal degeneracy noise. The starting point for this removal process is the descending list of searched local maxima, ordered by their $\mathcal{F}$-statistic values. Beginning with the local maxima possessing the highest $\mathcal{F}$-statistic value, indicative of potential authenticity as a real binary, we generate its LISA response GW signal as the initial waveform.
	Having obtained the initial waveform, we proceed to scrutinize the second-largest $\mathcal{F}$-statistic value to ascertain whether it corresponds to the degeneracy noise of the first. This involves computing the $\mathcal{F}$-statistic value using the parameters of the second local maxima on the first waveform. Subsequently, we deduct the computed $\mathcal{F}$-statistic value of the second local maxima on the first waveform from its own searched $\mathcal{F}$-statistic value. If the residue exceeds the threshold ($\Delta\mathcal{F}$-statistic $>$ 53), it is considered a new signal to retain. We then generate its waveform and incorporate it into the first waveform. Conversely, if the residue falls below the threshold, it is identified as noise and promptly removed. 
	The formula for computing the $\Delta\mathcal{F}$-statistic is given by:
	\begin{equation}
		\Delta\mathcal{F}\left(\theta_i\right) = \mathcal{F}\left(\theta_i, S\right) - \mathcal{F}\left(\theta_i, \sum_{m=1}^{i-1} h\left(\theta_m\right)\right),
	\end{equation}
	where $S$ represents the raw data, specifically the LDC1-4 residual data in this article. The term $\mathcal{F}\left(\theta_i, S\right)$ denotes the $\mathcal{F}$-statistic value for $i$ local maxima with parameters $\theta_i$ concerning the data $S$ (the searched $\mathcal{F}$-statistic value). Similarly, $\mathcal{F}\left(\theta_i, \sum\limits_{m=1}^{i-1} h\left(\theta_m\right)\right)$ represents the $\mathcal{F}$-statistic value for $i$ local maxima concerning the sum of the first $i-1$ retained waveforms (the computed $\mathcal{F}$-statistic value).
	
	This iterative procedure continues, processing all local maxima in descending order of $\mathcal{F}$-statistic.
	Ultimately, this step significantly reduces the number of initially searched local maxima from the total count of 646,120 (646,120 with $\mathcal{F}$-statistic $>$ 53, and the overall total is 849,576) to 25,615, which are the ones remaining within the acceptance zone after the removal of individual signal degeneracy noise. The flow chart of degeneracy noise removal in Fig.~\ref{fig:flowchart4} visually outlines the process.
	
	\subsection{Apply the astrophysical models}
	With diminishing SNR, the influence of instrument and confusion noise becomes increasingly conspicuous. Identifying instrument noise false alarm in one search is challenging. In order to mitigate the effects of instrument noise, we undertake a secondary search by fine-tuning the $\dot{f}$ parameter setting. Subsequently, we cross-validate the outcomes derived from the two distinct searches, thereby proficiently eliminating the impact of instrument noise from the dataset.
	
	\subsubsection{The second search based on astrophysical models}
	In the subsequent search, we maintain the configurations for all other parameters while specifically adjusting the range for $\dot{f}$.
	In the initial search, we provided an extensive range for $\dot{f}$,
	encompassing $[-10^{-16},~10^{-15}]$ when $f\le4$ mHz and $[-10^{-14},~10^{-13}]$ when $f>4$ mHz. For the second search, we apply the astrophysical model to restrict the $\dot{f}$ range for each frequency bin, employing a logarithmic scale of 10 for the search. The $\dot{f}$ range for Galactic binaries can be determined using the astrophysical model:
	\begin{equation}
		\label{format1}
		\dot{f}=\frac{96}{5} \pi^{8 / 3} \mathcal{M}^{5 / 3} f^{11 / 3}.
	\end{equation}
	Considering that the majority of Galactic binaries involve white dwarfs with masses ranging from $0.1~M_{\odot}$ \cite{kilic_lowest_2007} to $1.4~M_{\odot}$, we determine the range of the chirp mass $\mathcal{M} = (m_1
	m_2)^{3/5}/M^{1/5}$. Here, $m_1$ and $m_2$ represent the masses of the two white dwarfs, and $M = m_1+m_2$ denotes their total mass. The lower and upper limits of the chirp mass $\mathcal{M}$ are $\mathcal{M}_{\text{min}} = 0.06~M_{\odot}$ and $\mathcal{M}_{\text{max}} = 1.22~M_{\odot}$, respectively. In contrast to the first search, the $\dot{f}$ range for the second search is constrained to a narrower interval, with its magnitude being proportional to the frequency.
	
	\subsubsection{Cross validation}
	\label{sec:Cross_validation}
	In the second search, a substantial number of local maxima are identified, totaling 574,676 with $\mathcal{F}$-statistic $>$ 53 (out of a total of 822,730). Following the removal of individual degeneracy noise, 26,759 local maxima within the acceptance zone are retained. To evaluate the similarity between the gravitational waveforms of the two searches, we calculate the correlation coefficient $R_{ee}$ (Eq.~\ref{1}) between the 25,615 local maxima from the first search and the 26,759 local maxima from the second search. Local maxima with $R_{ee}>0.99$ in the first search are considered more reliable, resulting in the selection of 10,623 local maxima for the subsequent step. This may be attributed to the heightened sensitivity of instrument noise to $\dot{f}$ when imitating a signal, whereas real signals demonstrate greater robustness to variations in $\dot{f}$.
	
	\subsubsection{Restrict the Galactic Latitude}
	\label{sec:Restrict_the_Galactic_latitude}
	The distribution of degeneracy noise spans the entire celestial sphere. However, leveraging the well-established concentration of Galactic binaries in the Galactic Disk, we can enhance the precision of the search results by introducing constraints on their sky locations. By confining the Galactic Latitude of the outcomes within the range of $-$0.5 rad to 0.5 rad (as illustrated in Fig.~\ref{fig:galacticdistribution}), we effectively narrow down the selection to 9,143 remaining local maxima.

	Restricting the latitude range results in the omission of sources at higher latitudes. However, since most Galactic binaries are concentrated near the Galactic disk, the number of sources at high latitudes is significantly lower than at low latitudes (Fig.~\ref{fig:galacticdistribution}). This scarcity increases the relative impact of degeneracy noise in high-latitude regions. By directly applying a latitude constraint to focus on the Galactic disk, we can effectively mitigate such noise.
	
	\subsection{Remove overlapping signals degeneracy noise}
	\label{sec:Comprehensive_analysis2}
	In the final phase, our aim is to eliminate the overlapping degeneracy noise originating from the simultaneous presence of two or more binaries (as depicted in Fig.~\ref{fig:F_statistic_distribution8}). Since the $\mathcal{F}$-statistic value of such noise exceeds that of any individual binary it emanates from, it cannot be effectively addressed in the initial step designed to remove individual signal degeneracy noise. To tackle these instances of overlapping degeneracy noise, we take out the sources for which the difference of the  $\mathcal{F}$-statistic values before and after the first step is less than 30\%  (Sec.~\ref{sec:Comprehensive_analysis1}) and repeat the first step in ascending order of the searched local maxima, as illustrated in the flow chart of degeneracy noise removal depicted in Fig.~\ref{fig:flowchart4}. The waveform is initially generated by the signal with the lowest $\mathcal{F}$-statistic value and progressively incorporates signals with higher $\mathcal{F}$-statistic value. The objective is to discern whether the signal with the higher $\mathcal{F}$-statistic value is degeneracy noise resulting from overlapping with signals with lower $\mathcal{F}$-statistic values.
	
	Following this step, 6,508 local maxima persist.

	\section{Search results}
	\label{sec:Search_results}
	To determine the accuracy of the searched sources, we need to calculate the correlation coefficient ($R$) of the results with the injection sources:
	\begin{equation}
		\label{1}
		\begin{aligned}
			R\left(\theta, \theta^{\prime}\right) &=\frac{C\left(\theta, 
				\theta^{\prime}\right)}{\left[C(\theta, \theta) C\left(\theta^{\prime}, 
				\theta^{\prime}\right)\right]^{1 / 2}},\\
			C\left(\theta, \theta^{\prime}\right) &= 
			{\sum_{I}(\bar{s}^{I}(\theta)*\bar{s}^{I}(\theta^{\prime}))/S_n^I(f)},
		\end{aligned}
	\end{equation}
	where $\theta$ and $\theta^{\prime}$ represent the parameters, and $I \in \{A,E,T\}$. 
	We selected injection sources with 3 $<$ SNR $<$ 15 over a two-year observation period for calculation, with a total count of 55,714. A source is considered accurate if $R$ $\ge$ 0.9, relatively accurate if 0.8 $\le$ $R$ $<$ 0.9, contaminated if 0.5 $\le$ $R$ $<$ 0.8, and incorrect if $R$ $<$ 0.5.
	
	\subsection{False source fraction and SNR-ratio}
	\label{sec:our_result}
\begin{figure}
		\centering
		\includegraphics[width=1\linewidth]{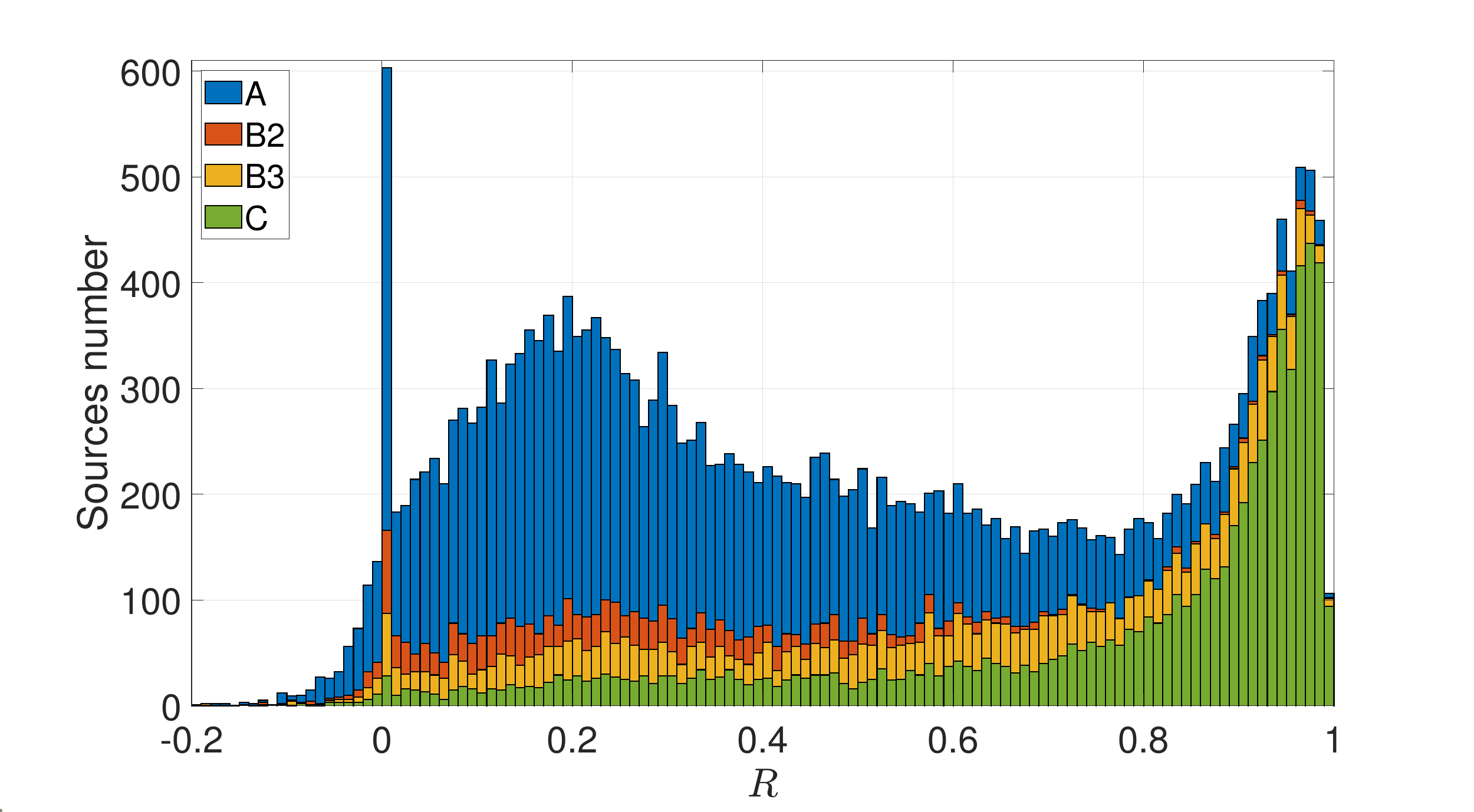}
		\caption{The histogram depicts the distribution of correlation coefficients in the results after each step. A, B2, B3, and C correspond to the four steps outlined in Sec.~\ref{sec:Comprehensive_analysis}. A signifies the outcomes after eliminating individual signal degeneracy noise (Sec.~\ref{sec:Comprehensive_analysis1}). B2 represents the results after cross-validation with the second search (Sec.~\ref{sec:Cross_validation}). B3 indicates the outcomes after restricting the Galactic Latitude (Sec.~\ref{sec:Restrict_the_Galactic_latitude}), and C denotes the reported sources after removing overlapping degeneracy noise (Sec.~\ref{sec:Comprehensive_analysis2}).}
		\label{fig:finalresults}
	\end{figure}

\begin{figure}
	\centering
	\includegraphics[width=1\linewidth]{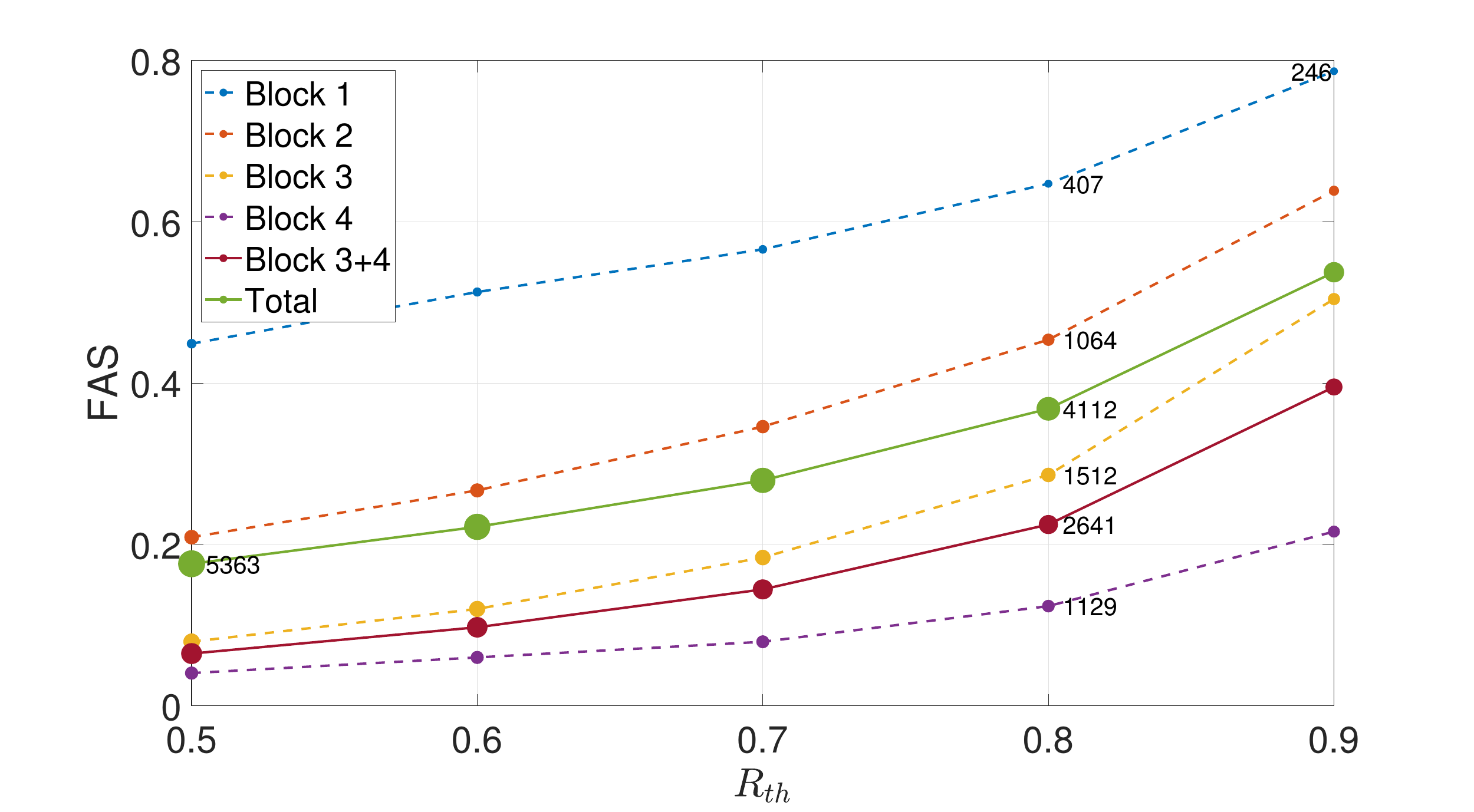}
	\caption{When $R$ is set to different thresholds, the corresponding FAS values vary. A higher $R_{th}$ results in a higher FAS value for the same reported sources. The area of each circle represents the number of sources that exceed the corresponding $R_{th}$. For clarity, we highlight the number of sources for the largest and smallest circles. Additionally, we provide the number of sources exceeding the threshold across all cases when the $R_{th}$ is set to 0.8.}
	\label{fig:drr2}
	\end{figure}

	We employ the reported sources ($N_r$) to represent the outcomes following the find-real-$\mathcal{F}$-statistic-analysis, and confirmed sources ($N_c$) to signify those reported sources with an $R$ value exceeding the established threshold ($R_\text{th}$). The false source fraction (FAS) is succinctly formulated as follows:
	\begin{equation}
		\text{FAS} = 1-\frac{N_c}{N_r}.
	\end{equation}
	This metric quantifies the ratio of sources with $R$ values that fail to meet the specified threshold. This definition is exactly complementary to the definition of detection rate in Ref.~\cite{zhang_resolving_2021}, meaning that the sum of the two equals 1. The threshold is a variable, and its value ranges from 0.5 to 0.9 (Fig.~\ref{fig:drr2}). In the subsequent discussion, we focus on results with $R_\text{th}=0.8$ as the primary reference threshold, and the corresponding FAS is expressed as $\text{FAS}_{0.8}$. The evolutionary trajectory of our methodology through the various stages of find-real-$\mathcal{F}$-statistic-analysis (elaborated in Sec.~\ref{sec:Comprehensive_analysis}) is vividly depicted in the histogram of $R$ presented in Fig.~\ref{fig:finalresults}. In the initial step, a significant reduction in individual degeneracy noise was achieved, resulting in 25,615 searched sources with $\text{FAS}_{0.8}$ = 76.8\% (A). The second step concentrated on diminishing noise through cross-validation with a secondary search, yielding 10,623 searched sources with $\text{FAS}_{0.8}$ = 54.8\% (B2). Subsequent refinement in the third step, which confined Galactic Latitude within $\pm 0.5$ rad, led to a reduction to 9,143 searched sources with $\text{FAS}_{0.8}$ = 48.0\% (B3). Finally, the fourth step addressed the difference in $\mathcal{F}$-statistic value and overlapping degeneracy noise with $\mathcal{F}$-statistic values surpassing those of any individual signal, resulting in 6,508 searched sources with $\text{FAS}_{0.8}$ = 36.8\% (C). In the following discussion, we refer to the 6,508 sources identified in C as reported sources.
	
	In Tab.~\ref{table:11}, we categorize the reported sources into four blocks based on frequency and SNR and analyze the proportion of sources with different $R$ values in each block. Additionally, we compute the proportion of sources within various $R$ ranges relative to the total number of reported sources across all blocks. Blocks 1, 2, and 3 correspond to $f \leq 3$ mHz, with SNR ranges of (7, 10), (10, 13), and $\geq 13$, respectively. By evaluating the FAS for each block, we determine which ones can be included in the final catalog. Block 4, which represents $f > 3$ mHz, achieves an $\text{FAS}_{0.8}$ of 12.3\%, with 1,129 sources having $R \geq 0.8$. Due to its strong performance, we do not further subdivide it by SNR in Tab.~\ref{table:11}. However, for $f \leq 3$ mHz, SNR grouping is crucial, as only sources with SNR $\geq 13$ maintain an $\text{FAS}_{0.8}$ of 28.6\%, with 1,512 sources having $R \geq 0.8$. The need for a higher SNR threshold at lower frequencies arises from the significant influence of foreground noise. When accounting for unresolved double white dwarf gravitational waves, the SNR—initially calculated using only instrumental noise—drops significantly, leading to a substantial increase in FAS. 
	Considering both Blocks 3 and 4 together, we identify a total of 3,406 sources, yielding an $\text{FAS}_{0.8}$ of 22.5\%, with 2,641 sources having $R \geq 0.8$. When considering Blocks 2, 3, and 4 together, we identify a total of 5,354 sources, resulting in an $\text{FAS}_{0.8}$ of 30.8\%, with 3,705 sources having $R \geq 0.8$. Across all 6,508 reported sources, the $\text{FAS}_{0.8}$ is 36.8\%, with 4,112 sources exhibiting $R \geq 0.8$.
	Fig.~\ref{fig:drr2} illustrates the FAS values for different categories when various $R$ thresholds are applied, where the circle area represents the number of sources exceeding the given threshold. As frequency and SNR decrease, the FAS of the blocks increases.
	\begin{table*}
		\renewcommand{\arraystretch}{1.5}
		\begin{center}  
			\begin{tabular}{|p{3cm}<{\centering}|p{1.5cm}<{\centering}|p{1.5cm}<{\centering}|p{1.5cm}<{\centering}|p{1.5cm}<{\centering}|p{1.5cm}<{\centering}|p{1.5cm}<{\centering}|p{1.5cm}<{\centering}|p{1.5cm}<{\centering}|}
				\bottomrule[1.5pt]
				Total reported & \multicolumn{8}{c|}{6508} \\
				\hline
				Reported & \multicolumn{2}{c|}{1154} & \multicolumn{2}{c|}{1948} & \multicolumn{2}{c|}{2118} & \multicolumn{2}{c|}{1288} \\
				\bottomrule[1.5pt]
				\multirow{3}{*}{$R<0.5$} & \multicolumn{2}{c|}{Block 1} & \multicolumn{2}{c|}{Block 2} & \multicolumn{2}{c|}{Block 3} & \multicolumn{2}{c|}{Block 4} \\ \cline{2-9}
				& $f$ & SNR & $f$ & SNR & $f$ & SNR & $f$ & SNR \\ \cline{2-9}
				& $\le$ 3 mHz & 7 $\sim$ 10 & $\le$ 3 mHz & 10 $\sim$ 13 & $\le$ 3 mHz & $\ge$ 13 & $>$ 3 mHz & $\ge$ 7 \\
				\hline
				Number & \multicolumn{2}{c|}{518} & \multicolumn{2}{c|}{407} & \multicolumn{2}{c|}{168} & \multicolumn{2}{c|}{52} \\
				\hline
				Proportion & \multicolumn{2}{c|}{44.9\%} & \multicolumn{2}{c|}{20.9\%} & \multicolumn{2}{c|}{7.9\%} & \multicolumn{2}{c|}{4.0\%} \\
				\hline
				Total Number & \multicolumn{8}{c|}{1145} \\
				\hline
				Total Proportion & \multicolumn{8}{c|}{17.6\%} \\
				\bottomrule[1.5pt]
				\multirow{3}{*}{$0.5\le R<0.8$} & \multicolumn{2}{c|}{Block 1} & \multicolumn{2}{c|}{Block 2} & \multicolumn{2}{c|}{Block 3} & \multicolumn{2}{c|}{Block 4} \\ \cline{2-9}
				& $f$ & SNR & $f$ & SNR & $f$ & SNR & $f$ & SNR \\ \cline{2-9}
				& $\le$ 3 mHz & 7 $\sim$ 10 & $\le$ 3 mHz & 10 $\sim$ 13 & $\le$ 3 mHz & $\ge$ 13 & $>$ 3 mHz & $\ge$ 7 \\
				\hline
				Number & \multicolumn{2}{c|}{229} & \multicolumn{2}{c|}{477} & \multicolumn{2}{c|}{438} & \multicolumn{2}{c|}{107} \\
				\hline
				Proportion & \multicolumn{2}{c|}{19.8\%} & \multicolumn{2}{c|}{24.5\%} & \multicolumn{2}{c|}{20.7\%} & \multicolumn{2}{c|}{8.3\%} \\
				\hline
				Total Number & \multicolumn{8}{c|}{1251} \\
				\hline
				Total Proportion & \multicolumn{8}{c|}{19.2\%} \\
				\bottomrule[1.5pt]
				\multirow{3}{*}{$0.8\le R<0.9$} & \multicolumn{2}{c|}{Block 1} & \multicolumn{2}{c|}{Block 2} & \multicolumn{2}{c|}{Block 3} & \multicolumn{2}{c|}{Block 4} \\ \cline{2-9}
				& $f$ & SNR & $f$ & SNR & $f$ & SNR & $f$ & SNR \\ \cline{2-9}
				& $\le$ 3 mHz & 7 $\sim$ 10 & $\le$ 3 mHz & 10 $\sim$ 13 & $\le$ 3 mHz & $\ge$ 13 & $>$ 3 mHz & $\ge$ 7 \\
				\hline
				Number & \multicolumn{2}{c|}{161} & \multicolumn{2}{c|}{360} & \multicolumn{2}{c|}{462} & \multicolumn{2}{c|}{119} \\
				\hline
				Proportion & \multicolumn{2}{c|}{14.0\%} & \multicolumn{2}{c|}{18.5\%} & \multicolumn{2}{c|}{21.8\%} & \multicolumn{2}{c|}{9.2\%} \\
				\hline
				Total Number & \multicolumn{8}{c|}{1102} \\
				\hline
				Total Proportion & \multicolumn{8}{c|}{16.9\%} \\
				\bottomrule[1.5pt]
				\multirow{3}{*}{$R\ge0.9$} & \multicolumn{2}{c|}{Block 1} & \multicolumn{2}{c|}{Block 2} & \multicolumn{2}{c|}{Block 3} & \multicolumn{2}{c|}{Block 4} \\ \cline{2-9}
				& $f$ & SNR & $f$ & SNR & $f$ & SNR & $f$ & SNR \\ \cline{2-9}
				& $\le$ 3 mHz & 7 $\sim$ 10 & $\le$ 3 mHz & 10 $\sim$ 13 & $\le$ 3 mHz & $\ge$ 13 & $>$ 3 mHz & $\ge$ 7 \\
				\hline
				Number & \multicolumn{2}{c|}{246} & \multicolumn{2}{c|}{704} & \multicolumn{2}{c|}{1050} & \multicolumn{2}{c|}{1010} \\
				\hline
				Proportion & \multicolumn{2}{c|}{21.3\%} & \multicolumn{2}{c|}{36.1\%} & \multicolumn{2}{c|}{49.6\%} & \multicolumn{2}{c|}{78.4\%} \\
				\hline
				Total Number & \multicolumn{8}{c|}{3010} \\
				\hline
				Total Proportion & \multicolumn{8}{c|}{46.3\%} \\
				\bottomrule[1.5pt]
			\end{tabular}
			\caption{We categorize the reported sources into four blocks based on SNR and frequency ($f$) and analyze the proportion of sources with $R$ values within different ranges in each block. We also calculate the proportion of sources within different $R$ ranges relative to the total number of reported sources when all blocks are considered together. We utilize injection sources with SNR ranging from 3 to 15 to compute the correlation coefficient ($R$), ensuring that our reported sources fall within this range and do not correspond to injection sources with higher SNR.} 
			\label{table:11} 
		\end{center}
	\end{table*}

	Tab.~\ref{table:31} presents the overlapping signals identified in the reported sources, with their definitions following Sec.~\ref{sec:overlap_signal}. Our results reveal up to three consecutive overlapping signals, while non-overlapping signals (sole) account for 85.7\%. The $\text{FAS}_{0.8}$ for non-overlapping signals is 35.2\%, increasing to 45.3\% for two overlapping signals and further rising to 58.0\% for three overlapping signals. 
	As the number of overlapping signals increases, the FAS increases, though this effect can be mitigated through multi-source fitting (see Tab.~\ref{table:4} in the Appendix \ref{sec:case_study}). However, non-overlapping signals still constitute the majority of our reported sources, yet the FAS remains relatively high and cannot be ignored. This suggests that the detection challenges are not solely due to source overlap. A more significant factor is likely foreground noise, originating from the gravitational waves of unresolved double white dwarfs. This noise leads to an overestimation of all SNR values, which were initially calculated based only on instrument noise.
	\begin{table}
		\renewcommand{\arraystretch}{1.5}
		\begin{center}  
			\begin{tabular}{|m{2cm}<{\centering}|m{2cm}<{\centering}|m{1cm}<{\centering}|m{1cm}<{\centering}|m{1cm}<{\centering}|}
				\hline
				& & Sole & Two & Three \\ \cline{2-5}
				& Reported & 5579 & 860 & 69 \\
				\hline
				\multirow{2}{*}{$R<0.5$} & Number & 916 & 210 & 19 \\ \cline{2-5}
				& Proportion & 16.4\% & 24.4\% & 27.5\% \\
				\hline
				\multirow{2}{*}{$0.5\le R<0.8$} & Number & 1050 & 180 & 21 \\ \cline{2-5}
				& Proportion & 18.8\% & 20.9\% & 30.4\% \\
				\hline
				\multirow{2}{*}{$0.8\le R<0.9$} & Number & 930 & 164 & 8 \\ \cline{2-5}
				& Proportion & 16.7\% & 19.1\% & 11.6\% \\
				\hline
				\multirow{2}{*}{$R\ge0.9$} & Number & 2683 & 306 & 21 \\ \cline{2-5}
				& Proportion & 48.1\% & 35.6\% & 30.4\% \\
				\hline
			\end{tabular}
			\caption{The overlapping signals from reported sources.} 
			\label{table:31}
		\end{center}   
	\end{table}

\begin{figure}
		\centering
		\includegraphics[width=1\linewidth]{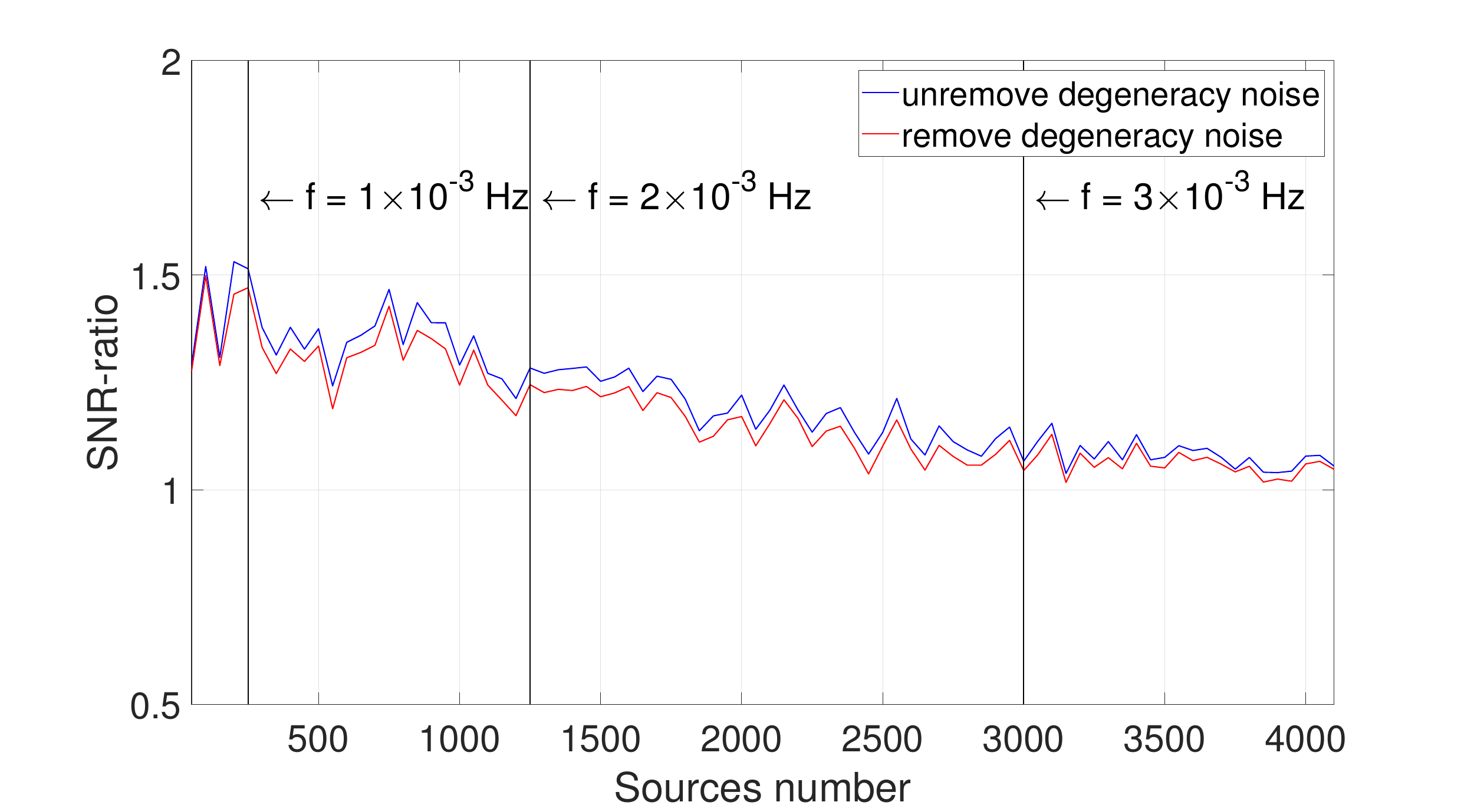}
		\caption{The SNR-ratio is calculated for reported sources with $R\ge0.8$ in comparison to their corresponding injection sources. The term ``unremove degeneracy noise" signifies the SNR before the find-real-$\mathcal{F}$-statistic analysis, whereas ``remove degeneracy noise" indicates the SNR after the find-real-$\mathcal{F}$-statistic analysis. A substantial number of unresolved Galactic binaries contribute energy to the sources, causing $\text{SNR}_\text{{con}}$ being significantly larger than their corresponding $\text{SNR}_\text{{inj}}$ as the frequency decreases.}
		\label{fig:snrratio}
	\end{figure}

	Owing to signal overlap, the $\mathcal{F}$-statistic value for each source may receive contributions from other sources, resulting in an inflated SNR. To illustrate this effect, we depict the SNR-ratio between reported sources with $R\ge0.8$ and their corresponding injection sources in Fig.~\ref{fig:snrratio}.
	\begin{equation}
		\text{SNR-ratio} = \frac{\text{SNR}_\text{{con}}}{\text{SNR}_\text{{inj}}}.
	\end{equation}
	The sources are arranged in ascending frequency order, and to achieve a smoother curve, we take an average for every 50 sources. In Fig.~\ref{fig:snrratio}, ``unremove degeneracy noise" refers to the SNR before the find-real-$\mathcal{F}$-statistic analysis, while ``remove degeneracy noise" corresponds to the SNR after the find-real-$\mathcal{F}$-statistic analysis (the steps for removing degeneracy noise are outlined in Sec.~\ref{sec:Comprehensive_analysis1} and Sec.~\ref{sec:Comprehensive_analysis2}). The average SNR-ratio is 1.21 and 1.17 for ``unremove degeneracy noise" and ``remove degeneracy noise", respectively. After the find-real-$\mathcal{F}$-statistic analysis, the fact that the average SNR-ratio approaches 1 indicates a closer match between the SNR of the sources and their corresponding injection sources, as the $\mathcal{F}$-statistic component from other sources has been eliminated.
	As the frequency increases, the SNR-ratio gradually converges to 1. However, at lower frequencies ($f < 2\times10^{-3}$ Hz), a substantial number of unresolved Galactic binaries contribute energy (degeneracy noise) to the sources, leading to $\text{SNR}_\text{{con}}$ being significantly larger than their corresponding $\text{SNR}_\text{{inj}}$.

	\subsection{Compared with other methods}
	\label{sec:Compared}
\begin{figure}
		\centering
		\includegraphics[width=1\linewidth]{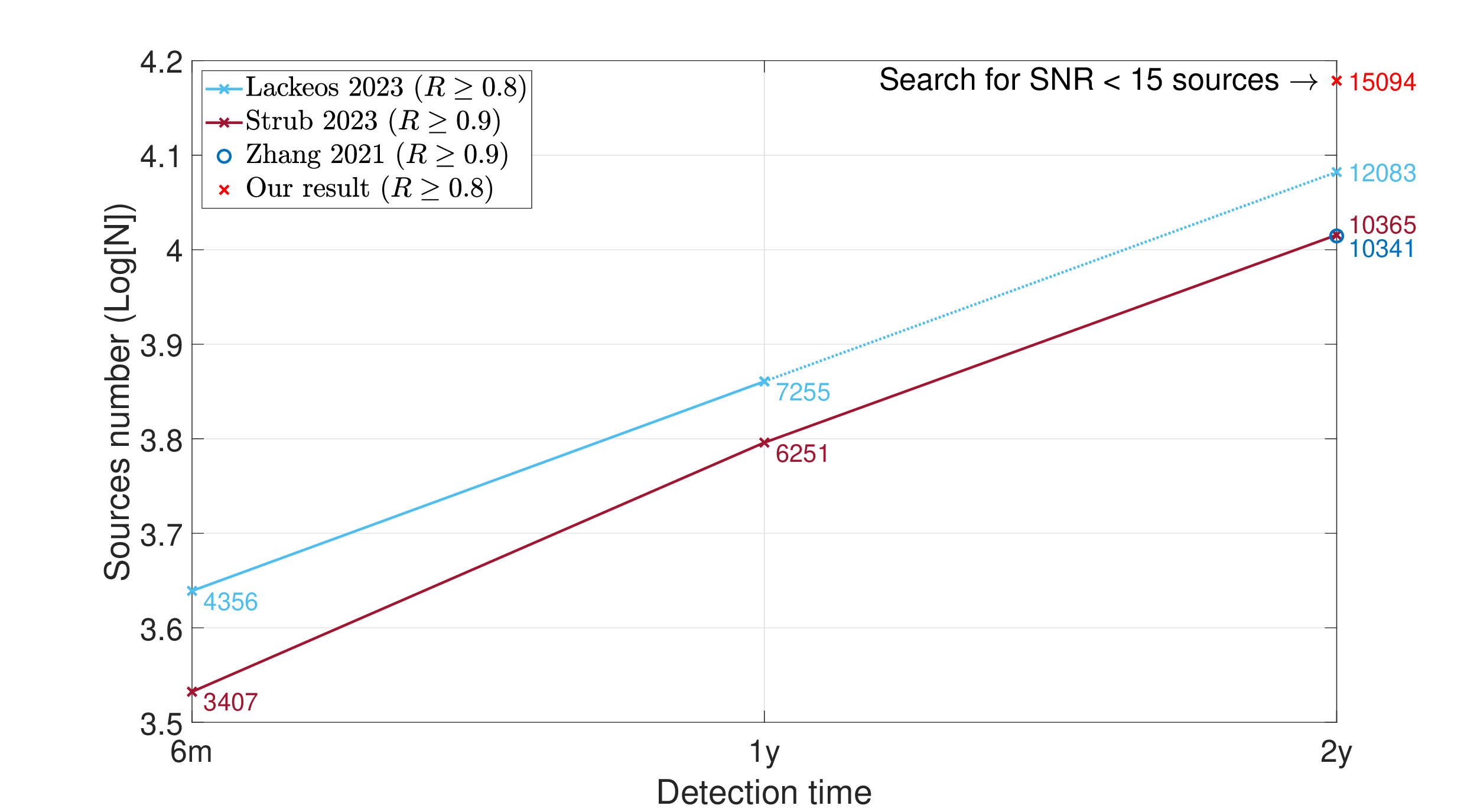}
		\caption{This figure compares the number of sources (with the ordinate in base-10 logarithm) identified by four different search methods: Lackeos 2023 from Ref.~\cite{lackeos_lisa_2023}, Strub 2023 from Ref.~\cite{2023PhRvD.108j3018S}, Zhang 2021 from Ref.~\cite{zhang_resolving_2021}, and our result. Since our search only targets sources with SNR $<$ 15, we have specifically marked this in the figure. Other results correspond to searches for sources without SNR restrictions. We highlight the number of results at various search methods and detection time.}
		\label{fig:snr_time}
	\end{figure}

    Recent progress in the search for LDC1-4 data has been reported in Refs.~\cite{zhang_resolving_2021, 2023PhRvD.108j3018S, lackeos_lisa_2023}. Since our study primarily targets sources with SNR $< 15$, a direct comparison with these results poses challenges. Nevertheless, we can roughly evaluate the effectiveness of our method by referring to Fig.~\ref{fig:snr_time}, which compares four methods: Lackeos 2023 (Table 2 in Ref.~\cite{lackeos_lisa_2023}), Strub 2023 (Table IV in Ref.~\cite{2023PhRvD.108j3018S}), Zhang 2021 (Table I in Ref.~\cite{zhang_resolving_2021}), and our result.
	
	Both Refs.~\cite{2023PhRvD.108j3018S, lackeos_lisa_2023} focus on identifying sources with detection SNR $\geq 7$. By extrapolating the Lackeos 2023 result using a dotted line in Fig.~\ref{fig:snr_time}, we estimate that their method would yield approximately 12,083 sources with $R \geq 0.8$ over a two-year detection time. Based on data in Ref.~\cite{lackeos_lisa_2023}, we compute the FAS for sources with $R \ge 0.8$ as $\mathrm{FAS}_{0.8\text{-}6m} = 29.7\%$ ($1 - \frac{4356}{6196}$) over six months, and $\mathrm{FAS}_{0.8\text{-}1y} = 27.6\%$ ($1 - \frac{7255}{10027}$) over one year. The two-year $\mathrm{FAS}_{0.8\text{-}2y}$ is expected to be even lower.
	Ref.~\cite{2023PhRvD.108j3018S} reports 10,365 detections over two years with $\mathrm{FAS}_{0.9\text{-}2y} = 13.3\%$ ($1 - \frac{10365}{11953}$), while Ref.~\cite{zhang_resolving_2021} reports 10,341 detections with $\mathrm{FAS}_{0.9\text{-}2y} = 15.7\%$ ($1 - \frac{10341}{12270}$). The FAS in our result is discussed in Sec.~\ref{sec:our_result}.
	Our analysis is based on the LDC1-4 residual data, from which sources with injection $\mathrm{SNR} \geq 15$ have been removed. In Fig.~\ref{fig:snr_time}, we therefore use the number of injected sources with $\mathrm{SNR} \geq 15$ (10,982) as a reference. Our method identifies 4,112 confirmed sources, resulting in a total of 15,094 sources.
	
\begin{figure}
	\centering
	\includegraphics[width=1\linewidth]{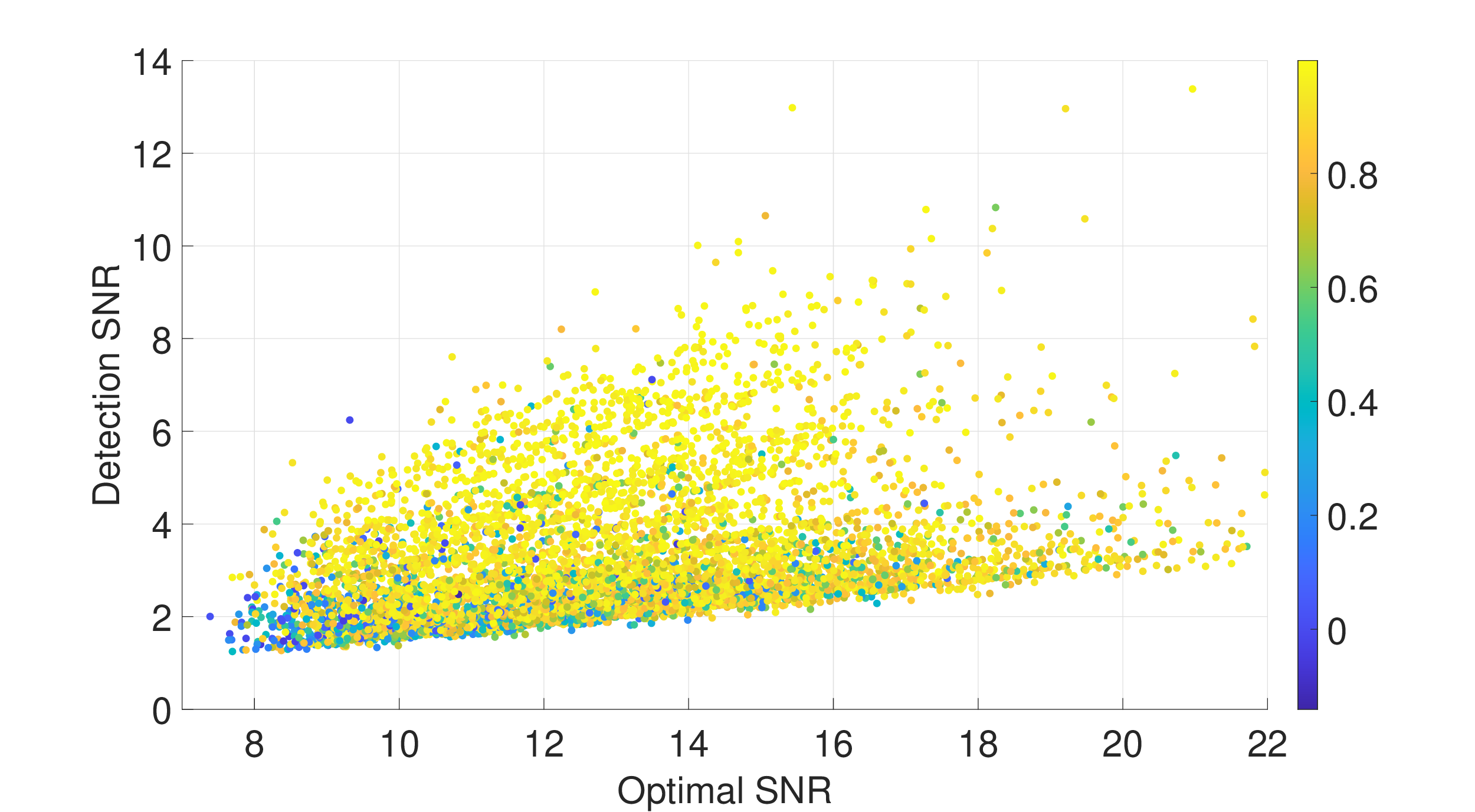}
	\caption{A scatter plot comparing the optimal SNR (X-axis) and detection SNR of reported sources. Each point represents a reported source, with its color indicating the correlation coefficient $R$.}
	\label{fig:detectionsnr}
	\end{figure}

	To enable fair comparison with Refs.~\cite{2023PhRvD.108j3018S, lackeos_lisa_2023}, which focus on detection $\mathrm{SNR} \geq 7$, we convert the optimal SNR of our reported sources to detection SNR in Fig.~\ref{fig:detectionsnr}. 
	Tab.~\ref{table:32} summarizes the number of reported sources across different detection SNR intervals and the proportion with various $R$ values. Detection SNR are calculated using the combined noise PSD of instrument noise and foreground noise in Eq.~\ref{eq:SNR}. Since our method operates without waveform subtraction, we adopt the PSD derived from the searched data (the ``Residual SNR $\ge 15$" curve in Fig.~\ref{fig:residual}).
	For sources with detection SNR $\ge 7$, our method achieves $\mathrm{FAS}_{0.9\text{-}2y} = 13.1\%$ ($1 - \frac{139}{160}$), comparable to that reported in Ref.~\cite{2023PhRvD.108j3018S} (13.3\%). Additionally, we obtain $\mathrm{FAS}_{0.8\text{-}2y} = 5.6\%$ ($1 - \frac{151}{160}$), which represents a substantial improvement over Ref.~\cite{lackeos_lisa_2023}, where $\mathrm{FAS}_{0.8\text{-}2y} \approx 27.6\%$.
	In the detection SNR range $3 \leq \mathrm{SNR} < 7$, our method achieves $\mathrm{FAS}_{0.8\text{-}2y} = 19.0\%$ ($1 - \frac{2111}{2605}$). While Refs.~\cite{2023PhRvD.108j3018S, lackeos_lisa_2023} do not report results in this range, our findings suggest that $\mathrm{FAS}_{0.8\text{-}2y}$ in the $3 \leq \mathrm{SNR} < 7$ interval is likely lower than that in the $\mathrm{SNR} \geq 7$ range reported in Ref.~\cite{lackeos_lisa_2023}. These results indicate that, within the same detection SNR range, our method yields comparable or lower FAS relative to other methods.

	By referring to Table I in Ref.~\cite{zhang_resolving_2021}, we can make a rough assessment of the effectiveness of our method, too. They present the main outcomes of their method applied to the LDC1-4 data, categorizing sources' frequencies into three distinct regions. Similarly, we categorize our result into the same regions, as shown in Tab.~\ref{table:11}. The Ref.~\cite{zhang_resolving_2021} has 2,778 reported sources with $f<3$ mHz and SNR $<$ 25. Since there are approximately 3,000 injection sources within the SNR range of 15 $\sim$ 25 and $f<3$ mHz, it appears that Ref.~\cite{zhang_resolving_2021} might not have extensively searched for sources with $f<3$ mHz and SNR $<15$. Conversely, our method identifies 5,220 reported sources within the range. 
	For $f>4$ mHz, there are approximately 1000 injection sources with SNR $<15$, where our method does not perform as effectively as that of Ref.~\cite{zhang_resolving_2021} (5 reported sources). We provide an explanation for this discrepancy in Sec.~\ref{sec:discussion}. Hence, our approach may perform better in the low frequency and low SNR range.

	\begin{table}
		\renewcommand{\arraystretch}{1.5}
		\begin{center}  
			\begin{tabular}{|m{2cm}<{\centering}|m{2cm}<{\centering}|m{1cm}<{\centering}|m{1cm}<{\centering}|m{1cm}<{\centering}|}
				\hline
				& & $\ge 7$ & 3 - 7 & $<3$ \\ \cline{2-5}
				& Reported & 160 & 2605 & 3743 \\
				\hline
				\multirow{2}{*}{$R<0.5$} & Number & 1 & 148 & 996 \\ \cline{2-5}
				& Proportion & 0.6\% & 5.7\% & 26.6\% \\
				\hline
				\multirow{2}{*}{$0.5\le R<0.8$} & Number & 8 & 346 & 897 \\ \cline{2-5}
				& Proportion & 5.0\% & 13.3\% & 24.0\% \\
				\hline
				\multirow{2}{*}{$0.8\le R<0.9$} & Number & 12 & 390 & 700 \\ \cline{2-5}
				& Proportion & 7.5\% & 15.0\% & 18.7\% \\
				\hline
				\multirow{2}{*}{$R\ge0.9$} & Number & 139 & 1721 & 1150 \\ \cline{2-5}
				& Proportion & 86.9\% & 66.1\% & 30.7\% \\
				\hline
			\end{tabular}
			\caption{The detection SNR for reported sources.}
			\label{table:32}
		\end{center}   
	\end{table}

	Our pipeline can also be applied to high SNR sources (e.g., SNR $\ge$ 15), but the significant amount of degeneracy noise originating from high SNR sources will impose greater demands on computing resources. Therefore, we believe it is preferable to initially employ the Ref.~\cite{zhang_resolving_2021} or other methods to process high SNR sources and subsequently utilize our pipeline to search for low SNR sources in the residual data. Our pipeline is primarily designed for scenarios where numerous sources with similar SNR coexist, making it particularly effective for detecting low SNR and low-frequency sources.
	
	\subsection{Deviation from injected parameters}
\begin{figure}
		\centering
		\includegraphics[width=1\linewidth]{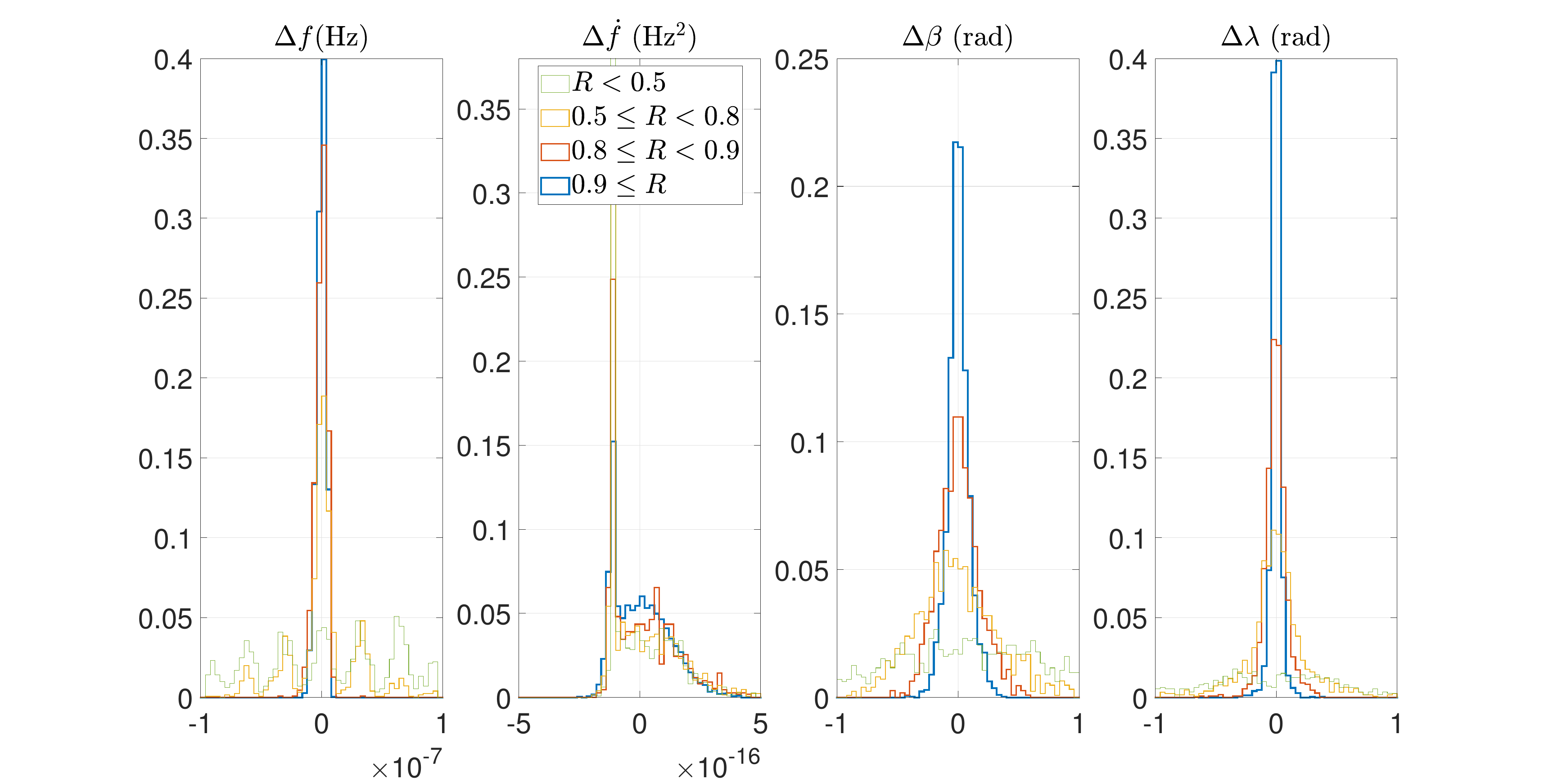}
		\caption{The parameter errors for the intrinsic parameters when $R$ takes different ranges. The second histogram has a high peak at $1\times10^{-16}$ $\text{Hz}^{2}$ because some sources have $\dot{f}$ falling on the negative boundary of the search range, resulting in similar errors.}
		\label{fig:absoluteerror}
	\end{figure}

	When considering different $R$ ranges, the reported sources exhibit varying levels of parameter errors. We categorize the reported sources into four groups based on $R$: $R \geq 0.9$, $0.8 \leq R < 0.9$, $0.5 \leq R < 0.8$, and $R < 0.5$, and investigate the magnitude of deviation from the injected parameters for each category, as shown in Fig.~\ref{fig:absoluteerror}.
	For sources with $R \geq 0.9$, characterized by high correlation coefficients, the deviation from the injected parameters is relatively smaller. Their average deviations in $f$, $\dot{f}$, $\beta$, and $\lambda$ are $3.11 \times 10^{-9}$ Hz, $9.45 \times 10^{-17}$ $\text{Hz}^2$, 0.07 rad, and 0.03 rad, respectively.
	For sources with $0.8 \leq R < 0.9$, the average deviations in $f$, $\dot{f}$, $\beta$, and $\lambda$ are $3.90 \times 10^{-9}$ Hz, $1.10 \times 10^{-16}$ $\text{Hz}^2$, 0.13 rad, and 0.09 rad, respectively.
	For sources with $0.5 \leq R < 0.8$, the average deviations in $f$, $\dot{f}$, $\beta$, and $\lambda$ are $1.77 \times 10^{-8}$ Hz, $1.18 \times 10^{-16}$ $\text{Hz}^2$, 0.28 rad, and 0.28 rad, respectively. The secondary peaks on either side of the main peak of $\Delta f$ in Fig.~\ref{fig:absoluteerror} for sources with $0.5 \leq R < 0.8$ suggest that some sources, whose $f$ values deviate from those of the real binary by more than the minimum frequency resolution $\mathrm{d}f = 1.59 \times 10^{-8}$ Hz for the two-year detection, have been searched in the degeneracy noise region.
	For sources with $R < 0.5$, characterized by low correlation coefficients, the deviations from the injected parameters are higher. Their average deviations in $f$, $\dot{f}$, $\beta$, and $\lambda$ are $4.45 \times 10^{-8}$ Hz, $1.09 \times 10^{-16}$ $\text{Hz}^2$, 0.68 rad, and 1.85 rad, respectively. Including such sources in the final catalog would significantly increase the deviations in frequency and positional parameters.
	
	For sources with $0.8 \leq R < 0.9$, the rate of change in the average deviation of $f$, $\dot{f}$, $\beta$, and $\lambda$ parameters, relative to sources with $R \geq 0.9$, is 25.4\%, 16.4\%, 85.7\%, and 200.0\%, respectively. Therefore, the main difference between relatively accurate sources and accurate sources lies in the accuracy of the sky position.
	For sources with $0.5 \leq R < 0.8$, the rate of change in the average deviation of $f$, $\dot{f}$, $\beta$, and $\lambda$ parameters, relative to sources with $0.8 \leq R < 0.9$, is 353.8\%, 7.3\%, 115.4\%, and 211.1\%, respectively. Except for $\dot{f}$, the errors in the other three parameters show a significant increase.

	\subsection{Parameter estimation}
	To estimate the parameters of each reported source, we adopt a rejection sampling approach \cite{luengo_survey_2020}, utilizing the void region in parameter space surrounding each source and its corresponding maximum $\mathcal{F}$-statistic value.
	Within the parameter space defined by this void, we assume a uniform prior and take the maximum $\mathcal{F}$-statistic value ($\mathcal{F}_\text{max}$) of the reported source as a reference. We then uniformly sample points in the void and assign each point a retention probability ($P_\text{retained}$) proportional to its $\mathcal{F}$-statistic value ($\mathcal{F}_\text{point}$):
	\begin{equation}
		P_\text{retained} = \frac{\mathcal{F}_\text{point}}{\mathcal{F}_\text{max}}.
	\end{equation}
	We ensure that 10,000 points are retained. These retained samples are subsequently used to perform kernel density estimation to reconstruct the parameter distribution of the reported source.
	
\begin{figure*}
		\centering  %图片全局居中
		\subfigbottomskip=2pt %两行子图之间的行间距
		\subfigcapskip=0pt %设置子图与子标题之间的距离
		\subfigure[\label{fig:Posterior_distribution_0.9}]{
			\includegraphics[width=0.48\linewidth]{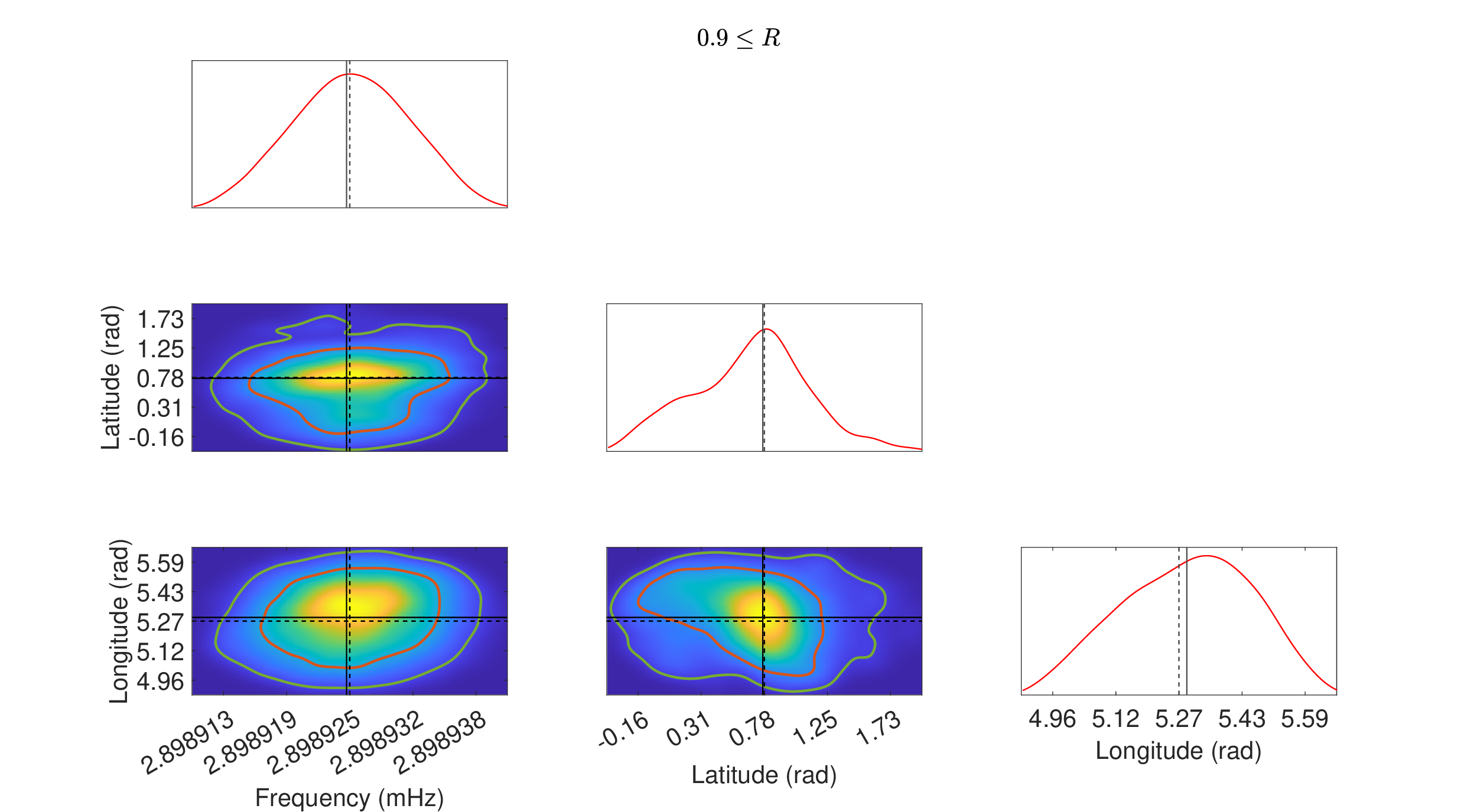}}
		\subfigure[\label{fig:Posterior_distribution_0.8_0.9}]{
			\includegraphics[width=0.48\linewidth]{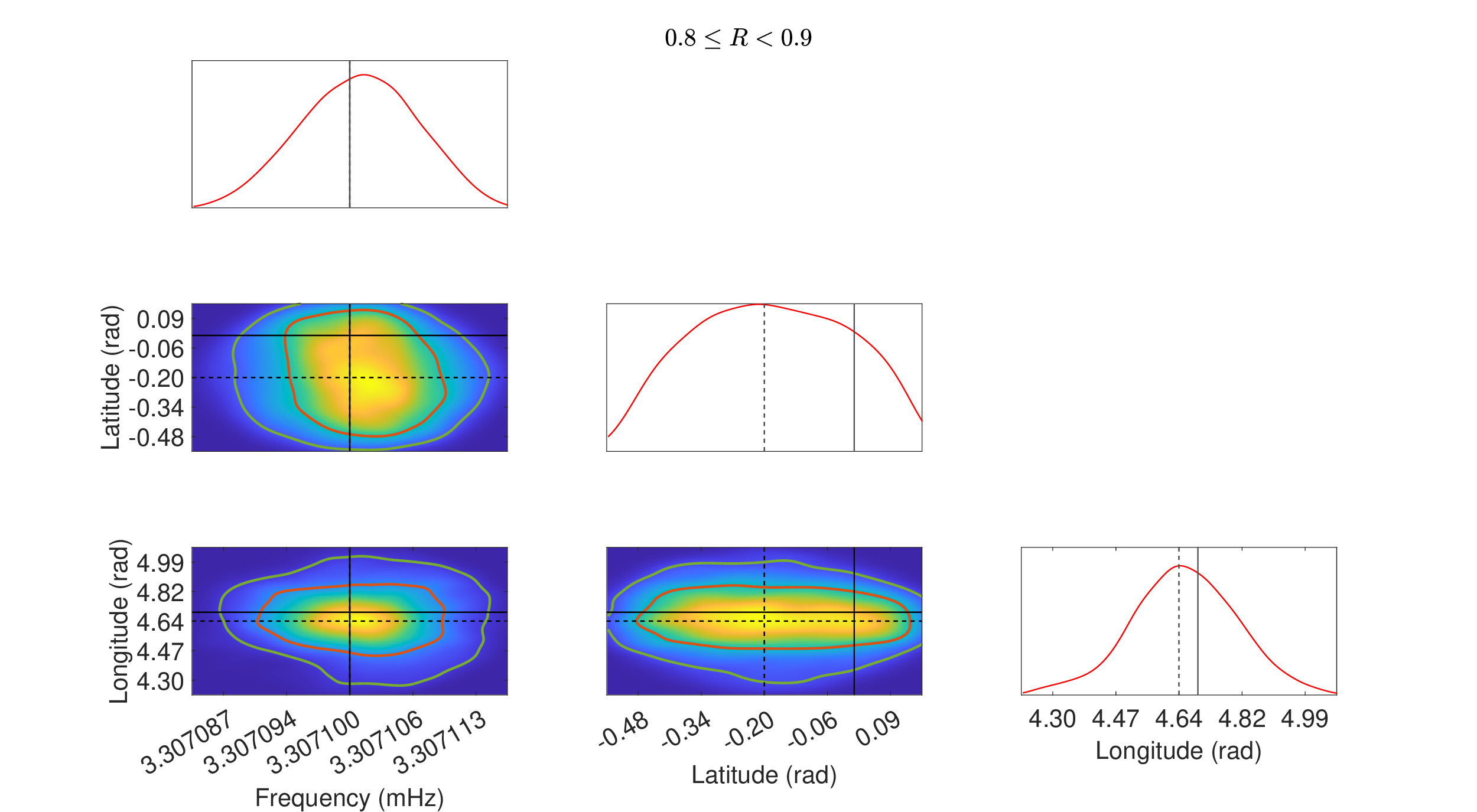}}\\
		\subfigure[\label{fig:Posterior_distribution_0.5_0.8}]{
			\includegraphics[width=0.48\linewidth]{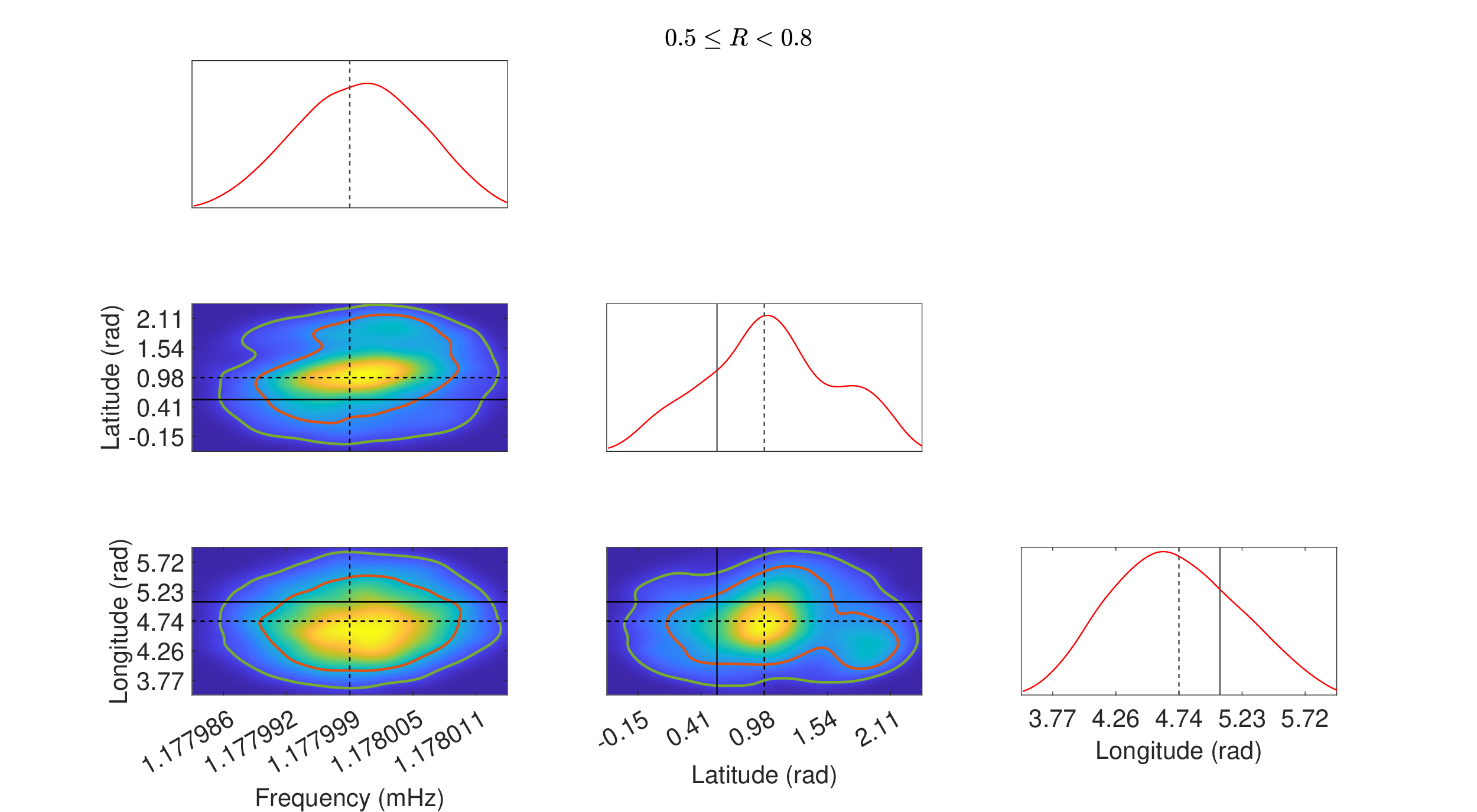}}
		\subfigure[\label{fig:Posterior_distribution_0.5}]{
			\includegraphics[width=0.48\linewidth]{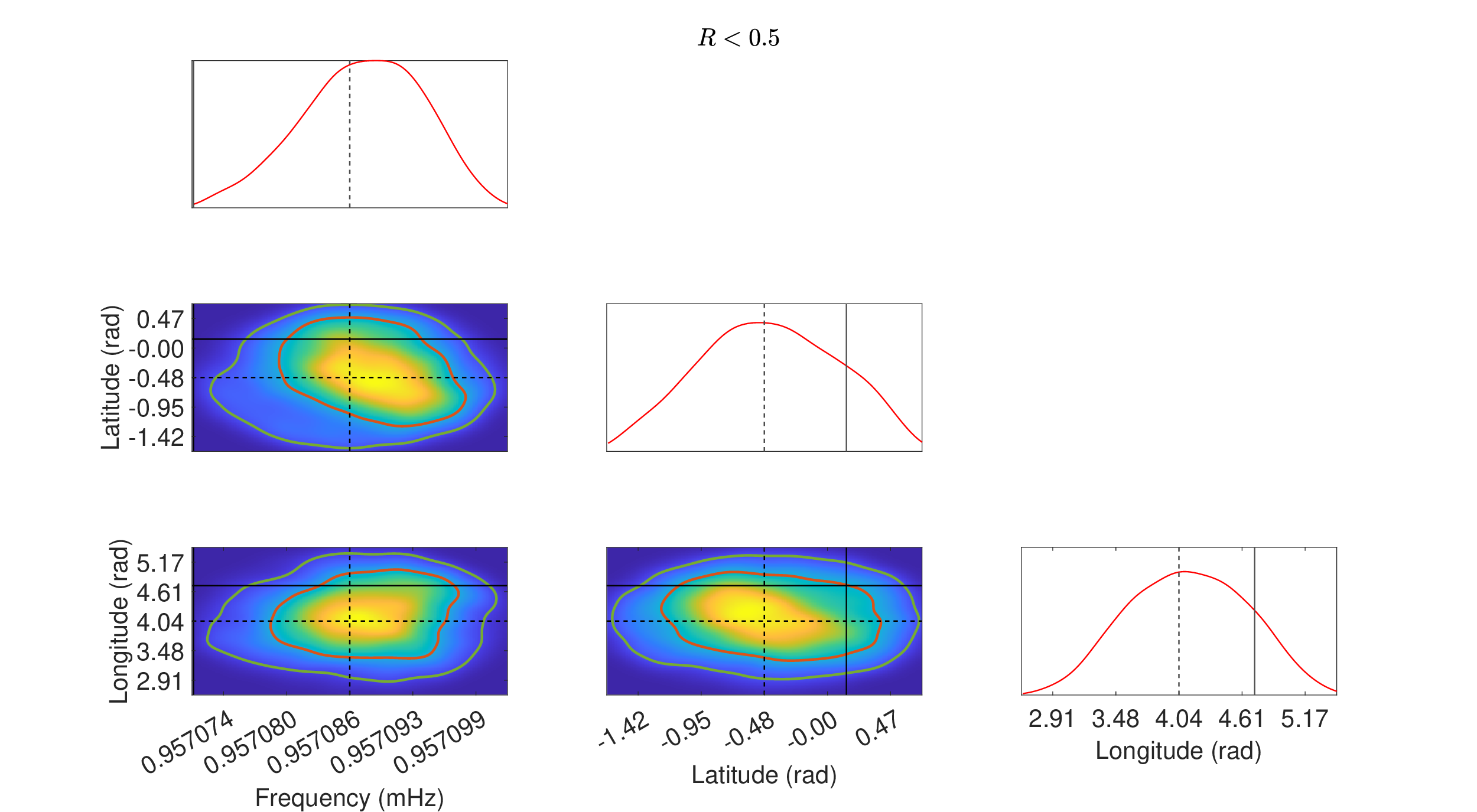}}
		\caption{\label{fig:Posterior_distribution}Parameter estimation results for four representative sources are presented. 
		Fig.~\ref{fig:Posterior_distribution_0.9} corresponds to a source with $R > 0.9$; 
		Fig.~\ref{fig:Posterior_distribution_0.8_0.9} to a source with $0.8 < R < 0.9$; 
		Fig.~\ref{fig:Posterior_distribution_0.5_0.8} to a source with $0.5 < R < 0.8$; 
		and Fig.~\ref{fig:Posterior_distribution_0.5} to a source with $R < 0.5$.
		In each panel, the black dotted lines indicate the parameter values of the reported source, 
		while the black solid lines represent the parameters of the corresponding injection source used in the calculation of $R$. 
		The red and green contours show the 68\% and 95\% credible regions of the parameter estimation, respectively.}
	\end{figure*}

	To demonstrate the effectiveness of this approach, we randomly selected four sources, each with a correlation coefficient $R$ falling into a different range: greater than 0.9, between 0.8 and 0.9, between 0.5 and 0.8, and below 0.5. The corresponding parameter distributions are shown in Fig.~\ref{fig:Posterior_distribution}.
	Fig.~\ref{fig:Posterior_distribution_0.9} corresponds to a source with $R > 0.9$, whose estimated parameter means are $2.90 \times 10^{-3}$ Hz ($f$), 0.67 rad ($\beta$), and 5.29 rad ($\lambda$), with standard deviations of $5.71 \times 10^{-9}$ Hz ($f$), 0.43 rad ($\beta$), and 0.16 rad ($\lambda$), respectively.
	Fig.~\ref{fig:Posterior_distribution_0.8_0.9} corresponds to a source with $0.8 < R < 0.9$, with parameter means of $3.31 \times 10^{-3}$ Hz ($f$), –0.18 rad ($\beta$), and 4.66 rad ($\lambda$), and standard deviations of $5.69 \times 10^{-9}$ Hz ($f$), 0.17 rad ($\beta$), and 0.14 rad ($\lambda$).
	Fig.~\ref{fig:Posterior_distribution_0.5_0.8} shows a source with $0.5 < R < 0.8$, having parameter means of $1.18 \times 10^{-3}$ Hz ($f$), 1.02 rad ($\beta$), and 4.70 rad ($\lambda$), and standard deviations of $6.05 \times 10^{-9}$ Hz ($f$), 0.58 rad ($\beta$), and 0.48 rad ($\lambda$).
	Fig.~\ref{fig:Posterior_distribution_0.5} presents a source with $R < 0.5$, whose parameter means are $9.57 \times 10^{-4}$ Hz ($f$), –0.45 rad ($\beta$), and 4.11 rad ($\lambda$), with standard deviations of $5.98 \times 10^{-9}$ Hz ($f$), 0.50 rad ($\beta$), and 0.53 rad ($\lambda$).
	
	In each panel, the black dotted lines indicate the parameter values of the reported source, while the black solid lines mark those of the corresponding injection source used to compute $R$. The red and green contours represent the 68\% and 95\% credible regions, respectively.  
	For the two cases with $R > 0.9$ and $0.8 < R < 0.9$, the injection source parameters lie within the 68\% credible region, indicating good agreement and successful parameter estimation.
	
	\subsection{Computational time}
	In the first search (without astrophysical models), a total of 849,576 local maxima were explored, necessitating an average of 310 PSO iterations per search. The average number of $\mathcal{F}$-statistic calculations for a local maximum is $310 \times 40 = 12,400$, including particles that are deemed invalid due to entering the voids. When determining the radii of the ``void" for a local maximum, an additional 120 particles are considered for $\mathcal{F}$-statistic calculations. The computation time totaled 816,955 seconds, utilizing 206 Intel(R) Xeon(R) Platinum 8378HC CPU @ 2.60GHz cores concurrently. This corresponds to $816,855 \times 206$ core-seconds during the search for these local maxima. Furthermore, the time required for removing individual degeneracy noise with the same configuration is approximately 5275 seconds (matlab code).
	
	\subsection{Discussion}
	\label{sec:discussion}
\begin{figure}
		\centering
		\includegraphics[width=1\linewidth]{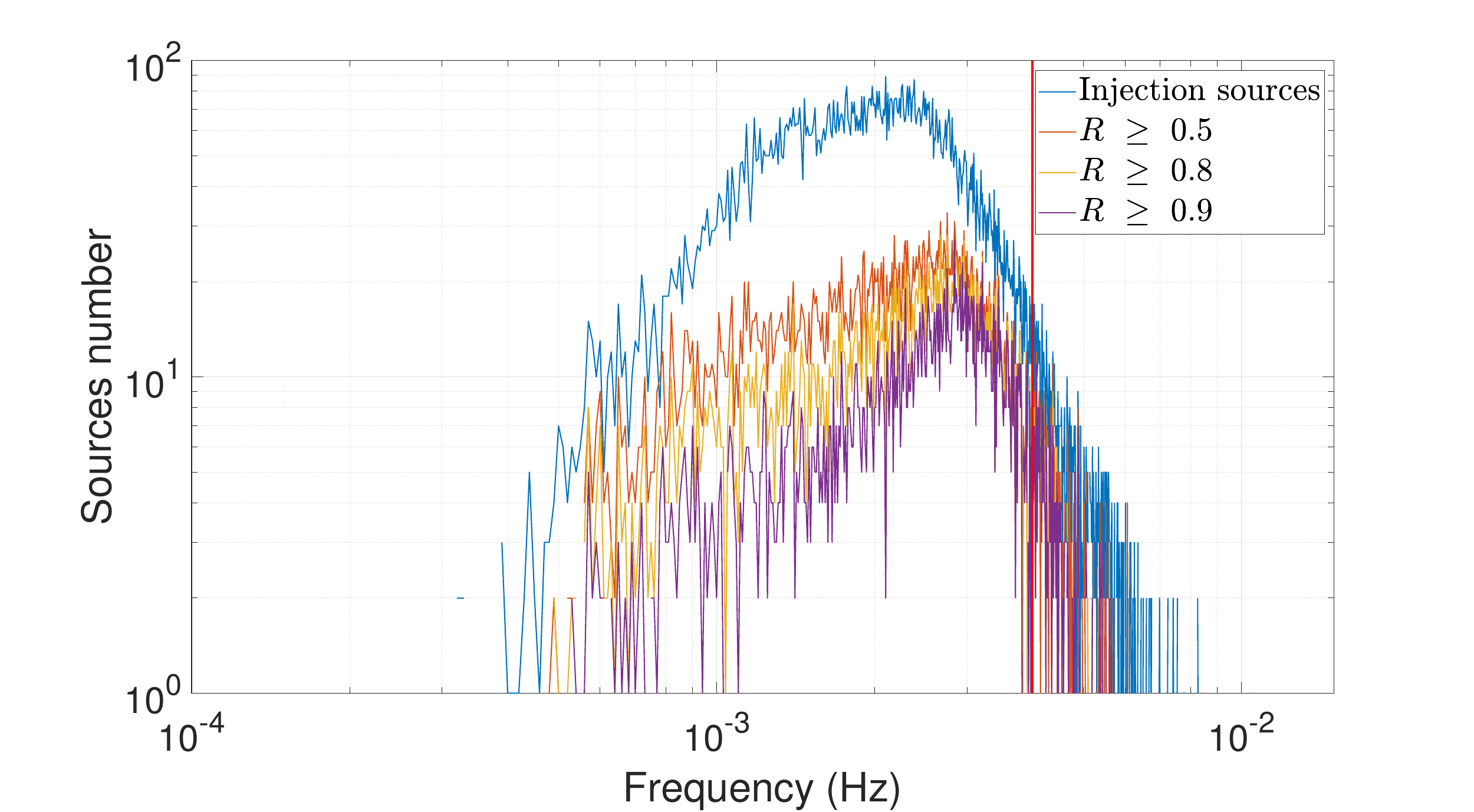}
		\caption{The number of sources in different frequency bins. The injection sources denoting those with 7 $<$ SNR $<$ 15 in LDC1-4 data. The other three colored lines illustrate the count of sources under different $R$ ranges to the left of the red vertical line. The red line serves as the cutoff at $4\times 10^{-3}$ Hz. On the right side of the line, we utilize the outcomes after removing individual degeneracy noise, restricting Galactic Latitude, and removing overlapping degeneracy noise in the second search.}
		\label{fig:numberdistribution}
	\end{figure}

	The numbers of sources with $R\ge0.8$ in different frequency bins are depicted in Fig.~\ref{fig:numberdistribution}. The injection sources refer to signals with 7 $<$ SNR $<$ 15 in the LDC1-4 data. With $R \ge0.5$, it becomes feasible to detect a substantial portion or approximately half of the injection sources in medium ($1 \times 10^{-3}$ Hz $ < f < 4 \times 10^{-3}$ Hz) and low ($f < 1 \times 10^{-3}$ Hz) frequency ranges. However, in the frequency bins with the highest signal density ($\sim 2\times 10^{-3}$ Hz), a notable number of sources remain undetected ($> 50\%$). This observation may be related to our predefined stop-search criteria or potential interference between the signals. However, a more significant factor is likely the influence of foreground noise, which causes the SNR calculated using instrument noise to be higher than the actual SNR in the search. At higher frequencies ($f > 4 \times 10^{-3}$ Hz), our method demonstrates poor performance. This is attributed to the widening frequency distribution of degeneracy noise as the signal frequency increases. Consequently, it becomes easier to identify the ``secondary peaks" (degeneracy noise) produced by the signal rather than the ``primary peak" (real signal). To enhance performance at high frequencies, allocating additional computing resources is advisable, and adjustments to our stop search criteria may be necessary.
	However, by constraining the $\dot{f}$ range and conducting a logarithmic search without altering other settings, we successfully identify the sources at higher frequencies. In Fig.~\ref{fig:numberdistribution}, the red line serves as the cutoff at $4\times 10^{-3}$ Hz. On the right side of the line, we utilize the outcomes after removing individual degeneracy noise, restricting the Galactic Latitude, and removing overlapping degeneracy noise in the second search (no cross-validation). On the left side of the line, we employ the reported sources with $R\ge0.8$ obtained through complete find-real-$\mathcal{F}$-statistic-analysis.
	
\begin{figure}
		\centering
		\includegraphics[width=1\linewidth]{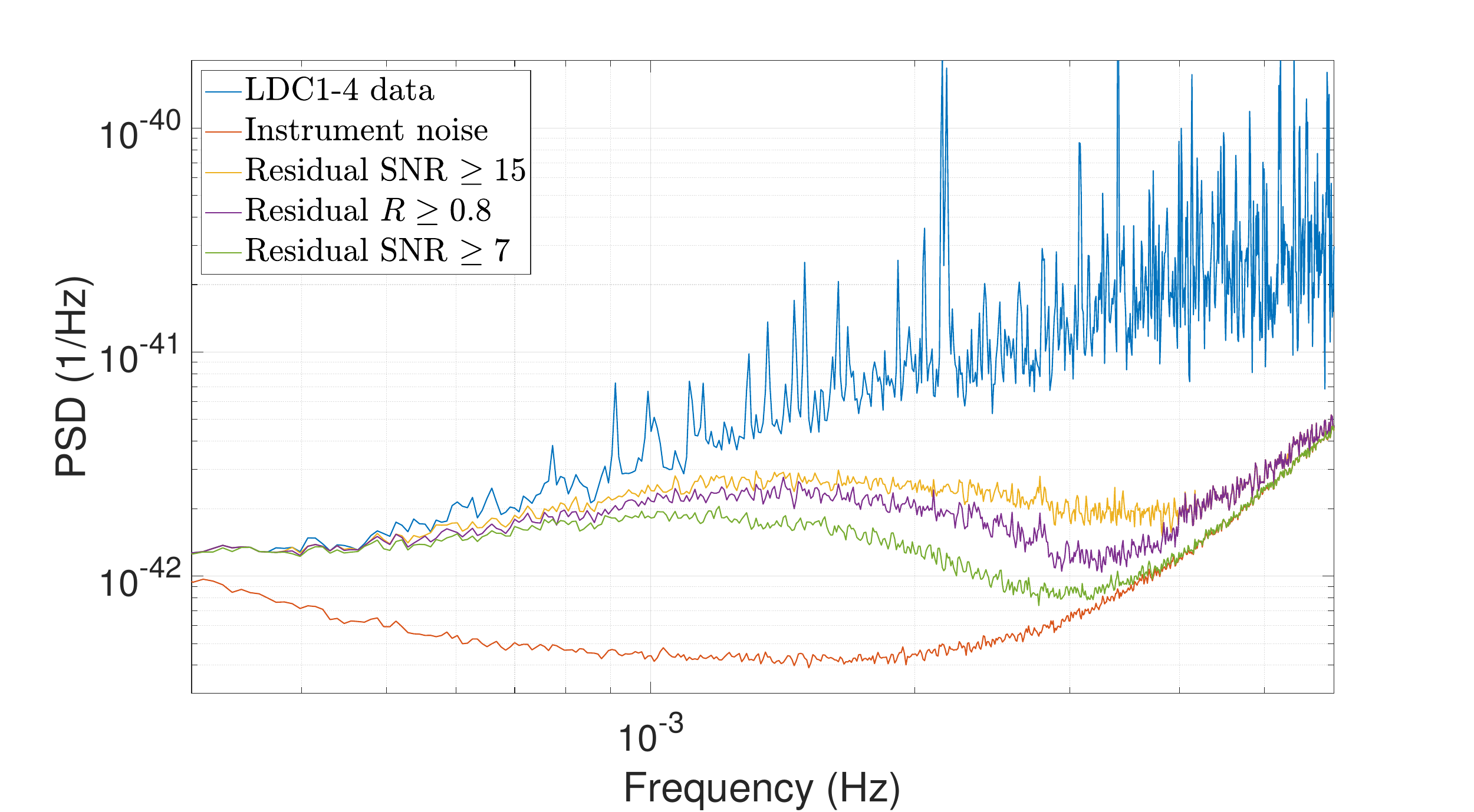}
		\caption{Power spectral density of the different datasets.}
		\label{fig:residual}
	\end{figure}

	In Fig.~\ref{fig:residual}, we present the Power Spectral Density (PSD) of various datasets. The term ``LDC1-4 data'' refers to raw data from the LISA Data Challenge, while ``Instrument noise'' represents simulated noise for the LISA mission. ``Residual SNR $\ge15$'' denotes LDC1-4 residual data, where injection sources with SNR $\ge$ 15 have been removed—representing the dataset on which our method operates. The term ``Residual $R\ge0.8$'' signifies data from which all reported sources with $R\ge0.8$ have been subtracted from ``Residual SNR $\ge15$'' (as depicted in ``remove degeneracy noise'' in Fig.~\ref{fig:snrratio}). We update the amplitude $\mathcal{A}$ of each source in proportion to the square root of the difference in their $\mathcal{F}$-statistic values before and after the removal of degeneracy noise. Finally, ``Residual SNR $\ge7$'' corresponds to LDC1-4 residual data with injection sources removed based on SNR $\ge$ 7, making it an ideal target for our search methodology. There are instances where some sources are not detected, or the reported sources in the results lack accuracy ($R$ $<$ 0.8).
	This phenomenon is attributed to the limitations of the current create voids and find-real-$\mathcal{F}$-statistic-analysis methods, particularly in resolving signals with very close parameters. For instance, when the frequency difference between two signals is less than $1\times \mathrm{d}f$, it becomes challenging to treat them as individual signals. In such scenarios, effectively eliminating degeneracy noise originating from both signals becomes intricate. Even in situations where two or three signals overlap, despite the frequency difference exceeding $1\times \mathrm{d}f$ (as illustrated in Fig.~\ref{fig:F_statistic_distribution9} and Fig.~\ref{fig:F_statistic_distribution14}), inconsistencies between the $\mathcal{F}$-statistic peak and the binary position add complexity to the subsequent process of removing degeneracy noise. These complexities contribute to the increase in the FAS. To overcome these challenges, other approaches, such as multi-source fitting or leveraging machine learning to analyze the distribution of $\mathcal{F}$-statistic values directly in the local maxima, represent promising avenues.

	\section{Conclusion}
	\label{sec:Conclusion}
	We introduce the LMPSO-CV approach for the simultaneous detection of Galactic binary gravitational waves with a space-based GW detector. To handle the numerous local maxima of the $\mathcal{F}$-statistic, most of which are degeneracy noise, we employ a find-real-$\mathcal{F}$-statistic analysis to effectively remove both degeneracy and instrument noise. We present the flowchart of our method in Fig.~\ref{fig:flowchart4}. Notably, our approach mitigates the challenges of inaccurate subtraction contamination, especially when resolving low SNR signals (e.g., SNR $<$ 15), a common difficulty in traditional iterative-subtraction methods.
	The find-real-$\mathcal{F}$-statistic analysis consists of four key steps: 1) eliminating individual signal degeneracy noise, 2) applying astrophysical models, and 3) restricting the $\mathcal{F}$-statistic difference and removing overlapping signal degeneracy noise. After completing these steps, our method identifies 6,508 signals with an $\text{FAS}_{0.8}$ of 36.8\% when applied to the LDC1-4 residual data (with 10,982 waveforms of injection sources with SNR $\ge$ 15 removed). When considering both Blocks 3 and 4, a total of 3,406 sources are identified, resulting in a $\text{FAS}_{0.8}$ of 22.5\%. For Block 4 alone, we identify 1,288 sources, with an $\text{FAS}_{0.8}$ of 12.3\%.

	Within the same detection SNR range, our method achieves a comparable or lower $\text{FAS}$ than other existing methods. However, due to the limitation imposed by the Galactic latitude constraint, our pipeline primarily focuses on low-SNR Galactic binaries (mainly double white dwarfs) located near the Galactic disk.
	
	In cases where two or more signals exhibit similar parameters, especially in terms of frequency, overlapping signals are inevitable. The situation becomes more complex when the frequency separation between two signals with comparable SNR is less than five times the minimum frequency resolution achievable during a two-year observation ($\Delta f < 5\times \mathrm{d}f$). In such cases, the $\mathcal{F}$-statistic peaks corresponding to these overlapping signals become intricate, posing significant challenges in the signal resolution process. This complexity further complicates the task of effectively reducing the FAS.

	%%%%%%%%%%%%%%%%%%%%%%%%%%%%%%%%%%%%%%%%%%%%%%%%%%%%%%%%%%%%%%

	\section*{Acknowledgments}
	We thank the anonymous referees for valuable comments and suggestions that helped us to significantly improve the manuscript. P. Gao thanks S. D. Mohanty, X. Zhang, J. Ming, S. Zhao and Y. Guo for their  invaluable comments and suggestions. Our  Local Maxima Particle Swarm Optimization (LMPSO) algorithm is developed based on the PSO script in GBSIEVER method \cite{zhang_resolving_2021} (this PSO code is actually taken from the public repository https://github.com/mohanty-sd/SDMBIGDAT19.git). 
	This work is supported by National Key R\&D Program of China (2020YFC2201400), and by the NSFC 
	(No.~11920101003, No.~11922303, No.~12021003 and No.~12005016). X. Fan. is supported by the Fundamental Research Funds for the Central Universities (No.2042022kf1182). Z. Cao was supported by ``the Interdiscipline Research Funds of Beijing Normal University" and CAS Project for Young Scientists in Basic Research YSBR-006. We gratefully appreciate the Wuhan Supercomputing Center for the generous computing resources to support our research efforts. We also thank LISA Consortium’s LDC
	working group for curating and supporting the simulated
	data used in this study.

	%%%%%%%%%%%%%%%%%%%%%%%%%%%%%%%%%%%%%%%%%%%%%%%%%%%%%%%%%%%%%%

	\appendix	
	\section{The case study of the method}
	\label{sec:case_study}
\begin{figure*}
		\centering  %图片全局居中
		\subfigbottomskip=2pt %两行子图之间的行间距
		\subfigcapskip=-5pt %设置子图与子标题之间的距离
		\subfigure[\label{fig:casestudy1}]{
			\includegraphics[width=0.48\linewidth]{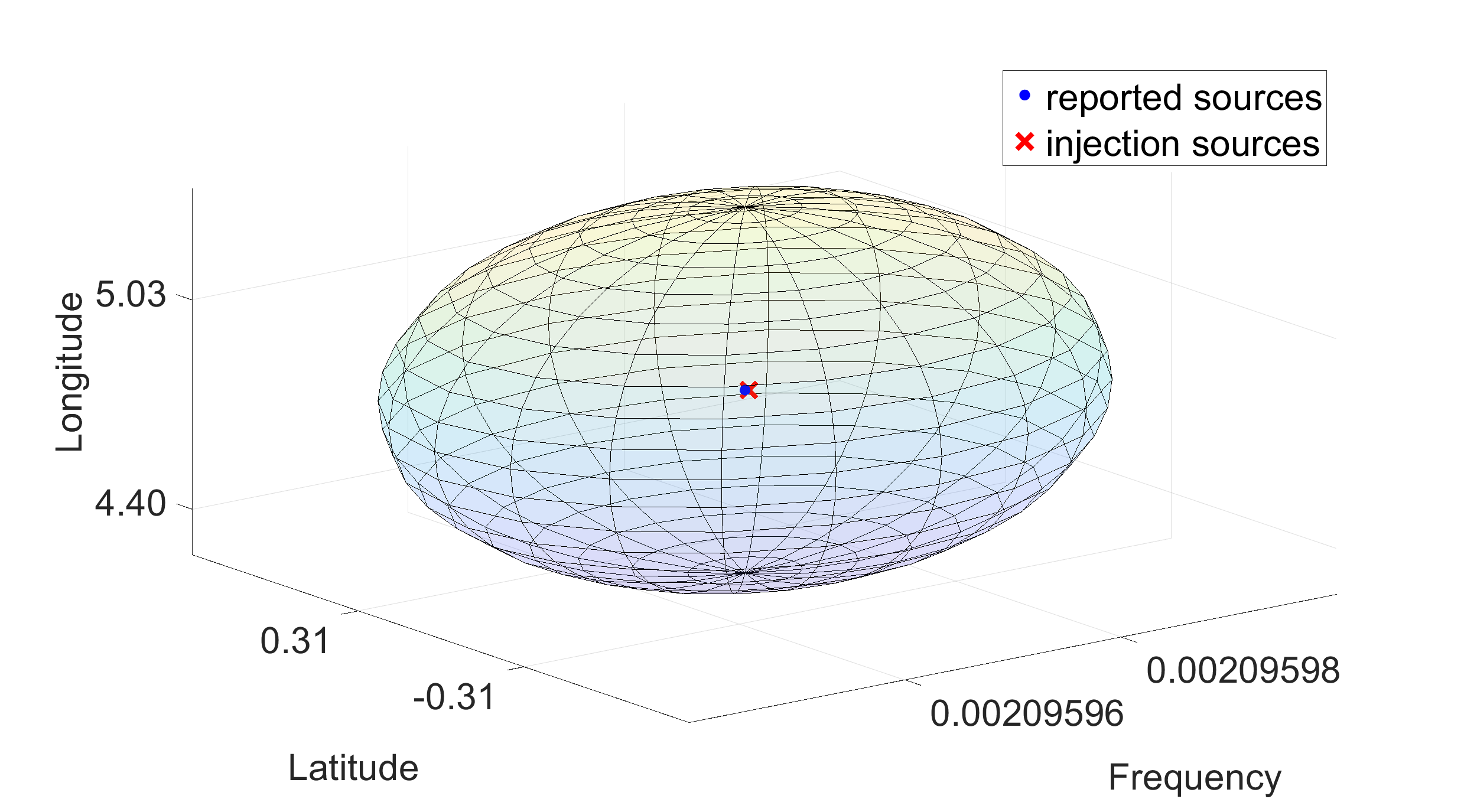}}
		\subfigure[\label{fig:casestudy4}]{
			\includegraphics[width=0.48\linewidth]{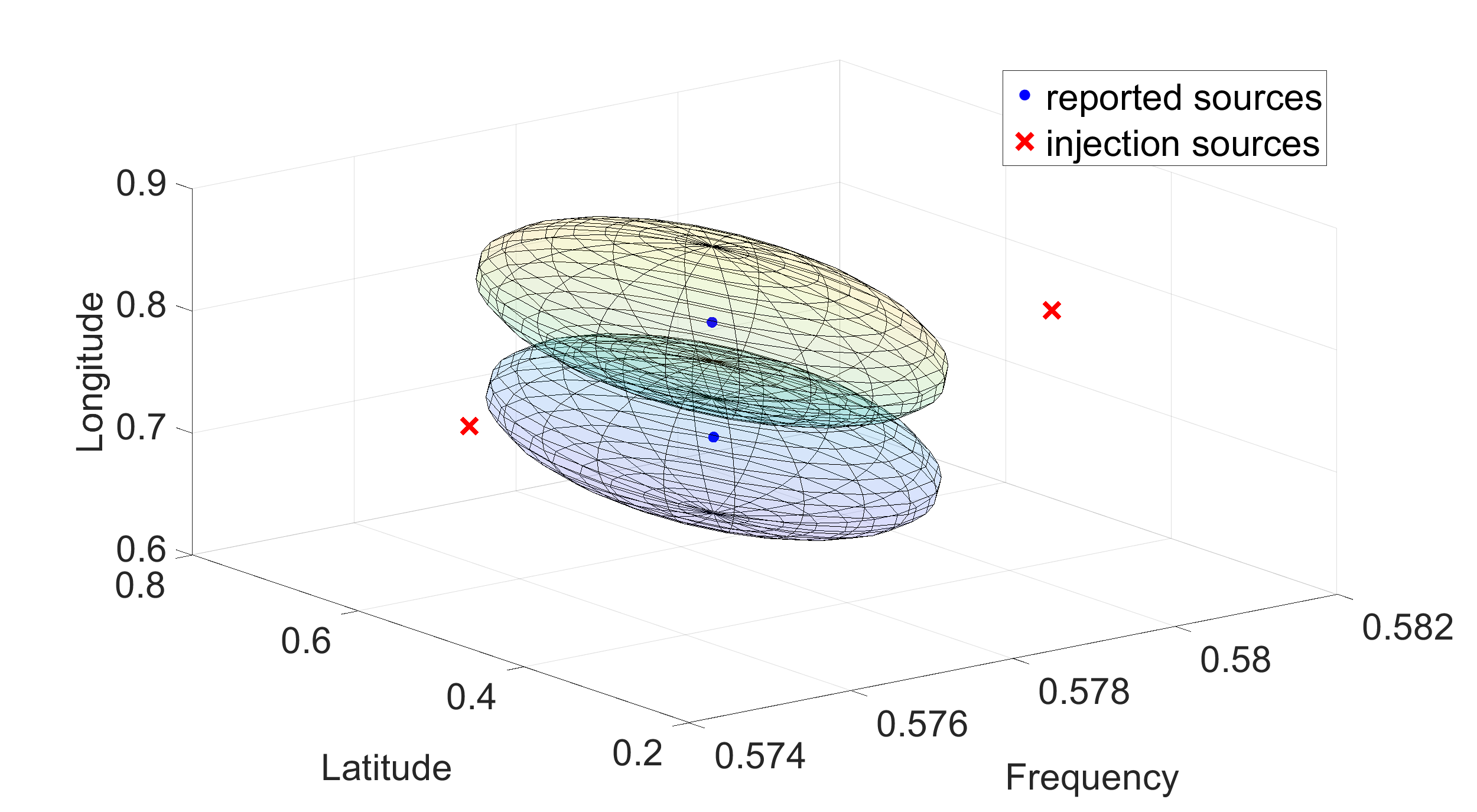}}\\
		\subfigure[\label{fig:casestudy3}]{
			\includegraphics[width=0.48\linewidth]{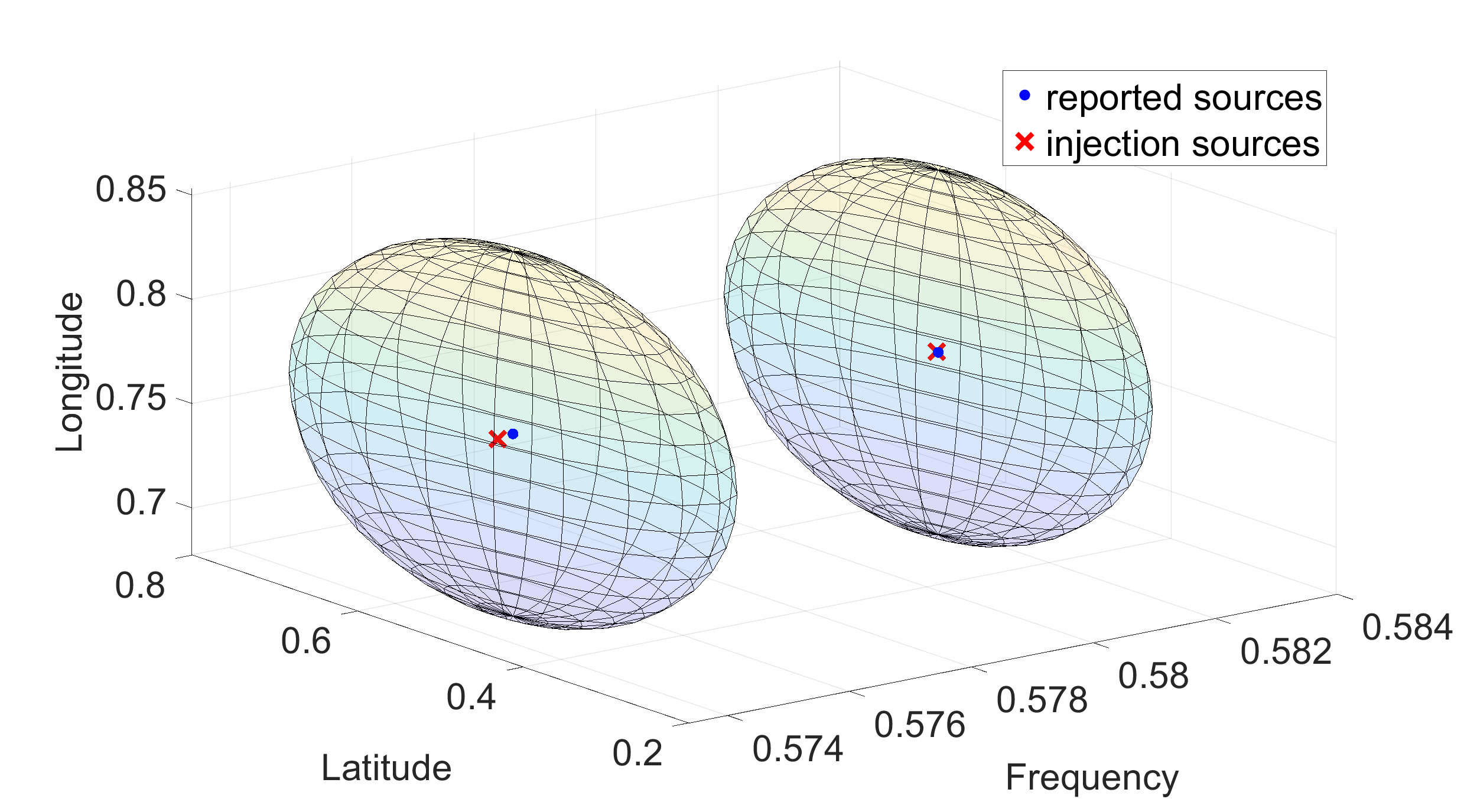}}
		\caption{The red crosses signify the position of injection sources in the parameter space, while the blue dots represent the reported sources identified by our method. The ellipsoids depict the voids with the reported sources as the center.}
	\end{figure*}

	We employed our method to analyze specific sources and observed its effects. In Tab.~\ref{table:2}, we present the parameters of binary $A$, which were used to generate Fig.~\ref{fig:fstatisticdistribution} and Fig.~\ref{fig:distribution41}. For the test, we first generated the gravitational waveform of binary $A$ in the LISA data, then introduced instrument noise to create mock data. Since there is only one gravitational wave signal in the data, identifying the signal and removing noise is straightforward. In Fig.~\ref{fig:casestudy1}, we present the parameter positions of the injection source (binary $A$) and the source identified by our method in normalized parameter space. It is evident that the positions of the two sources coincide, confirming the accuracy and effectiveness of our search. The correlation coefficient for the identified source is 0.9998, as shown in Tab.~\ref{table:4}.

	\begin{table*}
		\renewcommand{\arraystretch}{1.5}
		\begin{center}   
			\begin{tabular}{|c|c|c|c|c|c|c|c|c|c|}
				\hline
				& $f$ & $\dot{f}$ & $\beta$ & $\lambda$ & $\mathcal{A}$ & $\iota$ & $\psi$ & $\phi_{0}$ & SNR \\
				\hline
				Binary $A$ & 0.002090825 & $5.798272 \times 10^{-17}$ & 0.034327 & 4.809826 & $3.127627\times 10^{-22}$ & 1.036248 & 0.228909 & 3.227355 & 139.3 \\
				\hline
			\end{tabular}
			\caption{The parameters of the binary $A$.} 
			\label{table:2} 
		\end{center}   
	\end{table*}
	
\begin{figure*}
		\centering
		\includegraphics[width=1\linewidth]{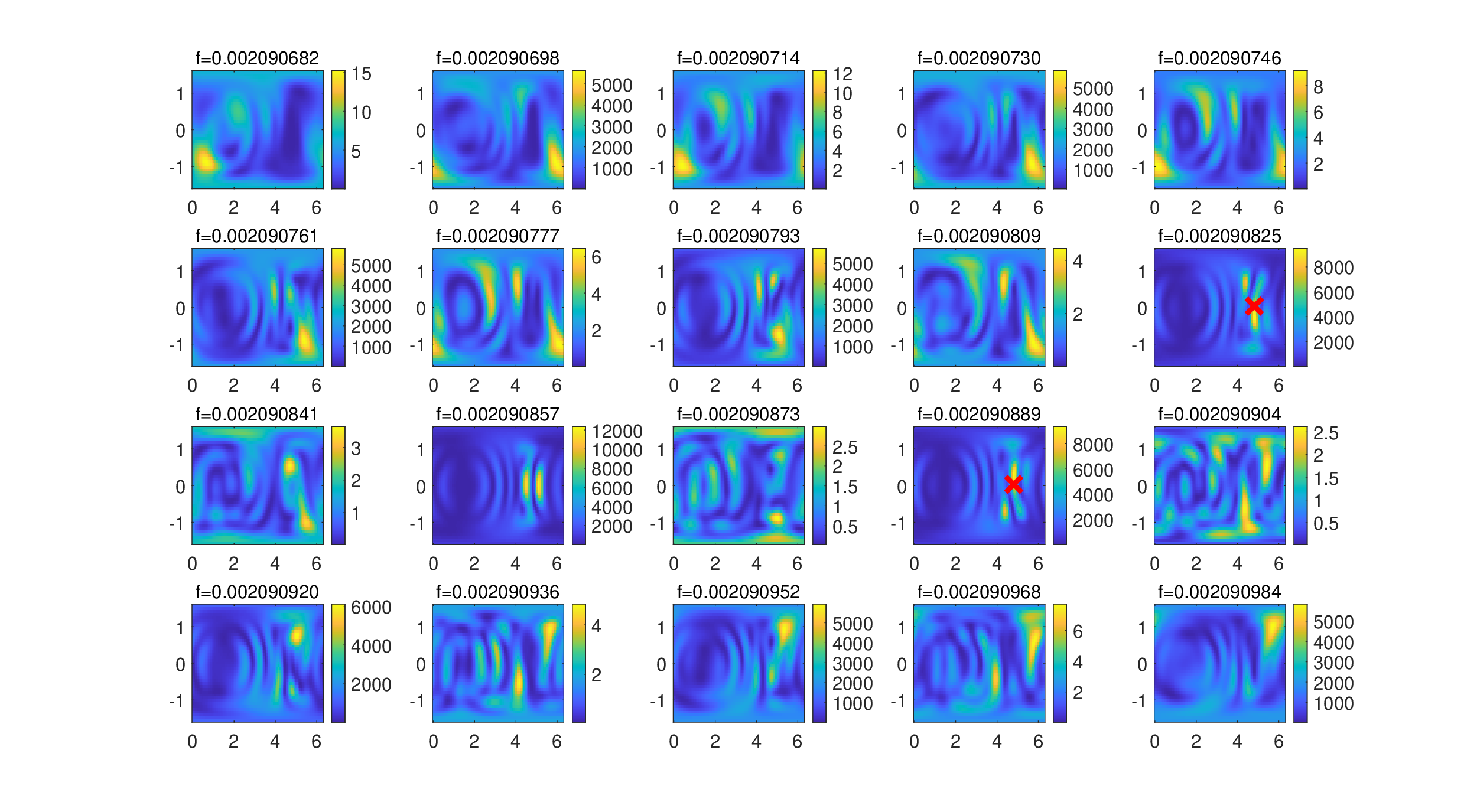}
		\caption{Slices of the $\mathcal{F}$-statistic in three-dimensional parameter space for binary $A$ and $B$. The parameter position of binary $A$ is marked in the subfigure (2, 5) with a red cross, and the parameter position of binary $B$ is marked in the subfigure (3, 4) with a red cross. The frequency intervals between two adjacent subfigures correspond to the minimum frequency resolution $\mathrm{d}f=1.59\times10^{-8}$ Hz. In each subfigure, the horizontal coordinate represents Ecliptic Longitude $\lambda$, and the vertical coordinate is Ecliptic Latitude $\beta$. The figure does not encompass the entire frequency range of the degeneracy noise distribution.}
		\label{fig:fstatisticdistribution16}
	\end{figure*}

	We also assess the performance of our method in scenarios involving overlapping signals. Gravitational waveforms are generated for binaries $A$ and $B$ in the LISA detector, with the frequency of binary $B$ increased by $4 \times \mathrm{d}f$ relative to binary $A$, corresponding to the configurations shown in Fig.~\ref{fig:F_statistic_distribution8} and Fig.~\ref{fig:F_statistic_distribution9}. These waveforms, combined with instrument noise, form the mock data. Fig.~\ref{fig:fstatisticdistribution16} illustrates the $\mathcal{F}$-statistic distribution in the three-dimensional parameter space for this case, where the two crosses represent the parameter positions of binaries $A$ and $B$.
	In the first scenario, we searched for all local maxima that met the specified criteria. This search identified two reported sources with nearly identical frequencies, corresponding to two local maxima in subgraph (3, 2) of Fig.~\ref{fig:fstatisticdistribution16}. Both maxima had $\mathcal{F}$-statistic values exceeding those at the true parameter positions, with correlation coefficients of 0.3976 and 0.4128. In the second scenario, we use the search results from the first scenario (including the number of sources and the range of frequency values) as a prior and perform multi-source fitting. This yields two reported sources that are directly aligned with the true parameter positions, with correlation coefficients of 0.9970 and 0.9925, as shown in Tab.~\ref{table:4}. For the first scenario, we present the parameter positions of the injected sources (binaries $A$ and $B$) and those identified by our method in the normalized parameter space, as shown in Fig.~\ref{fig:casestudy4}.

\begin{figure*}
		\centering
		\includegraphics[width=1\linewidth]{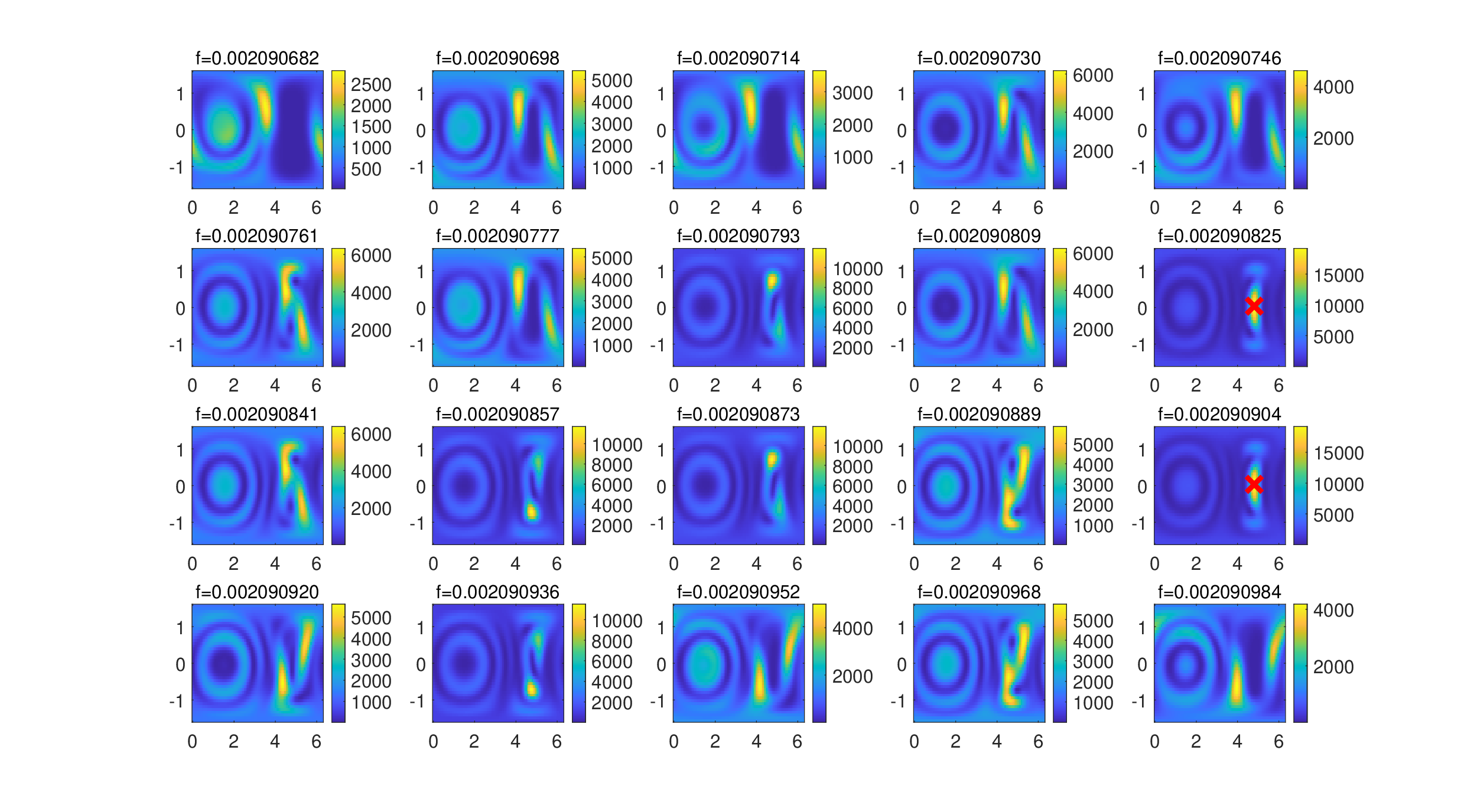}
		\caption{Slices of the $\mathcal{F}$-statistic in three-dimensional parameter space for binary $A$ and $B$. The parameter position of binary $A$ is marked in the subfigure (2, 5) with a red cross, and the parameter position of binary $B$ is marked in the subfigure (3, 5) with a red cross. The frequency intervals between two adjacent subfigures correspond to the minimum frequency resolution $\mathrm{d}f=1.59\times10^{-8}$ Hz. In each subfigure, the horizontal coordinate represents Ecliptic Longitude $\lambda$, and the vertical coordinate is Ecliptic Latitude $\beta$. The figure does not encompass the entire frequency range of the degeneracy noise distribution.}
		\label{fig:fstatisticdistribution17}
	\end{figure*}

	We generated gravitational waveforms for binaries $A$ and $B$, with the frequency of binary $B$ increased by $5 \times \mathrm{d}f$ relative to binary $A$. Fig.~\ref{fig:fstatisticdistribution17} shows the distribution of the $\mathcal{F}$-statistic in three-dimensional parameter space for the two sources, which is corresponding to the situation in Fig.~\ref{fig:F_statistic_distribution6} and Fig.~\ref{fig:F_statistic_distribution7}. Since no local maximum in the $\mathcal{F}$-statistic exceeds the value at the true positions of the two injection sources, the find-real-$\mathcal{F}$-statistic analysis proceeds smoothly, and the two sources are successfully identified. In Fig.~\ref{fig:casestudy3}, we present the parameter positions of the injection sources (binaries $A$ and $B$) and the sources identified by our method in normalized parameter space. The correlation coefficients for the two reported sources are 0.9847 and 0.9927, respectively, as shown in Tab.~\ref{table:4}.

	\begin{table*}
		\renewcommand{\arraystretch}{1.5}
		\begin{center}   
			\begin{tabular}{|c|c|c|c|c|c|c|c|c|c|c|}
				\hline
				& $f$ & $\dot{f}$ & $\beta$ & $\lambda$ & $\mathcal{A}$ & $\iota$ & $\psi$ & $\phi_{0}$ & SNR & $R$ \\
				\hline
				\multirow{1}{*}{Fig.~\ref{fig:casestudy1}} & 0.002090825 & $6.345725 \times 10^{-17}$ & 0.037617 & 4.810014 & $3.041765 \times 10^{-22}$ & 1.011543 & 0.249035 & 3.182173 & 139.5 & 0.9998 \\
				\hline
				\multirow{2}{*}{Fig.~\ref{fig:casestudy4}} & 0.002090858 & $4.438414 \times 10^{-17}$ & 0.028935 & 4.508666 & $2.389844 \times 10^{-22}$ & 1.027265 & 0.271220 & 4.485963 & 107.8 & 0.3976 \\
				\cline{2-11}
				 & 0.002090857 & $7.105742 \times 10^{-17}$ & 0.006471 & 5.117904 & $2.474984 \times 10^{-22}$ & 1.036071 & 0.587262 & 0.837439 & 111.6 & 0.4128 \\
				\hline
				\multirow{2}{*}{\makecell{Fig.~\ref{fig:casestudy4} \\\textcolor{red}{multi-source fitting}}} & 0.002090822 & $1.417738 \times 10^{-16}$ & 0.035298 & 4.806297 & $3.038836 \times 10^{-22}$ & 1.016045 & 0.259410 & 3.350101 & 138.8 & 0.9970 \\
				\cline{2-11}
				 & 0.002090884 & $1.992336 \times 10^{-16}$ & 0.037537 & 4.804885 & $3.121834 \times 10^{-22}$ & 1.046375 & 0.247762 & 3.509833 & 137.8 & 0.9925 \\
				\hline
				\multirow{2}{*}{Fig.~\ref{fig:casestudy3}} & 0.002090829 & $-2.099675 \times 10^{-17}$ & 0.058268 & 4.805761 & $3.129668 \times 10^{-22}$ & 1.031005 & 0.219305 & 2.782279 & 140.4 & 0.9847 \\
				\cline{2-11}
				 & 0.002090904 & $2.530915 \times 10^{-17}$ & 0.028884 & 4.809350 & $3.180976 \times 10^{-22}$ & 1.051616 & 0.222846 & 3.364739 & 139.5 & 0.9927 \\
				\hline
			\end{tabular}
			\caption{The parameters of reported sources identified by our method in the case study.} 
			\label{table:4} 
		\end{center}   
	\end{table*}

	\bibliography{Reference}
	
\end{document}